\documentclass[reprint, twocolumn, superscriptaddress]{revtex4-2}

\usepackage[utf8]{inputenc}
\usepackage[ngerman, english]{babel}
\usepackage[T1]{fontenc}

\usepackage{amsthm}
\usepackage{graphicx, color}
\usepackage{amsmath}
\usepackage{amssymb}
\usepackage{mathtools}
\usepackage{caption}

\usepackage{tikz}
\usetikzlibrary{quantikz2}

\usepackage{adjustbox}

\usepackage{url}
\usepackage[colorlinks=false, linkbordercolor=blue]{hyperref}
\usepackage{orcidlink}

\addto\captionsenglish{}

\begin{document}

\title{A Quantum Algorithm for Solving the Poisson Equation for Free Field Conditions via the Hockney Method}

\author{Hans A. K\"osel\orcidlink{0009-0005-7138-996X}}

\email[corresponding author -- e-mail address: ]{hans.koesel@dlr.de}
\affiliation{Institute of Aerodynamics and Flow Technology, German Aerospace Center (DLR), Lilienthalplatz 7, 38108 Braunschweig, Germany}

\author{Roland Ewert\orcidlink{0009-0004-4331-041X}}
\affiliation{Institute of Aerodynamics and Flow Technology, German Aerospace Center (DLR), Lilienthalplatz 7, 38108 Braunschweig, Germany}

\author{Jan W. Delfs\orcidlink{0000-0001-8893-1747}}
\affiliation{Institute of Aerodynamics and Flow Technology, German Aerospace Center (DLR), Lilienthalplatz 7, 38108 Braunschweig, Germany}

\date{\today}

\begin{abstract}

For the often encountered problem of the Poisson equation, this work presents a quantum algorithm solving it based on the quantum Fourier transform (QFT) for periodic boundary conditions as well as free field conditions, where the latter is realized via the Hockney method. 
Besides the QFT and an initialization procedure for amplitude encoding, the algorithm just uses a procedure for multiplying the state vector by a diagonal matrix w.r.t. amplitude encoding. 
For the latter, two alternative implementations are considered here. 
The first variant is a version of the LCU method and the second is a sequence of multi-controlled rotation gates that represents a factoring of the multiplied values into absolute values and complex phase factors. 
The functionality of the algorithm is verified via comparing the results obtained from state vector simulations for one- and two-dimensional test examples with their analytical solutions. 
For the considered test examples, it is found that the success probability for obtaining the desired ancilla qubit subspace in the LCU version is a factor of around two higher than that for the sequence of multi-controlled rotation gates. 
However, the LCU version requires a number of ancilla qubits up to the number of qubits that is set to store the discretized source term of the Poisson equation in amplitude encoding, whereas the sequence of multi-controlled rotation gates demands only one ancilla qubit. 
Computations of the success probabilities for both variants furthermore indicate that the success probability converges for a specific problem with increasing resolution. 
Concerning the required computational resources for the quantum algorithm, the conclusion is drawn that while the QFT is a more efficient procedure than its classical counterpart, the current implementations of the other necessary steps in the algorithm diminish the efficiency w.r.t. the runtime.

\end{abstract}

\maketitle

\section{Introduction}

The Poisson equation reads
\begin{align}
 {{\underline{\nabla}}^2} {\varphi}({\underline{x}}) &= S({\underline{x}})  
\label{eq:Poisson_eq}
\end{align}
for a scalar solution function $ {\varphi} $ and a given scalar function $ S $ w.r.t. a vector of Cartesian coordinates $ \underline{x} $, for which $ \underline{\nabla} $ denotes the gradient operator. 
It is an often encountered problem in science and engineering and here, the perspective regarding computational fluid dynamics (CFD) is taken. 
E.g., the solving of the Navier-Stokes equations (NSEs), by which the dynamics of fluids can be described, often features a Poisson equation for the special case of an incompressible fluid. 
The mass equation and the momentum equation of the NSEs for an incompressible flow are often solved thereby that derived equations are solved, involving in general a transport equation that is accompanied by a Poisson equation. 
E.g., for keeping the momentum equation, a Poisson equation can be derived for the pressure that is needed in it or alternatively, for considering the vorticity, a linear advection-diffusion equation can be derived as the transport equation for it and in two dimensions, the velocity required in it can be obtained from a Poisson equation for a stream function giving the $z$-component of a vector potential for the velocity. 
Such a system of differential equations can be solved numerically by alternately solving the two equations in the framework of a time-marching procedure.

The use of the discrete Fourier transform (DFT) for numerically realizing a solving of the Poisson equation via the Fourier transform (FT) is usually a relatively efficient approach \cite{Steijl_and_Barakos_Q_algs_for_CFD, Steijl_Q_algorithms_for_fluid_simulations}, however, its applicability is in principle limited to problems that have periodic boundary conditions (BCs). 
The Hockney method \cite{Hockney_and_Eastwood_Computer_simulation_Book} extends this procedure to the situation without boundaries, i.e. free space -- a situation that can be of interest for many problems: 
E.g. for problems in computational aeroacoustics (CAA) that involve only low Mach numbers, where the consideration of an incompressible medium yields usually sufficiently reasonable results for the fluctuations in flows due to turbulence, such free field BCs alone can be set for studying the pressure fluctuations on a wall in a turbulent boundary layer \cite{Hu_et_al_Simulation_of_turbulent_boundary_layer_wall_pressure_fluctuations}. 
But moreover, more complex setups can often be represented as a free field problem. 
E.g. for the presence of objects, where $ {\varphi}({\underline{x}}) $ is defined only for the space without the objects but with BCs on the surfaces of the objects, an approach for linear differential equations in general is to multiply the solution field $ {\varphi}({\underline{x}}) $ by a Heaviside distribution $ \theta $ of a field $ f({\underline{x}}) $ that is smaller than zero inside objects, equal to zero on the surface of objects and greater than zero outside of objects, defining the field $ {\tilde{\varphi}} = {\theta}(f({\underline{x}})) \cdot {\varphi}({\underline{x}}) $, which is set to be zero inside the objects and represents a field in free space. 
This procedure is common mostly for differential equations with additional terms like the wave equation \cite{Delfs_Lecture_notes} but can be adopted also for the Poisson equation. 
The consideration of $ {{\underline{\nabla}}^2} {\tilde{\varphi}}({\underline{x}}) $ leads then to a Poisson equation for $ {\tilde{\varphi}} $ according to $ {{\underline{\nabla}}^2} {\tilde{\varphi}}({\underline{x}}) = {\tilde{Q}} $, in which the {\emph{source}} field, i.e. the given field $ {\tilde{Q}} $ on the right side, is an expression of the quantities from the original formulation ({\hyperref[eq:Poisson_eq]{\ref*{eq:Poisson_eq}}}), i.e. $ Q $ and the expressions for the BCs w.r.t. $ \varphi $. 
Further, the defining relation of $ \varphi $ and $ {\tilde{\varphi}} $ can be inserted into these expressions for the BCs in $ \tilde{Q} $, allowing to compute the solution $ \tilde{\varphi} $ via a Picard iteration, where however, a Poisson equation has to be solved even more times. 
However, the solving of a Poisson equation remains in general a relatively time-consuming step in classical simulations \cite{Steijl_and_Barakos_Q_algs_for_CFD, Steijl_Q_algorithms_for_fluid_simulations, Wang_et_al_Q_alg_for_Poisson_eq}.

Also for CFD, quantum computing (QC) is currently considered as a potential technology to achieve more efficient algorithms \cite{Succi_et_al_Review}. 
In particular, the quantum Fourier transform (QFT), which implements a DFT w.r.t. the state vector of the qubit system, is a QC procedure for which a lower number of computational operations has been found compared to its currently best classically implemented counterpart, which is the fast Fourier transform (FFT) \cite{Nielsen_and_Chuang_QC_Book, Pfeffer_Multi-dim_QFT_arxiv_v1}: 
While this measure for the computational effort scales like $ {N} \log_{2}({N}) $ for the FFT w.r.t. the application to $ N = {2^n} $, $ n \in \mathbb{N} $ values, the QFT as a building block, i.e. without the efforts for initializing and reading out the quantum state, has just a scaling like $ {( \log_{2}({N}) )}^2 $.

A quantum algorithm that was proposed for solving specifically the Poisson equation is the work by Y. Cao et al. \cite{Cao_et_al_Q_alg_for_Poisson_eq}, which was continued by S. Wang et al. \cite{Wang_et_al_Q_alg_for_Poisson_eq}. 
These works resort to the representation of the equation via finite differences with Dirichlet BCs and use the HHL algorithm \cite{Harrow_et_al_HHL-alg} (which also contains the QFT as a building block) for solving the resulting system of linear equations. 
Similarly, the HHL algorithm was also used in the work of M. Mandelt Buxad\'{e} et al. \cite{Mandelt_Buxade_et_al_Hybrid_Newton_method_arxiv_v1}, where a Poisson equation with a source term that contains the solution function itself is considered as an exemplary test case. 
Another method by which the Poisson equation can be treated is the lattice Boltzmann method (LBM). 
This has been done by L. Budinski \cite{Budinski_Q_alg_by_streamfunction_vorticity_LBM} in the simulation of a quantum algorithm for the LBM, which was applied to the more complex test setup of a lid-driven cavity, using the stream function-vorticity formulation.

The present work from us presents a quantum algorithm for solving the Poisson equation for multiple dimensions with periodic BCs or free field conditions directly via DFT and a matrix-vector multiplication. 
The approach presented here can therefore be regarded as an extension of the works of R. Steijl and G. N. Barakos \cite{Steijl_and_Barakos_Q_algs_for_CFD, Steijl_Q_algorithms_for_fluid_simulations}, in which the QFT was considered for doing the DFT step in the solving of a Poisson equation for a periodic domain in the context of a vortex-in-cell method. 
Concerning the numerical method used here to implement a situation different from periodic BCs, which is based on a doubling of the computational domain for which the source field is given, we note furthermore that the works \cite{Over_et_al_Q_alg_for_AD_eq} and \cite{Bengoechea_et_al_Q_algs_BCs} employ a similar technique to realize Dirichlet and Neumann BCs for the linear advection-diffusion equation and the heat equation.

During the writing process, we also became aware of the works \cite{Steijl_VKI_lecture_notes_2026} and \cite{Dewitte_et_al_QFT_harmonic_balance_solver}. 
In these works, similar quantum circuits as the variant with the sequence of multi-controlled rotation gates inspected here are applied. 
However, in these works, other equations and just one dimension are considered. 
In \cite{Steijl_VKI_lecture_notes_2026}, R. Steijl explains how such a circuit can be used to implement the solving of the linear advection equation and the heat equation for one time step and in \cite{Dewitte_et_al_QFT_harmonic_balance_solver}, L. Dewitte et. al use such a circuit to solve the Burgers equation for temporally periodic BCs, where they discuss even extensions for such a circuit, like e.g. procedures for inferring the complex phases of the amplitudes of the resulting state vector.

Our paper has the following structure: 
At first, the numerical method is reviewed in section {\hyperref[sec:Method]{{\hyperref[sec:Method]{\ref*{sec:Method}}} -- \emph{Method}}}. 
Then, section {\hyperref[sec:Q_alg]{{\hyperref[sec:Q_alg]{\ref*{sec:Q_alg}}} -- \emph{Quantum Algorithm}}} presents its specific realization as a quantum algorithm proposed here, discussing alternative implementations for the individual building blocks. 
To verify the functionality of the quantum algorithm, it was simulated by means of IBM's quantum computing simulation framework {\emph{Qiskit}} \cite{qiskit} for one- and two-dimensional test problems, which is documented and discussed in section {\hyperref[sec:Functionality_tests]{{\hyperref[sec:Functionality_tests]{\ref*{sec:Functionality_tests}}} -- \emph{Functionality Tests}}} via a selection of results. 
The calculation of the analytical solutions to which the results obtained from the simulated state vector are compared as well as complementary results are outsourced to the appendix {\hyperref[App_sec:Additional_material_for_functionality_tests]{\ref*{App_sec:Additional_material_for_functionality_tests}}}. 
Furthermore, section {\hyperref[sec:Estimations_of_required_computational_resources]{{\hyperref[sec:Estimations_of_required_computational_resources]{\ref*{sec:Estimations_of_required_computational_resources}}} -- \emph{Estimations of Required Computational Resources}}} gives a discussion of the computational effort that is expected for the quantum algorithm. 
At last, the section {\hyperref[sec:Conclusion]{{\hyperref[sec:Conclusion]{\ref*{sec:Conclusion}}} -- \emph{Conclusion}}} gives a summary of our work and perspectives for potential next steps.

\section{Method}
\label{sec:Method}

This section explains the numerical procedure considered in this work for solving the Poisson equation for periodic BCs or free space, starting with the analytical formulation:

\subsection{Analytical Consideration}
\label{subsec:Analytical_consideration}

For the Poisson equation according to
\begin{align}
{{\underline{\nabla}}^2} {\varphi}({\underline{x}}) &= {{S}}({\underline{x}}) ,
\end{align}
where BCs shall not yet be specified, a solution can in principle be directly obtained by applying a FT to the source field $ S $, dividing the transformed source field by the value resulting from the FT of $ {{\underline{\nabla}}^2} $ and transforming back. 
Using here a definition of the one-dimensional FT $ {{\mathcal{F}}}[f] $ w.r.t. a function $ f(x) $ with an exponential factor $ {e^{{\text{i}} {{k}} {x} }} $, for which the correspondence
\begin{align}
{\frac{\partial}{{\partial}{x} }} & \overset{FT}{\longrightarrow}  -{\text{i}} {{k}}
 \label{eq:Def_FT_relation_derivative}
\end{align}
follows, this reads
\begin{align}
 {{\underline{\nabla}}^2} {\varphi}({\underline{x}}) = {{S}}({\underline{x}}) 
 ~~ & \Rightarrow ~~
 -{{\underline{k}}^2} {\hat{\varphi}}({\underline{k}}) = {\hat{{S}}}({\underline{k}}) 
 \label{eq:step1_Poisson_eq_FT_applied} \\
 & \Rightarrow ~~ {\hat{\varphi}}({\underline{k}}) = -{ \frac{ {\hat{{S}}}({\underline{k}}) }{ {{\underline{k}}^2} }}  
 \label{eq:step2_solution_varphi_FT_applied} \\
 & \Rightarrow ~~ {\varphi}({\underline{x}}) = {\mathcal{F}^{-1}}{\biggl[ -{ \frac{ {\hat{{S}}}({\underline{k}}) }{ {{\underline{k}}^2} }} \biggr]}{({\underline{x}})} 
 \label{eq:step3_solution_varphi_via_inverse_FT} 
 ,
\end{align}
where a FT of a function that depends on multiple variables means that a one-dimensional FT has to be performed for all variables, respectively. 
Specifically, the definition of the one-dimensional FT $ {{\mathcal{F}}_{ent}} $ for the entire space and its inverse $ {{\mathcal{F}}^{-1}_{ent}} $ according to
\begin{alignat}{2} 
 {\hat{f}}({{k}}) 
 &= {{\mathcal{F}}_{ent}[ {f} ]({{k}})} 
 & &:= {\int_{-{\infty}}^{+{\infty}}} {f({x})} {e^{{\text{i}} {{k}} {x} }} ~{d{x}} \label{eq:Def_FT_ent} 
 \\[0.15cm]
 \Leftrightarrow  ~~
 f(x) &= 
 {{\mathcal{F}}_{ent}^{-1}[ {\hat{f}} ]({x})} & &:= {\frac{1}{2{\pi}}} {\int_{-{\infty}}^{+{\infty}}} {{\hat{f}}({{k}})} {e^{-{\text{i}} {{k}} {x}}} ~{d{{k}}}  \label{eq:Def_FT_ent_inverse} 
\end{alignat}
and the definition of the one-dimensional FT $ {{\mathcal{F}}_{per}} $ for periodic functions and its inverse $ {{\mathcal{F}}^{-1}_{per}} $ according to
\begin{alignat}{2} 
 {\hat{f}}({{k}})
 &= {{\mathcal{F}}_{per}[ {f} ]({{k}})} 
 & &:= {\int_{{x_0}}^{{x_0} + L}} {f({x})} {e^{{\text{i}} {{k}} {x} }} ~{d{x}} \label{eq:Def_FT_per} 
 \\[0.15cm]
 \Leftrightarrow  ~~
 f(x) &= 
 {{\mathcal{F}}_{per}^{-1}[ {\hat{f}} ]({x})} & &:= {\frac{1}{L}} {\sum_{{k}}} 
 {{\hat{f}}({{k}})} 
 {e^{-{\text{i}} {{k}} {x}}}   \label{eq:Def_FT_per_inverse} 
\end{alignat}
w.r.t. the interval $ [{x_0}, {x_0} + L] $, implying the discretized $ {k} $-values
\begin{align}
 {{k}} &= q \cdot {\Delta}{k} , \quad {\Delta}{k} = {\frac{2{\pi}}{L}} , \quad q \in \mathbb{Z} ,
 \label{eq:Def_FT_per_discretized_k-values}
\end{align}
are used here.

Since it holds
\begin{align}
{{\mathcal{F}}^{-1}}{ \left[ {\hat{G}}({\underline{k}}) \cdot {\hat{S}}({\underline{k}}) \right] }({\underline{x}}) &= 
 {\int}
 G({\underline{x}}, {\underline{\xi}}) \cdot S({\underline{\xi}}) ~ dV({\underline{\xi}})
 \label{eq:Relation_Convolution_and_inverse_FT_of_product}
\end{align}
for the FT, where $ G $ has the form $ G({\underline{x}}, {\underline{\xi}}) = G({\underline{x}} - {\underline{\xi}}) = G({{x_1} - {{\xi}_1}}, {{x_2} - {{\xi}_2}}, \dots) $ and $ dV({\underline{\xi}}) $ means the domain for which the functions $ S $ and $ G $ are defined, the evaluation of eq. ({\hyperref[eq:step3_solution_varphi_via_inverse_FT]{\ref*{eq:step3_solution_varphi_via_inverse_FT}}}) is equivalent to the calculation of $ {\varphi}({\underline{x}}) $ via the Green function of the Poisson equation, defined via the solution for a $ {\delta} $-distribution as the source, i.e.:
\begin{align}
 {{\underline{\nabla}}^2} G({\underline{x}}, {\underline{\xi}}) &= {\delta}({\underline{x}} -  {\underline{\xi}}) 
 \label{eq:Def_GF_of_Poisson_eq}
\end{align}
Accordingly, for the consideration of the full space or a bounded space with periodic BCs, $ -{\frac{1}{{\underline{k}}^2}} $ can be identified as the FT of the Green function in spatial representation w.r.t. $ {\underline{x}} - {\underline{\xi}} $. 
In general, i.e. for arbitrary BCs, the solution can be calculated in principle by
\begin{align}
 {\varphi}({\underline{x}}) = 
 {\int}
 G({\underline{x}}, {\underline{\xi}}) \cdot S({\underline{\xi}}) ~ dV({\underline{\xi}})  
\label{eq:Def_solution_varphi_via_GF_in_general}
\end{align}
if a Green function that satisfies these BCs is determined.

For periodic BCs w.r.t. the domain $ [ {x_{i,0}}, {x_{i,0}} + {L_i} ] $ of the Cartesian direction $ x_i $ according to
\begin{align}
 & {\varphi}({x_1}, {x_2}, \dots, {x_{i,0}}, \dots) \nonumber \\
 & \quad \overset{!}{=} {\varphi}({x_1}, {x_2}, \dots, {x_{i,0}} + {L_i}, \dots) , \\[0.3cm]
 & \left[{\frac{\partial}{{\partial}{x_i}}}{\varphi} \right] ({x_1}, {x_2}, \dots, {x_{i,0}}, \dots)  \nonumber \\
 & \quad \overset{!}{=} \left[{\frac{\partial}{{\partial}{x_i}}}{\varphi} \right] ({x_1}, {x_2}, \dots, {x_{i,0}} + {L_i}, \dots)
\end{align}
for all $ i $, it is to note that integrating the Poisson equation over the domain and applying the fundamental theorem of calculus implies that the Poisson equation has only then a solution if the integral of the source field over the domain is zero:
\begin{align} 
 {\overline{S}} := {\int} S({\underline{x}}) ~ dV({\underline{x}}) & \overset{!}{=} 0
 \label{eq:Per_BCs_requirement_source_integral_over_domain_zero}
\end{align}
Therefore, the defining equation for the Green function ({\hyperref[eq:Def_GF_of_Poisson_eq]{\ref*{eq:Def_GF_of_Poisson_eq}}}) has to be modified for periodic BCs according to eq. (7) in \cite{Marshall_Periodic_Green_functions}
\begin{align}
 {{\underline{\nabla}}^2} 
 {G_{per}}({\underline{x}}, {\underline{\xi}}) 
 &= {\delta}({\underline{x}} -  {\underline{\xi}}) - {{\prod}_{i}} {\frac{1}{ L_i }} .
 \label{eq:Def_GF_of_Poisson_eq_for_per_BCs}
\end{align}
The Green functions of the Poisson equation for periodic BCs in one to three dimensions are given in \cite{Marshall_Periodic_Green_functions} and the free field Green functions for these dimensionalities can be found e.g. in appendix B of \cite{Delfs_Lecture_notes}.

For a source function $ S(x) $ that is approximately zero outside of a certain spatial interval $ [ {x_0}, {x_0} + L ] $, the spatial integrals $ \int_{-{\infty}}^{+{\infty}} $ in the consideration of the free field problem according to ({\hyperref[eq:Def_FT_ent]{\ref*{eq:Def_FT_ent}}}) and ({\hyperref[eq:Def_solution_varphi_via_GF_in_general]{\ref*{eq:Def_solution_varphi_via_GF_in_general}}}) can be approximated by the integral over this finite domain $ \int_{{x_0}}^{{x_0} + L} $, like for the problem with periodic BCs, and the solution $ {\varphi}(x) $ is given via
\begin{align}
 {\varphi}(x) &= {\int_{{x_0}}^{{x_0} + L}} G(x - {\xi}) \cdot S({\xi}) ~ d{\xi} .
 \label{eq:Def_solution_varphi_via_GF_finite_interval}
\end{align}
If the solution is of interest only at those positions for which the source $ S $ does not vanish, the consideration can be restricted as well only to this interval for $ x $, i.e. it is considered just $ {\xi}, x  \in [ {x_0}, {x_0} + L ] $. 
However, according to formula ({\hyperref[eq:Def_solution_varphi_via_GF_finite_interval]{\ref*{eq:Def_solution_varphi_via_GF_finite_interval}}}), $ {\varphi}(x) $ is therefore not only determined by $ S $ w.r.t. this interval of length $ L $ but also by the information about $ G $ over the entire resulting value range for $  x- {\xi} $, which is $ [-L, L] $. 
Since the Green function for periodic BCs $ G_{per} $ itself is a solution of a problem with period $ L $, the information about $ G_{per} $ is already completely contained in an interval of length $ L $ but for the Green function $ G_{ent} $ of a problem that is referred to the entire space, this does in general not apply. 
For the free field problem, therefore, the information about $ G $ for the entire interval $ [-L, L] $ is needed, which has to be kept in mind for the numerical solving via the DFT:

\subsection{Numerical Realization}
\label{subsec:Numerical_realization}

For a numerical treatment with discretized variables, an integral over a finite interval, like for the FT of periodic functions or functions that vanish outside of an interval, can be approximated by a sum over $ N \in \mathbb{N} $ sample function values. 
For a FT, this can be done as
\begin{align}
{\int_{ { {{x}}_{0} } }^{ {{x}}_{0} + L } }
 {e^{ {\text{i}} {k} x }} {f(x)} ~ dx 
 & \approx  {\sum_{j=0}^{N-1}} {e^{ {\text{i}} {k} ( {{\tilde{x}}_{0}} + j{\Delta}x ) }} {f_j} ~ {\Delta}x 
 \label{eq:FT_integral_over_finite_interval_approximated_as_sum}
 , \\
 {f_j} &:= {f( {{\tilde{x}}_{0}} + j{\Delta}x )} , ~ {\Delta}x = {\frac{L}{N}} , \nonumber
\end{align}
where it is used here $ {{\tilde{x}}_{0}} := {{{x}}_{0}} + {\frac{{\Delta}x}{2}} $ in order to set the sample positions as the centers of the $ N $ subintervals into which the interval $ [{x_0}, {x_0} + L] $ is divided. 
However, the solving procedure according to the three steps in eqs. ({\hyperref[eq:step1_Poisson_eq_FT_applied]{\ref*{eq:step1_Poisson_eq_FT_applied}}}) - ({\hyperref[eq:step3_solution_varphi_via_inverse_FT]{\ref*{eq:step3_solution_varphi_via_inverse_FT}}}) involves also the inverse FT. 
To be able to numerically treat it as well via a finite number of values, it is considered the modified problem that $ f(x) $ as well as $ {\hat{f}}(k) $ shall be periodic, since the condition of periodicity $ f(x) = f(x + L) $ restricts the $ k $-values to $ k = q \cdot {\Delta}k, {\Delta}k = {\frac{2{\pi}}{L}} , q \in \mathbb{Z} $ (cf. ({\hyperref[eq:Def_FT_per_discretized_k-values]{\ref*{eq:Def_FT_per_discretized_k-values}}})) and vice versa, $ {\hat{f}}(k) = {\hat{f}}(k + N \cdot {\Delta}k) $ implies $ x = j \cdot {\Delta}x, {\Delta}x = {\frac{2{\pi}}{N {\Delta}k}} , j \in \mathbb{Z} $. 
The FT for periodic functions ({\hyperref[eq:Def_FT_per]{\ref*{eq:Def_FT_per}}}), ({\hyperref[eq:Def_FT_per_inverse]{\ref*{eq:Def_FT_per_inverse}}}), applied to this discretized problem thus reads
\begin{align}
 {{\hat{f}}_q} = {\frac{L}{{N}}} {\sum_{j = 0}^{N - 1}} ~ {e^{{\text{i}} 2 {\pi} {\frac{jq}{N}} }} {f_j} ,  
 \quad 
 {{{f}}_j} = {\frac{1}{L}} {\sum_{q = 0}^{N - 1}} ~ {e^{-{\text{i}} 2 {\pi} {\frac{jq}{N}} }} {{\hat{f}}_q} . 
 \label{eq:solution_approach_resulting_form_of_DFT}
\end{align}
This just corresponds to the DFT, for which here the convention as a unitary matrix $ {U_{DFT}} $ according to
\begin{alignat}{2} 
 {{\hat{f}}_q} &= {{\left[ {U_{DFT}} \cdot {\underline{f}} \right]}_{q}} & &:= {\frac{1}{\sqrt{N}}} {\sum_{j = 0}^{N - 1}} ~ {e^{{\text{i}} 2 {\pi} {\frac{jq}{N}} }} {f_j}
 \label{eq:Def_DFT} 
 \\[0.15cm]
 \Leftrightarrow  ~~
 {{{f}}_j} &= {{\left[ {U^{-1}_{DFT}} \cdot {\underline{\hat{f}}} \right]}_{j}} & &:=  {\frac{1}{\sqrt{N}}} {\sum_{q = 0}^{N - 1}} ~ {e^{-{\text{i}} 2 {\pi} {\frac{jq}{N}} }} {{\hat{f}}_q} 
 \label{eq:Def_DFT_inverse} 
\end{alignat}
is used. 
In comparison to the DFT definition ({\hyperref[eq:Def_DFT]{\ref*{eq:Def_DFT}}}), ({\hyperref[eq:Def_DFT_inverse]{\ref*{eq:Def_DFT_inverse}}}), the forward transform in ({\hyperref[eq:solution_approach_resulting_form_of_DFT]{\ref*{eq:solution_approach_resulting_form_of_DFT}}}) is multiplied by the factor $ {\frac{L}{\sqrt{N}}} $ and the inverse transform is divided by this factor. 
Concerning the actual expression ({\hyperref[eq:FT_integral_over_finite_interval_approximated_as_sum]{\ref*{eq:FT_integral_over_finite_interval_approximated_as_sum}}}), shifts in the variables like $ {{\tilde{x}}_{0}} $ can be incorporated in the transforms in the framework of a redefinition according to $ {{\hat{f}}_{q}} \longrightarrow {e^{-{\text{i}} k {{\tilde{x}}_{0}} }} {{\hat{f}}_{q}}  $. 
While a different factor like $ {\frac{L}{\sqrt{N}}} $ has to be kept in mind for the numerical implementation of the discussed solving approach with use of the DFT of the Green function, this is not the case for a shifting factor like $ {e^{-{\text{i}} k {{\tilde{x}}_{0}} }} $ since it results just from the FT of the source function $ S $ w.r.t. $ [{x_0}, {x_0} + L] $ but not from the FT of the Green function w.r.t. $ [-L, L] $, so that this factor effectively cancels out via the inverse FT at the end.

The DFT is defined for a sequence of values that is extended periodically. 
However, for $ N \longrightarrow \infty $ while $ L $ is kept constant, the sum for $ {{\hat{f}}_q} $ in ({\hyperref[eq:solution_approach_resulting_form_of_DFT]{\ref*{eq:solution_approach_resulting_form_of_DFT}}}) assumes the integral representation ({\hyperref[eq:Def_FT_per]{\ref*{eq:Def_FT_per}}}) (for $ {x_0} = 0 $) since $ {\Delta}x = {\frac{L}{N}} \longrightarrow dx $ and the extension of the $ k $-interval goes to $ N \cdot {\Delta}k \longrightarrow \infty $. 
If it is then further considered $ {x_0} \longrightarrow -{\infty}, {x_0} + L \longrightarrow \infty $ while $ N $ is kept constant, the sum for $ f_j $ in ({\hyperref[eq:solution_approach_resulting_form_of_DFT]{\ref*{eq:solution_approach_resulting_form_of_DFT}}}) assumes the integral representation ({\hyperref[eq:Def_FT_ent_inverse]{\ref*{eq:Def_FT_ent_inverse}}}) since $ {\Delta}k = {\frac{2{\pi}}{({x_0} + L) - {x_0}}} \longrightarrow dk $. 
In the framework of this limit, the DFT can approximate also the analytical FT ({\hyperref[eq:Def_FT_ent]{\ref*{eq:Def_FT_ent}}}), considered for free field problems.

Concerning the numerical representation of the Green functions, it can be encountered that the analytical Green function in spatial representation has a singularity at $ {\underline{x}} - {\underline{\xi}} = {\underline{0}} $, like e.g. the free field Green function for two dimensions, which is given by
\begin{align}
 {G_{ent, 2D}}({\underline{x}} - {\underline{\xi}}) &= {\frac{1}{2{\pi}}}{\ln( |{\underline{x}} - {\underline{\xi}}| )} .
 \label{eq:GF_2D_free_field}
\end{align}
In this case, the Green function has to be regularized, which means that its value at the position of the singularity has to be set to a finite value. 
Here, this value is chosen as zero. 
Furthermore, also the component $ {{\hat{G}}_{{\underline{q}} = {\underline{0}}}} $ of the numerically represented FT of the Green function was set here to zero. 
Referred to the integral representation, this corresponds to setting $ { {\hat{G}}({\underline{k}} = {\underline{0}}) } $ and thus the integral of the Green function w.r.t. the domain considered for it to zero, i.e. formulated for one dimension
\begin{align}
 {{\hat{G}}({k = 0})} 
 &= {\int_{-{L_{GF}}}^{{L_{GF}}}} {G(r)} 
 ~ dr \overset{!}{=} 0   , \label{eq:Integral_over_GF_shall_be_0} \\
 r &= x - {\xi} , \quad {L_{GF}} =
  \left\{ 
 \begin{array}{ll}
 {L}, & G = {G_{ent}} \\
 {\frac{L}{2}}, & G = {G_{per}} 
 \end{array} 
  \right. . \nonumber
\end{align}
Since $ {{\hat{G}}({\underline{k}})} $ is a factor in the FT of the solution $ {{\hat{\varphi}}({\underline{k}})} $ (cf. ({\hyperref[eq:Relation_Convolution_and_inverse_FT_of_product]{\ref*{eq:Relation_Convolution_and_inverse_FT_of_product}}})), thereby also the integral of the computed result over the respectively considered domain is set to zero. 
For the case of periodic BCs, the result computed in this approach gives the solution $ \varphi $ in the domain $ [{x_{1, 0}}, {x_{1, 0}} + {L_1}] \times [{x_{2, 0}}, {x_{2, 0}} + {L_2}] \times \dots $ directly. 
For the free field case, however, it is obtained output for a larger domain that contains the $ \varphi $-values for the interval $ [{x_{1, 0}}, {x_{1, 0}} + {L_1}] \times [{x_{2, 0}}, {x_{2, 0}} + {L_2}] \times \dots $ as a subset (see next paragraph). 
For the situations considered here for the Poisson equation, the solution is only determined up to the addition of an arbitrary constant $ C \in \mathbb{C} $. 
With the choice of zero for the FT component of the computed result for $ {\underline{k}} = {\underline{0}} $, this constant is thus set such that the spatial integral of the computed function over the domain is zero. 
The choice that had to be made here to fix $ C $ should however be of minor importance for applications since $ \varphi $ often just describes a potential and the quantities that are actually of interest result from spatial derivatives of it. 
Concerning this aspect, it shall be noted that according to the relation ({\hyperref[eq:Def_FT_relation_derivative]{\ref*{eq:Def_FT_relation_derivative}}}), this step can be easily incorporated in the explained procedure by multiplying $ {{\hat{S}}_{\underline{q}}} \cdot {{\hat{G}}_{\underline{q}}} $ by the value $ -{\text{i}}{q_r}  {\Delta}{k_r} $ for the desired dimension $ r $.

\begin{figure*}[t]
\begin{center}
\includegraphics[width=\linewidth]{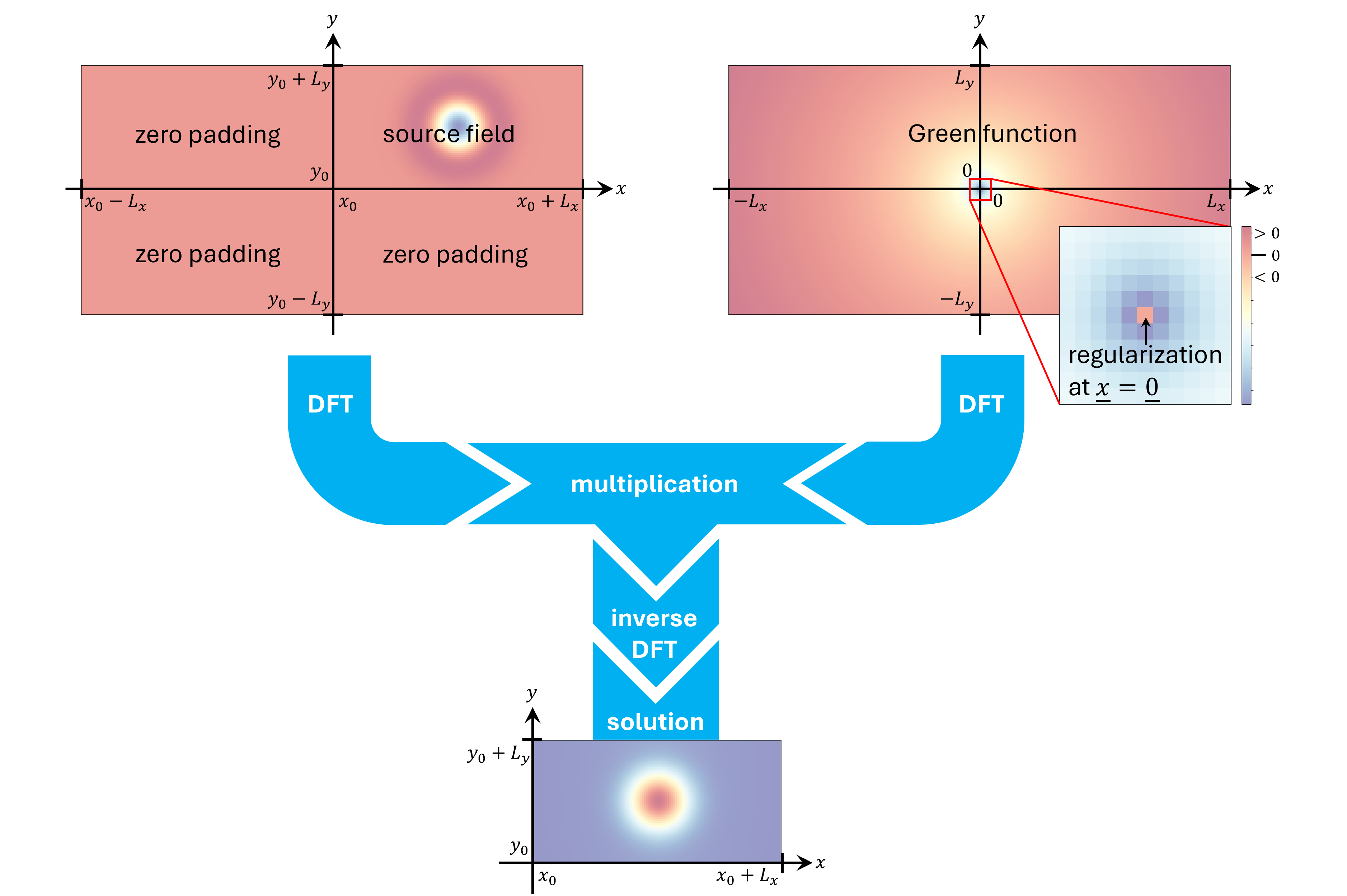}
\end{center}
\caption[]{Illustration of the Hockney method.}
\label{Fig:Hockney-method_scheme}
\end{figure*}
For the case of periodic BCs, where it is also resorted to periodic functions in the analytical consideration, it can be used alternatively the analytical values $ -{\frac{1}{{\underline{k}}^2}} $ instead of the DFT of $ G_{per} $ in spatial representation: 
As mentioned above, the values of the Green function $ G_{per} $ in spatial representation are given already by an interval of length $ L $ w.r.t. a specific dimension $ x $, so that the consideration of the interval for $ x - {\xi} $ in ({\hyperref[eq:Def_solution_varphi_via_GF_finite_interval]{\ref*{eq:Def_solution_varphi_via_GF_finite_interval}}}) can be restricted just to $ [-{\frac{L}{2}}, {\frac{L}{2}}] $. 
Hence, a DFT of $ {G_{per}} $ with $ N $ equidistant sample positions for this interval just gives the sample values $ -{\frac{1}{{{k}}^2}} $ in the $ k $-interval $ [-{\frac{N}{2}}{\Delta}k, {\frac{N}{2}}{\Delta}k ] $ with a resolution that is the same as for the DFT of $ S $ with $ N $ equidistant sample positions for the interval $ [{x_0}, {x_0} + L] $. 
W.r.t. the DFT components $ \{ {{\hat{S}}_{q=0}}, {{\hat{S}}_{q=1}}, \dots , {{\hat{S}}_{q=N-1}} \}  $ with even $ N $, the multiplication has to be done then with the regularized values
\begin{align}
 & \{ { 0 }, {-{\frac{1}{{(1 \cdot {\Delta}k)}^2}}}, {-{\frac{1}{{(2 \cdot {\Delta}k)}^2}}}, \dots , {-{\frac{1}{{({\frac{N}{2}} \cdot {\Delta}k)}^2}}}, \nonumber \\
 & ~~ {-{\frac{1}{{( (-{\frac{N}{2}} + 1) \cdot {\Delta}k)}^2}}} , {-{\frac{1}{{( (-{\frac{N}{2}} + 2) \cdot {\Delta}k)}^2}}}, \dots , \nonumber \\
 & ~~ {-{\frac{1}{{( (-1) \cdot {\Delta}k)}^2}}}  \} , 
 \label{eq:sequence_sampled_analytical_values}
\end{align}
respectively.

Since the DFT is formulated for a setup of finite extension that is periodically continued, the analytical values $ -{\frac{1}{{\underline{k}}^2}} $ do not represent the DFT of the Green function for the free field problem. 
Hence, this DFT has to be computed w.r.t. the full interval $ [-L, L] $. 
Since the DFT components of $ {G_{ent}} $ and $ S $ have to be multiplied value by value, it has to be taken also an interval of the length $ 2L $ for the DFT of $ S $. 
This is simply realized by extending the domain of the source function $ [{x_0}, {x_0} + L] $ to an interval of length $ 2L $, where it is irrelevant how this padding of the original interval into a larger one of extension $ 2L $ is done. 
The function values for $ S $ in the added regions are set to zero since it had to be assumed for the free field situation that the source function vanishes outside of the originally considered domain. 
The described procedure to compute the free field problem is referred to as the {\emph{Hockney method}} and summarized in Fig. {\hyperref[Fig:Hockney-method_scheme]{\ref*{Fig:Hockney-method_scheme}}}.

\section{Quantum Algorithm}
\label{sec:Q_alg}

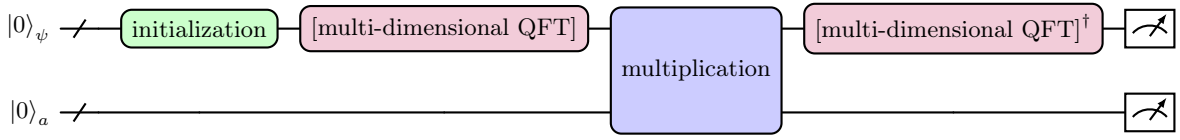
\begin{figure*}[t]
\begin{center}
\tikzset{invisible/.style={fill=none,draw=none,line width=0pt,inner xsep=0pt,inner ysep=0pt}}
\tikzset{transparent/.style={fill=none}}
\begin{quantikz}
  \lstick{$ {{\left| {0} \right\rangle}_{\psi}} $} & \qwbundle{ } &[-0.2cm] 
   \gate[1, style={fill=green!20, rounded corners}]{  \text{initialization}  } &[-0.2cm] 
    \gate[1, style={fill=purple!20, rounded corners}]{ \left[ \text{multi-dimensional QFT}  \right] } &[-0.2cm] \gate[2, style={fill=blue!20, rounded corners}]{  \text{multiplication}  } &[-0.2cm] 
     \gate[1, style={fill=purple!20, rounded corners}]{ {{\left[ \text{multi-dimensional QFT} \right] }^{\dag}} } &[-0.2cm] \meter{} \\
 \lstick{$ {{\left| {0} \right\rangle}_{a}} $} & \qwbundle{ } &  &  &   &   &  \meter{} 
\end{quantikz}
\end{center}
\caption[]{Quantum circuit schematics of the module structure of the quantum algorithm.}
\label{Fig:Q_alg_modules}
\end{figure*}

In correspondence to the three calculation steps according to the eqs. ({\hyperref[eq:step1_Poisson_eq_FT_applied]{\ref*{eq:step1_Poisson_eq_FT_applied}}}) -- ({\hyperref[eq:step3_solution_varphi_via_inverse_FT]{\ref*{eq:step3_solution_varphi_via_inverse_FT}}}), our quantum circuit for realizing the procedure explained in the previous section has the structure depicted in Fig. {\hyperref[Fig:Q_alg_modules]{\ref*{Fig:Q_alg_modules}}}.

The qubit system is grouped into two registers, for which the quantum states are denoted here in general as $ \left| {\psi} \right\rangle $ and $ \left| {a} \right\rangle $, respectively. 
Correspondingly, lines in the circuit diagrams with a slash at the beginning mean that there can be multiple lines, representing the time evolution of individual qubits from left to right, and $ {\left| 0 \right\rangle }_{r} $ means that all qubits of the register $ r \in \{ {\psi}, a \} $ are in the $ \left| 0 \right\rangle $-state, which is set here as the starting state without loss of generality. 
The quantum circuit proposed here resorts to the common amplitude encoding w.r.t. multiple qubits \cite{Nielsen_and_Chuang_QC_Book} to store the information about the discretized source field $ S $ as an initial state in the register $ \psi $, for which an appropriate further processing is enabled by the so-called {\emph{ancilla qubits}}, forming the register $ a $. 
This initial state is prepared via the building block called 'initialization'. 
Following that, the QFT is used to implement the DFT ({\hyperref[eq:Def_DFT]{\ref*{eq:Def_DFT}}}) of the source field in the amplitude encoding scheme, where this operation is labeled in Fig. {\hyperref[Fig:Q_alg_modules]{\ref*{Fig:Q_alg_modules}}} with '$ [ \text{multi-dimensional QFT} ] $'. 
Between this DFT and the inverse DFT, given by the operation $ {{[ \text{multi-dimensional QFT} ]}^{\dag}} $, where $ \dag $ indicates Hermitian conjugation, the module 'multiplication' mediates the multiplication of the components of the transformed source $ {\hat{S}}_{\underline{q}} $ by the components of the DFT of the Green function $ {\hat{G}}_{\underline{q}} $ or the analytical values ({\hyperref[eq:sequence_sampled_analytical_values]{\ref*{eq:sequence_sampled_analytical_values}}}). 
So, the specific realization of the quantum circuit is set up based on the values of the discretized source field in spatial representation $ {{S}}_{\underline{j}} $ as a first input data field and the values of the DFT of the Green function $ {\hat{G}}_{\underline{q}} $ or the values ({\hyperref[eq:sequence_sampled_analytical_values]{\ref*{eq:sequence_sampled_analytical_values}}}) as a second input. 
In the state resulting before the measurements of the qubits, the bitstrings formed by the combinations of the states $ \left| 0 \right\rangle $ and $ \left| 1 \right\rangle $ of the qubits in the $ \psi $-register encode the individual grid points just like for the initial state but in contrast, the associated amplitudes encode solution values only provided that the ancilla qubits are measured in a specific configuration. 
More details and specific implementations for the individual building blocks are given in the following subsections:

\subsection{Initialization}
\label{subsec:Initialization}

\begin{figure*}[t]
\begin{center}
\tikzset{invisible/.style={fill=none,draw=none,line width=0pt,inner xsep=0pt,inner ysep=0pt}}
\tikzset{transparent/.style={fill=none}}
\begin{align*}
\begin{quantikz}[align equals at=10]
\lstick{${q_0}$} & \gategroup[1,steps=1, style={invisible}, label style={label position=above,anchor=mid,yshift=-0.4cm, xshift=-0.2cm}]{LSB} &[-0.2cm]  \gate[19, steps=1, style={fill=green!20, inner ysep=-2pt}, background, label style={yshift=2.5cm}]{ \left[ INIT  \right]( {{\underline{S}}_{pad}} ) }   &[-0.2cm]  \\[-0.325cm]
 \lstick{${q_1}$} &  &  &  \\
 \lstick{${q_2}$}   &  &  &  \\
\lstick{$ \vdots ~~~ $} & \wave&&&&\\
 \lstick{${q_{{n_1}-2}}$}  &  &  &  \\
 \lstick{${q_{{n_1}-1}}$}   &  &  &  \\[1cm]
 \lstick{${q_{{n_1}}}$}   &  &  &  \\
 \lstick{${q_{{n_1}+1}}$}   &  &  &  \\
 \lstick{${q_{{n_1}+2}}$}   &  &  &  \\
\lstick{$ \vdots ~~~ $} & \wave&&\\
 \lstick{${q_{{n_1} + {n_2}-2}}$}   &  &  &  \\
 \lstick{${q_{{n_1} + {n_2}-1}}$}   &  &  &  \\[1cm]
\lstick{$ \vdots ~~~ $} & 
 \wave&&&\\[1cm]
\lstick{${q_{{n_1} + \dots + {n_{d-1}}}}$}   &  &  &  \\
 \lstick{${q_{{n_1} + \dots + {n_{d-1}}+1}}$}   &  &  &  \\
 \lstick{${q_{{n_1} + \dots + {n_{d-1}}+2}}$}   &  &  &  \\
\lstick{$ \vdots ~~~ $} & \wave&&\\
 \lstick{${q_{{n_1} + \dots + {n_{d}}-2}}$}   &  &  &  \\[-0.32cm]
 \lstick{${q_{{n_1} + \dots + {n_{d}}-1}}$} & 
 \gategroup[1,steps=1, style={invisible}, label style={label position=below,anchor=mid,yshift=-0.1cm, xshift=-0.2cm}]{MSB} 
  &  & 
\end{quantikz}
 & ~ \widehat{=} ~  
\begin{quantikz}[align equals at=10]
\lstick{${q_0}$} & \gategroup[19,steps=1, style={invisible}, label style={label position=above,anchor=mid,yshift=-0.4cm, xshift=-0.2cm}]{LSB} &[-0.2cm]  \gate[18, style={transparent, inner ysep=-2pt}, label style={yshift=2.25cm}]{ \left[ INIT  \right]( {{\underline{S}}} ) }
\gategroup[18,steps=1, style={line width=0pt, fill=green!20, inner ysep=-5pt, inner xsep=-3pt}, background]{}   &[-0.2cm]   \\[-0.3cm]
 \lstick{${q_1}$} &  &  &  \\
 \lstick{${q_2}$} &  &  &  \\
\lstick{$ \vdots ~~~ $} & \wave&&\\
 \lstick{${q_{{n_1}-2}}$} &  &  &  \\[-0.1cm]
 \lstick{${q_{{n_1}-1}}$}  &  & \linethrough  &  \\[0.89cm]
 \lstick{${q_{{n_1}}}$}  &  &  &  \\
 \lstick{${q_{{n_1}+1}}$}  &  &  &  \\
 \lstick{${q_{{n_1}+2}}$}  &  &  &  \\
\lstick{$ \vdots ~~~ $} & \wave&&\\
 \lstick{${q_{{n_1} + {n_2}-2}}$}  &  &  &  \\[-0.1cm]
 \lstick{${q_{{n_1} + {n_2}-1}}$}  &  & \linethrough
  &  \\[0.9cm]
\lstick{$ \vdots ~~~ $} &  \wave&&\\[1cm]
\lstick{${q_{{n_1} + \dots + {n_{d-1}}}}$}  &  &  &  \\
 \lstick{${q_{{n_1} + \dots + {n_{d-1}}+1}}$}  &  &  &  \\
 \lstick{${q_{{n_1} + \dots + {n_{d-1}}+2}}$}  &  &  &  \\
\lstick{$ \vdots ~~~ $} & \wave&&\\[-0.325cm]
 \lstick{${q_{{n_1} + \dots + {n_{d}}-2}}$}  &  &  &  \\[-0.285cm]
 \lstick{${q_{{n_1} + \dots + {n_{d}}-1}}$} & 
 \gategroup[1,steps=1, style={invisible}, label style={label position=below,anchor=mid,yshift=-0.31cm, xshift=-0.2cm}]{MSB} 
  &  &  
\end{quantikz} 
\end{align*}
\end{center}
\captionsetup{justification=raggedright, singlelinecheck=false}
\caption[]{Equivalence of an initialization procedure for its application to a data field $ {\underline{S}}_{pad} $, which explicitly includes a padding of the source field $ \underline{S} $, and the application to the source field $ \underline{S} $ alone. 
The procedure $ [INIT] $ prepares a data field of the form ({\hyperref[eq:discretized_source_field_long_vector]{\ref*{eq:discretized_source_field_long_vector}}}) as a corresponding state vector w.r.t. the involved qubits in amplitude encoding. 
LSB and MSB indicate the least and most significant bit, respectively, and qubit lines that go over the gate block for the initialization mean that the initialization procedure does not act on them. 
If the padding region shall be realized for a specific dimension before the region of the source (as in Fig. {\hyperref[Fig:Hockney-method_scheme]{\ref*{Fig:Hockney-method_scheme}}}), an X-gate has to be applied to the qubit of the subregister for this dimension that is not involved in $ [INIT]({\underline{S}}) $.}
\label{Fig:Initialization_padding}
\end{figure*}
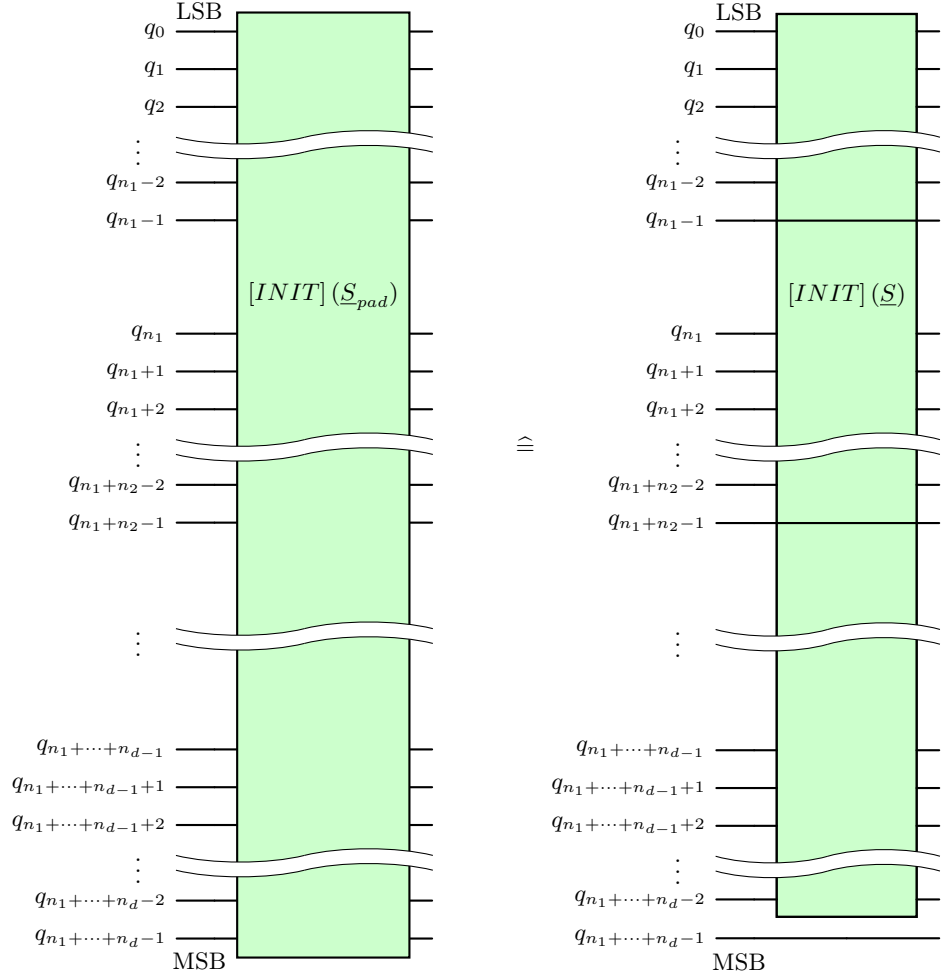

An initialization procedure w.r.t. amplitude encoding, i.e. a procedure for preparing an arbitrary state in this encoding format based on a specific starting state, is given e.g. in the work of V. V. Shende et al. \cite{Shende_et_al_Synthesis_of_q_circuits}.

The input that has to be initialized here as such a state vector is the discretized source field, given by the components $ S_{\underline{j}} $ according to the multi-dimensional case of the DFT ({\hyperref[eq:Def_DFT]{\ref*{eq:Def_DFT}}}), ({\hyperref[eq:Def_DFT_inverse]{\ref*{eq:Def_DFT_inverse}}}). 
The discretized source field is represented here as a long vector of the structure
\begin{align}
 & ( {{S}({x_{1,0}}, {x_{2,0}}, \dots, {x_{d,0}} )}, 
 {{S}({x_{1,1}}, {x_{2,0}}, \dots, {x_{d,0}} )}, \nonumber \\ 
 & ~ {{S}({x_{1,2}}, {x_{2,0}}, \dots, {x_{d,0}} )}, 
 \dots ,  {{S}({x_{1, {{\tilde{n}}_1}-1 }}, {x_{2,0}}, \dots, {x_{d,0}} )}, \nonumber \\
 & ~ {{S}({x_{1,0}}, {x_{2,1}}, \dots, {x_{d,0}} )},  {{S}({x_{1,1}}, {x_{2,1}}, \dots, {x_{d,0}} )}, \nonumber \\
 & ~ {{S}({x_{1,2}}, {x_{2,1}}, \dots, {x_{d,0}} )}, 
 \dots , 
 {{S}({x_{1, {{\tilde{n}}_1}-1 }}, {x_{2,1}}, \dots, {x_{d,0}} )}, \nonumber \\
 & ~ {{S}({x_{1,0}}, {x_{2,2}}, \dots, {x_{d,0}} )}, 
 {{S}({x_{1,1}}, {x_{2,2}}, \dots, {x_{d,0}} )}, \nonumber \\
 & ~ {{S}({x_{1,2}}, {x_{2,2}}, \dots, {x_{d,0}} )}, 
 \dots , 
 {{S}({x_{1, {{\tilde{n}}_1}-1 }}, {x_{2,2}}, \dots, {x_{d,0}} )}, \dots, \nonumber \\
 & ~ {{S}({x_{1,0}}, {x_{2, {{\tilde{n}}_2}-1 }}, \dots, {x_{d,0}} )}, 
 {{S}({x_{1,1}}, {x_{2, {{\tilde{n}}_2}-1 }}, \dots, {x_{d,0}} )}, \nonumber \\
 & ~ {{S}({x_{1,2}}, {x_{2, {{\tilde{n}}_2}-1 }}, \dots, {x_{d,0}} )}, 
 \dots , \nonumber \\
 & ~ {{S}({x_{1, {{\tilde{n}}_1}-1 }}, {x_{2, {{\tilde{n}}_2}-1 }}, \dots, {x_{d,0}} )}, 
 \dots, \nonumber \\
 & ~ {{S}({x_{1,0}}, {x_{2, {{\tilde{n}}_2}-1 }}, \dots, {x_{d, {{\tilde{n}}_d}-1 }} )}, \nonumber \\
 & ~ {{S}({x_{1,1}}, {x_{2, {{\tilde{n}}_2}-1 }}, \dots, {x_{d, {{\tilde{n}}_d}-1 }} )}, \nonumber \\
 & ~ {{S}({x_{1,2}}, {x_{2, {{\tilde{n}}_2}-1 }}, \dots, {x_{d, {{\tilde{n}}_d}-1 }} )}, 
 \dots , \nonumber \\
 & ~ {{S}({x_{1, {{\tilde{n}}_1}-1 }}, {x_{2, {{\tilde{n}}_2}-1 }}, \dots, {x_{d, {{\tilde{n}}_d}-1 }} )}
 {)}^{T}
 \label{eq:discretized_source_field_long_vector}
\end{align}
w.r.t. to the dimensionality $ d \in \mathbb{N} $ and $ {{\tilde{n}}_r} = {2^{{n_r}}} $ with $ {n_r} \in \mathbb{N} $ for all dimensions $ r $. 
This storing pattern corresponds to the ordering of the grid points in the tensor product of the state vectors of subregisters that are set just for the number of points in a specific dimension, respectively:
\begin{align}
 & \begin{pmatrix}
 {{a_{\psi}}({x_{1, 0 }})} \\
 {{a_{\psi}}({x_{1, 1 }})} \\
 \vdots \\
 {{a_{\psi}}({x_{1, {{{\tilde{n}}_1} - 1} }})}
\end{pmatrix}
  \otimes 
  \begin{pmatrix}
 {{a_{\psi}}({x_{2, 0 }})} \\
 {{a_{\psi}}({x_{2, 1 }})} \\
 \vdots \\
 {{a_{\psi}}({x_{2, {{{\tilde{n}}_2} - 1} }})}
\end{pmatrix}
\otimes \dots \nonumber \\ 
  & ~~ \otimes
\begin{pmatrix}
 {{a_{\psi}}({x_{d, {0}}})} \\
 {{a_{\psi}}({x_{d, {1}}})} \\
 \vdots \\
 {{a_{\psi}}({x_{d, {{{\tilde{n}}_d} - 1} }})}
\end{pmatrix}
\end{align}
Here, $ {n_r} $ is the number of qubits in the subregister of $ \psi $ for the dimension $ r $, which generate $ {{\tilde{n}}_r} $ basis states, where the basis state with the label $ {j_r} \in \{ 0, \dots, {{\tilde{n}}_r} - 1 \} $ is associated with the discretization point $ {x_{r, {j_r}}} $ in the respective dimension and has the amplitude $ {a_{\psi}}({x_{r, {j_r}}}) $. 
Since a quantum state vector has to be normalized to one, the normalized version of the actual vector ({\hyperref[eq:discretized_source_field_long_vector]{\ref*{eq:discretized_source_field_long_vector}}}) has to be computed and in terms of a state vector simulation of the circuit, the norm has to be kept in mind in order to multiply it in turn by the amplitudes of the state vector resulting at the end before the measurement operations.

\begin{figure*}[t]
\begin{adjustbox}{width=1.0\textwidth}
\tikzset{invisible/.style={fill=none,draw=none,line width=0pt,inner xsep=0pt,inner ysep=0pt}}
\begin{quantikz}
\lstick{${q_0}$} & \gategroup[1,steps=1, style={invisible}, label style={label position=above,anchor=mid,yshift=-0.3cm, xshift=-0.2cm}]{LSB} &[-0.2cm]  &[-0.2cm]  &[-0.2cm] \ \ldots\  &[-0.2cm]  &[-0.2cm]  &[-0.2cm] \ctrl{5} &[-0.2cm] \ \ldots\  &[-0.2cm] &[-0.2cm]  &[-0.2cm] \ctrl{2} &[-0.2cm] &[-0.2cm] \ctrl{1} &[-0.2cm]  \gate{H} & 
 \swap{5}\gategroup[6, steps=3, style={dashed, rounded corners}, background, label style={label position=below, anchor=north,yshift=-0.2cm}]{{\text{${n_1}$-qubit-SWAP}}} & & 
 & \\
 \lstick{${q_1}$} &  &  &   & \ \ldots\ &  & \ctrl{4} &  &  \ \ldots\  &  & \ctrl{1} &  &  \gate{H}  & \gate{P({\frac{2{\pi}}{2^{2}}})} &  &  & \swap{3} & &  \\
 \lstick{${q_2}$} &  &  &  & \ \ldots\ & \ctrl{3} &  &  &  \ \ldots\  & \gate{H}  & \gate{P({\frac{2{\pi}}{2^{2}}})} &  \gate{P({\frac{2{\pi}}{2^{3}}})} & &  &  & & & \swap{1} & \\
\lstick{$ \vdots ~~~ $} & \wave&&&&&&&&&&&&&&&& &\\
 \lstick{${q_{{n_1}-2}}$} &  &  & \ctrl{1} & \ \ldots\ &  &  &  & \ \ldots\  & &  &  &  &  &  & & \targX{} & &  \\
 \lstick{${q_{{n_1}-1}}$} &  & \gate{H} & \gate{P({\frac{2{\pi}}{2^{2}}})} & \ \ldots\ & \gate{P({\frac{2{\pi}}{2^{{n_1}-2}}})} & \gate{P({\frac{2{\pi}}{2^{{n_1}-1}}})} & \gate{P({\frac{2{\pi}}{2^{{n_1}}}})} & \ \ldots\  &  & & &  &  &  & \targX{} &    & & \\[1cm]
 \lstick{${q_{{n_1}}}$} &  &  &   & \ \ldots\  &  &  & \ctrl{5} & \ \ldots\  & &  & \ctrl{2} & & \ctrl{1} &  \gate{H} & 
 \swap{5}\gategroup[6, steps=3, style={dashed, rounded corners}, background, label style={label position=below, anchor=north,yshift=-0.2cm}]{{\text{${n_2}$-qubit-SWAP}}} & & 
 & \\
 \lstick{${q_{{n_1}+1}}$} &  &  &   & \ \ldots\ &  & \ctrl{4} &  &  \ \ldots\  &  & \ctrl{1} &  &  \gate{H}  & \gate{P({\frac{2{\pi}}{2^{2}}})} &  &  & \swap{3} & &  \\
 \lstick{${q_{{n_1}+2}}$} &  &  &  & \ \ldots\ & \ctrl{3} &  &  &  \ \ldots\  & \gate{H}  & \gate{P({\frac{2{\pi}}{2^{2}}})} &  \gate{P({\frac{2{\pi}}{2^{3}}})} & &  &  & & & \swap{1} & \\
\lstick{$ \vdots ~~~ $} & \wave&&&&&&&&&&&&&&&& &\\
 \lstick{${q_{{n_1} + {n_2}-2}}$} &  &  & \ctrl{1} & \ \ldots\ &  &  &  & \ \ldots\  & &  &  &  &  &  & & \targX{} & &  \\
 \lstick{${q_{{n_1} + {n_2}-1}}$} &  & \gate{H} & \gate{P({\frac{2{\pi}}{2^{2}}})} & \ \ldots\ & \gate{P({\frac{2{\pi}}{2^{{n_2}-2}}})} & \gate{P({\frac{2{\pi}}{2^{{n_2}-1}}})} & \gate{P({\frac{2{\pi}}{2^{{n_2}}}})} & \ \ldots\  &  & & &  &  &  & \targX{} &    & & \\[1cm]
\lstick{$ \vdots ~~~ $} & 
 \wave&&&&&&&&&&&&&&&&& &\\[1cm]
\lstick{${q_{{n_1} + \dots + {n_{d-1}}}}$} &  &  &   & \ \ldots\  &  &  & \ctrl{5} & \ \ldots\  & &  & \ctrl{2} & & \ctrl{1} &  \gate{H} & 
 \swap{5}\gategroup[6, steps=3, style={dashed, rounded corners}, background, label style={label position=below, anchor=north,yshift=-0.2cm}]{{\text{${n_d}$-qubit-SWAP}}} & & 
 & \\
 \lstick{${q_{{n_1} + \dots + {n_{d-1}}+1}}$} &  &  &   & \ \ldots\ &  & \ctrl{4} &  &  \ \ldots\  &  & \ctrl{1} &  &  \gate{H}  & \gate{P({\frac{2{\pi}}{2^{2}}})} &  &  & \swap{3} & &  \\
 \lstick{${q_{{n_1} + \dots + {n_{d-1}}+2}}$} &  &  &  & \ \ldots\ & \ctrl{3} &  &  &  \ \ldots\  & \gate{H}  & \gate{P({\frac{2{\pi}}{2^{2}}})} &  \gate{P({\frac{2{\pi}}{2^{3}}})} & &  &  & & & \swap{1} & \\
\lstick{$ \vdots ~~~ $} & \wave&&&&&&&&&&&&&&&& &\\
 \lstick{${q_{{n_1} + \dots + {n_{d}}-2}}$} &  &  & \ctrl{1} & \ \ldots\ &  &  &  & \ \ldots\  & &  &  &  &  &  & & \targX{} & &  \\
 \lstick{${q_{{n_1} + \dots + {n_{d}}-1}}$} & \gategroup[1,steps=1, style={invisible}, label style={label position=below,anchor=mid,yshift=-0.05cm, xshift=-0.2cm}]{MSB} & \gate{H} & \gate{P({\frac{2{\pi}}{2^{2}}})} & \ \ldots\ & \gate{P({\frac{2{\pi}}{2^{{n_d}-2}}})} & \gate{P({\frac{2{\pi}}{2^{{n_d}-1}}})} & \gate{P({\frac{2{\pi}}{2^{{n_d}}}})} & \ \ldots\  &  & & &  &  &  & \targX{} &    & & 
\end{quantikz}
\end{adjustbox}
\captionsetup{justification=raggedright, singlelinecheck=false}
\caption[]{Quantum circuit pattern of the multi-dimensional QFT according to \cite{Pfeffer_Multi-dim_QFT_arxiv_v1}. 
It is implemented for a state vector of the form ({\hyperref[eq:discretized_source_field_long_vector]{\ref*{eq:discretized_source_field_long_vector}}}) by performing one-dimensional QFT routines for the subregisters that are associated to the individual dimensions in parallel.}
\label{Fig:Multi-dim_QFT}
\end{figure*}
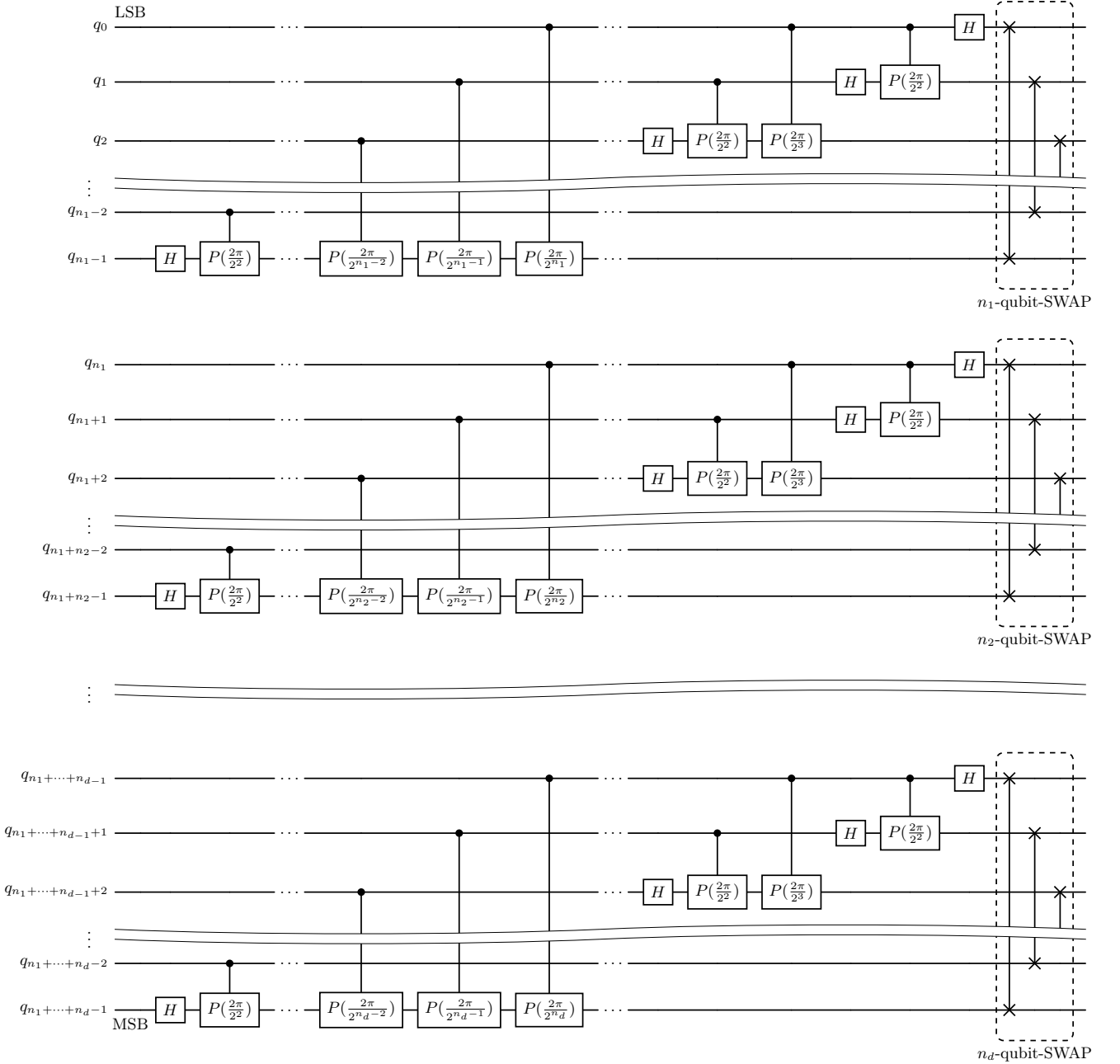

For the case of the free field problem, as explained in sec. {\hyperref[sec:Method]{\ref*{sec:Method}}}, the padding with the values of zero has to be taken into account into the input data field of the source function. 
However, it shall be explicitly noted that this padding step does not have to be done in the generation of this data field but can be implemented quite efficiently in a quantum algorithm for amplitude encoding since the padding can be set such that it results by adding the extension of the source domain before or after the source domain itself for all dimensions, respectively, and this is realized simply by adding a single qubit for all dimensions. 
More precisely, if $ {\underline{S}} $ denotes the normalized vector ({\hyperref[eq:discretized_source_field_long_vector]{\ref*{eq:discretized_source_field_long_vector}}})  for a source field without padding and $ {\underline{S}}_{pad} $ the corresponding vector for a source field with padding, then a further qubit has to be added to each of the subregisters w.r.t. $ r $ as the most significant bit (MSB), where this qubit has to be in the $ \left| 0  \right\rangle $-state to realize the padding region after the original source domain and it has to be in the $ \left| 1  \right\rangle $-state to get the padding before the original source domain (e.g. by applying an X-gate if the starting state of the added qubit shall be the $ \left| 0 \right\rangle $-state). 
This doubling of the state vector dimension is illustrated in Fig. {\hyperref[Fig:Initialization_padding]{\ref*{Fig:Initialization_padding}}} as a quantum circuit relation, where $ [ INIT ] $ is an arbitrary initialization procedure for amplitude encoding and qubit lines that go through it mean that it does not involve these qubits. 
In general, MSB and LSB indicate the most and least significant bits in the circuit diagrams and the individual qubits are labeled with 'q'.

\subsection{Quantum Fourier Transform}
\label{subsec:QFT}

In general, the source field $ \underline{S} $ depends on multiple discretized variables, for which a one-dimensional DFT has to be performed respectively. 
For this, the extension of the one-dimensional QFT to multiple dimensions from P. Pfeffer \cite{Pfeffer_Multi-dim_QFT_arxiv_v1} was used. 
For the structure of the state vector ({\hyperref[eq:discretized_source_field_long_vector]{\ref*{eq:discretized_source_field_long_vector}}}), it can be shown via the corresponding tensor product representation that the desired transformation w.r.t. a specific dimension can be implemented just by applying the circuit for the one-dimensional QFT to the subregister for the corresponding dimension, so that the multi-dimensional QFT corresponds to applying multiple one-dimensional QFT routines in parallel as shown in Fig. {\hyperref[Fig:Multi-dim_QFT]{\ref*{Fig:Multi-dim_QFT}}}. 
In Fig. {\hyperref[Fig:Multi-dim_QFT]{\ref*{Fig:Multi-dim_QFT}}}, the 2-qubit gates labeled with $ P({\phi}) $ stand for controlled phase shift gates, which are given by the matrix representation
\begin{align}
{U_{CP}}({\phi}) &= 
\begin{pmatrix}
1 & 0 & 0 & 0 \\
0 & 1 & 0 & 0 \\
0 & 0 & 1 & 0 \\
0 & 0 & 0 & {e^{{\text{i}} {\phi} }}
\end{pmatrix} , \quad {\phi} \in [0, 2{\pi} )
\label{eq:CP_matrix_form}
\end{align}
and are abbreviated here in the following as CP-gates. 
Furthermore, it is to note that the multi-qubit-SWAP gate blocks at the end of the one-dimensional QFT circuits are optional since their presence has just the effect that the ordering of the entries in the output state vector of the QFT is in fact as in the actual definition of the DFT ({\hyperref[eq:Def_DFT]{\ref*{eq:Def_DFT}}}), ({\hyperref[eq:Def_DFT_inverse]{\ref*{eq:Def_DFT_inverse}}}). 
However, for the further processing in the quantum algorithm, this specific ordering is not required.

\subsection{Matrix-Vector Multiplication}
\label{subsec:Matrix-vector_multiplication}

The multiplication of the DFT components of the source field $ S_{\underline{q}} $ by the DFT components of the Green function $ G_{\underline{q}} $ or the analytical values ({\hyperref[eq:sequence_sampled_analytical_values]{\ref*{eq:sequence_sampled_analytical_values}}}) can then be realized by multiplying the state vector $ \left| \psi \right\rangle $ obtained after the multi-dimensional QFT by a corresponding diagonal matrix. 
E.g., for an ordering of the discretized wave vectors $ \underline{k} $ like in ({\hyperref[eq:discretized_source_field_long_vector]{\ref*{eq:discretized_source_field_long_vector}}}) and the use of the analytical values, it has to be implemented
\begin{align}
 & 
\begin{pmatrix}
 0 & 0 & 0 & \dots & 0 \\
 0 & {\frac{-1}{ {{{k}}^2_{1,1}} + {{{k}}^2_{2,0}} + \dots  }} & 0 & \dots & 0 \\
 0 & 0 & {\frac{-1}{ {{{k}}^2_{1,2}} + {{{k}}^2_{2,0}} + \dots }} & & \vdots \\
 \vdots & \vdots & & \ddots &  \\
 0 & 0 & \dots &  & {\frac{-1}{ {{{k}}^2_{1,{{{\tilde{n}}_1} - 1}}} + {{{k}}^2_{1,{{{\tilde{n}}_2} - 1}}} + \dots }}
\end{pmatrix}
  \nonumber \\
  & ~~ \cdot 
 ( {{\hat{{S}}({k_{1,0}}, {k_{2,0}}, \dots )}} ,  {{\hat{{S}}({k_{1,1}}, {k_{2,0}}, \dots )}} ,  {{\hat{{S}}({k_{1,2}}, {k_{2,0}}, \dots )}} , \nonumber \\ 
 & \quad \dots , {{\hat{{S}}({k_{1,{{{\tilde{n}}_1} - 1}}}, {k_{2,{{{\tilde{n}}_2} - 1}}}, \dots )}} {)}^{T}
 .
\end{align}

This matrix is not unitary, which is why it cannot be represented via gates that act just on the register $ \psi $ but has to be embedded in a larger matrix that is set unitarily and realized via adding ancilla qubits to the system. 
The application of the non-unitary matrix results then however only in a specific subspace w.r.t. the ancilla qubit register $ a $, given by a specific measurement outcome for the qubits in it, so that a run of the quantum algorithm fails with a certain probability. 
In the quantum circuit diagrams shown in this subsection, the ancilla qubits are added as MSBs, however the ordering of $ \psi $- and $ a $-register is irrelevant. 
For implementing the matrix-vector multiplication required here in such a way, two alternative procedures were considered:

\subsubsection{Linear Combination of Unitary Matrices}
\label{subsubsec:Lin_combination_of_unitary_matrices}

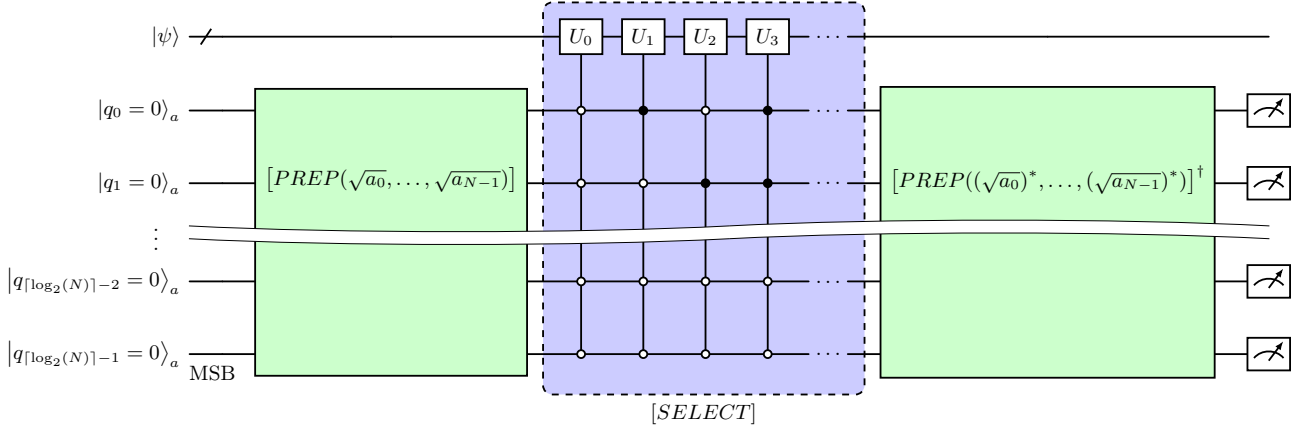
\begin{figure*}[t]
\begin{center}
\begin{adjustbox}{width=1.2\textwidth}
\tikzset{invisible/.style={fill=none,draw=none,line width=0pt,inner xsep=0pt,inner ysep=0pt}}
\begin{quantikz}
 \lstick{$ {{\left| {\psi} \right\rangle}} $} & \qwbundle{ } &  & \gate{{U_0}}\gategroup[6, steps=5, style={fill=blue!20, dashed, rounded corners}, background, label style={label position=below, anchor=north,yshift=-0.2cm}]{ $ \left[ SELECT \right] $ } &[-0.2cm] \gate{{U_1}} &[-0.2cm] \gate{{U_2}} &[-0.2cm] \gate{{U_3}} &[-0.2cm] \ \ldots\  & & \\
\lstick{$ {{\left| {q_0} = 0 \right\rangle}_{a}} $} &  & \gate[5, style={fill=green!20}, label style={yshift=0.8cm}]{ \left[ PREP( {\sqrt{a_0}}, \dots, {\sqrt{a_{N-1}}} )  \right] } & \octrl{-1} & \ctrl{-1} & \octrl{-1} & \ctrl{-1} & \ \ldots\ & \gate[5, style={fill=green!20}, label style={yshift=0.8cm}]{ {{\left[ PREP( {({\sqrt{a_0}})}^{*}, \dots, {({\sqrt{a_{N-1}}})}^{*} )  \right] }^{\dag}} }  & \meter{}  \\
 \lstick{$ {{\left| {q_1} = 0 \right\rangle}_{a}} $} &  & & \octrl{-1} & \octrl{-1} & \ctrl{-1} & \ctrl{-1}  & \ \ldots\ & & \meter{}  \\
\lstick{$ \vdots ~~~ $} & \wave&&&&&&&&&&&&&&&&& &\\
 \lstick{$ {{\left| {q_{{ \lceil \log_{2}(N) \rceil }-2}} = 0 \right\rangle}_{a}} $} &  &  &  \octrl{-2} & \octrl{-2} & \octrl{-2} & \octrl{-2}&  \ \ldots\ & & \meter{}  \\
 \lstick{$ {{\left| {q_{{ \lceil \log_{2}(N) \rceil }-1}} = 0 \right\rangle}_{a}} $} & \gategroup[1,steps=1, style={invisible}, label style={label position=below,anchor=mid,yshift=-0.05cm, xshift=-0.15cm}]{MSB}  &  &  \octrl{-1} & \octrl{-1} & \octrl{-1} & \octrl{-1}&  \ \ldots\ & & \meter{} 
\end{quantikz}
\end{adjustbox}
\end{center}
\captionsetup{justification=raggedright, singlelinecheck=false}
\caption[]{Quantum circuit pattern of the LCU procedure. 
There have to be as many ancilla qubits as needed to store a value for each non-vanishing coefficient in the expansion ({\hyperref[eq:LCU_decomposition]{\ref*{eq:LCU_decomposition}}}) in amplitude encoding, for which $ [PREP] $ is a corresponding gate sequence. 
The $ [SELECT] $-gate block involves all qubits. 
In it, the ancilla qubits act as control qubits, where the control bitstring is given by the binary representation of $ \kappa $ for the respective non-vanishing $ a_{\kappa} $. 
The ancilla qubits are added here as the MSBs, however, the ordering is irrelevant for the LCU procedure. 
In accordance to the definition of $ [PREP] $ in the text, all ancilla qubits shall have the starting state $ | 0 \rangle $ and have to be also measured in the state $ | 0 \rangle $ for mediating the multiplication of $ | \psi \rangle $ by the non-unitary matrix.}
\label{Fig:LCU_q_circuit}
\end{figure*}

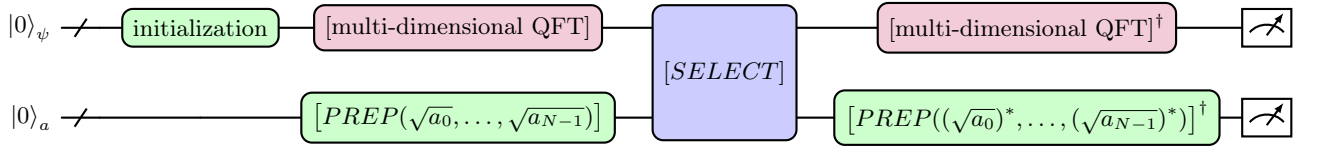
\begin{figure*}[t]
\begin{center}
\tikzset{invisible/.style={fill=none,draw=none,line width=0pt,inner xsep=0pt,inner ysep=0pt}}
\tikzset{transparent/.style={fill=none}}
\begin{quantikz}
  \lstick{$ {{\left| {0} \right\rangle}_{\psi}} $} & \qwbundle{ } &[-0.2cm] 
   \gate[1, style={fill=green!20, rounded corners}]{  \text{initialization}  } &[-0.2cm] 
    \gate[1, style={fill=purple!20, rounded corners}]{ \left[ \text{multi-dimensional QFT}  \right] } & \gate[2, style={fill=blue!20, rounded corners}]{ \left[ SELECT  \right] } & 
     \gate[1, style={fill=purple!20, rounded corners}]{ {{\left[ \text{multi-dimensional QFT} \right] }^{\dag}} } &[-0.2cm] \meter{} \\
 \lstick{$ {{\left| {0} \right\rangle}_{a}} $} & \qwbundle{ } &  & 
  \gate[1, style={fill=green!20, rounded corners}]{ \left[ PREP( {\sqrt{a_0}}, \dots, {\sqrt{a_{N-1}}} )  \right] } & 
   & 
 \gate[1, style={fill=green!20, rounded corners}]{ {{\left[ PREP( {({\sqrt{a_0}})}^{*}, \dots, {({\sqrt{a_{N-1}}})}^{*} )  \right] }^{\dag}} }  &  \meter{} 
\end{quantikz}
\end{center}
\captionsetup{justification=raggedright, singlelinecheck=false}
\caption[]{Quantum circuit schematics of the module structure of the quantum algorithm for the implementation of the multiplication step via the LCU procedure. 
The green colored gate blocks perform state preparation w.r.t. amplitude encoding.}
\label{Fig:LCU_in_q_alg}
\end{figure*}

The multiplication of a state vector by an arbitrary matrix can be implemented by the method called {\emph{linear combination of unitaries}} \cite{Childs_and_Wiebe_LCU}, abbreviated as LCU. 
In this method, the first step is to find a decomposition of the considered matrix $ A $ in terms of unitary matrices according to
\begin{align}
A &= {\sum_{{\kappa}=0}^{N-1}} {{a}_{\kappa}} {{U}_{\kappa}} , \quad {a_{\kappa}} \in \mathbb{C} , ~ {{U}^{\dag}_{\kappa}} = {{U}^{-1}_{\kappa}} , ~ N \in \mathbb{N} .
\label{eq:LCU_decomposition}
\end{align}
This decomposition is required to be given or represents in principle also a computation task in this approach. 
For the unitary basis matrices $ {{U}_{\kappa}} $, tensor products of the $ 2 $\texttimes $ 2 $-identity matrix $ {\bf{1}}_{2} = \left( \begin{smallmatrix} 1 & 0 \\ 0 & 1  \end{smallmatrix} \right) $ and the three Pauli matrices $ {{\sigma}_x} = \left( \begin{smallmatrix} 0 & 1 \\ 1 & 0  \end{smallmatrix} \right) $, $ {{\sigma}_y} = \left( \begin{smallmatrix} 0 & -{\text{i}} \\ {\text{i}} & 0  \end{smallmatrix} \right) $, $ {{\sigma}_z} = \left( \begin{smallmatrix} 1 & 0 \\ 0 & -1  \end{smallmatrix} \right) $ were considered here, for which the decomposition according to ({\hyperref[eq:LCU_decomposition]{\ref*{eq:LCU_decomposition}}}) was obtained with {\emph{Qiskit}}'s built-in routine for this task {\texttt{SparsePauliOp.from\_operator(A)}} via the outputs for {\texttt{.paulis}} and {\texttt{.coeffs}}, respectively.

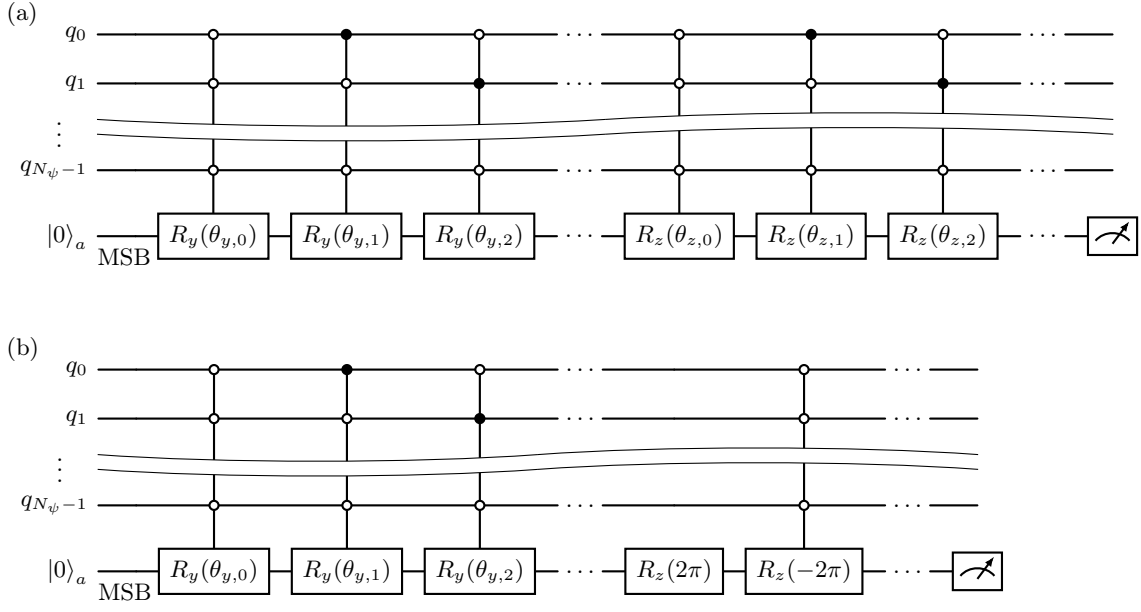
\begin{figure*}[t]
\begin{center}
\tikzset{invisible/.style={fill=none,draw=none,line width=0pt,inner xsep=0pt,inner ysep=0pt}}
\begin{align*}
 \qquad \qquad & 
\begin{quantikz}
\lstick{$ {q_0} $} & \gategroup[1,steps=1, style={invisible}, label style={label position=below,anchor=mid,yshift=0.3cm, xshift=-1.5cm}]{(a)} &[-0.2cm] \octrl{1} &[-0.2cm] \ctrl{1} &[-0.2cm] \octrl{1}  &[-0.2cm] \ \ldots\ &[-0.2cm]  \octrl{1} &[-0.2cm] \ctrl{1} &[-0.2cm] \octrl{1} &[-0.2cm] \ \ldots\ &[-0.2cm]   \\
 \lstick{$ {q_1} $} &  & \octrl{2} & \octrl{2} & \ctrl{2} &  \ \ldots\ & \octrl{2} & \octrl{2} & \ctrl{2} & \ \ldots\ &  \\
\lstick{$ \vdots ~~~ $} & \wave&&&&&&&&&&&&&&&&&& &\\
 \lstick{$ {q_{{N_{\psi}}-1}} $} &  & \octrl{1} &  \octrl{1} & \octrl{1} &  \ \ldots\ & \octrl{1} &  \octrl{1} & \octrl{1} & \ \ldots\ &   \\
 \lstick{$ {{\left| 0 \right\rangle}_{a}} $} & \gategroup[1,steps=1, style={invisible}, label style={label position=below,anchor=mid,yshift=-0.1cm, xshift=-0.15cm}]{MSB}  & \gate{ {R_y}({{\theta}_{y,0}})} & \gate{ {R_y}({{\theta}_{y,1}})} & \gate{ {R_y}({{\theta}_{y,2}})} &  \ \ldots\ & \gate{ {R_z}({{\theta}_{z,0}})} & \gate{ {R_z}({{\theta}_{z,1}})} & \gate{ {R_z}({{\theta}_{z,2}})} & \ \ldots\ & \meter{} 
\end{quantikz}
 \nonumber \\[0.6cm]
 \qquad  \qquad & 
\begin{quantikz}
\lstick{$ {q_0} $} & \gategroup[1,steps=1, style={invisible}, label style={label position=below,anchor=mid,yshift=0.3cm, xshift=-1.5cm}]{(b)}  &[-0.2cm] \octrl{1} &[-0.2cm] \ctrl{1} &[-0.2cm] \octrl{1}  &[-0.2cm] \ \ldots\ &[-0.2cm]   &[-0.2cm] \octrl{1} &[-0.2cm]  \ \ldots\ &[-0.2cm]   \\
 \lstick{$ {q_1} $} &  & \octrl{2} & \octrl{2} & \ctrl{2} &  \ \ldots\ &  & \octrl{2} & \ \ldots\ &  \\
\lstick{$ \vdots ~~~ $} & \wave&&&&&&&&&&&&&&&&& &\\
 \lstick{$ {q_{{N_{\psi}}-1}} $} &  & \octrl{1} &  \octrl{1} & \octrl{1} &  \ \ldots\ &  &  \octrl{1} &  \ \ldots\ &   \\
 \lstick{$ {{\left| 0 \right\rangle}_{a}} $} & \gategroup[1,steps=1, style={invisible}, label style={label position=below,anchor=mid,yshift=-0.1cm, xshift=-0.15cm}]{MSB}  & \gate{ {R_y}({{\theta}_{y,0}})} & \gate{ {R_y}({{\theta}_{y,1}})} & \gate{ {R_y}({{\theta}_{y,2}})} &  \ \ldots\ & \gate{ {R_z}({2{\pi}})} & \gate{ {R_z}({-2{\pi}})} &  \ \ldots\ & \meter{} 
\end{quantikz}
\end{align*}
\end{center}
\captionsetup{justification=raggedright, singlelinecheck=false}
\caption[]{Quantum circuit pattern that can be used to implement specifically the multiplication of a state vector by a diagonal matrix, requiring one ancilla qubit. 
The ancilla qubit is added here as the MSB, however, the ordering is irrelevant for the procedure. 
In accordance to the definition of the gate angles in the text, the ancilla qubit shall have the starting state $ | 0 \rangle $ and has to be measured in the state $ | 1 \rangle $ for mediating the multiplication of the state vector formed by the 'q'-qubits by the desired matrix. 
(a) depicts the general circuit pattern and (b) a simplification that can be made if all multiplied values have a negative sign except for the value associated to the first entry, which shall have a positive sign. 
The latter applies here for the case of periodic BCs.}
\label{Fig:Sequence_multi-controlled_R_gates_q_circuit}
\end{figure*}

The implementation of the LCU method consists of a so-called {\emph{preparation}} gate $ [PREP] $ and a {\emph{selection}} gate $ [SELECT] $. 
The preparation gate is defined via the action on an ancilla qubit register $ a $ for which the starting initial state shall be the state in which all ancilla qubits are in the $ \left| 0  \right\rangle $-state, i.e. the basis state $ {{\left| \kappa = 0  \right\rangle}_{a}} $:
\begin{align}
{\left[ PREP( \sqrt{ {a_0} }, \dots, \sqrt{ {a_{N-1}} } ) \right]} {{\left| 0 \right\rangle}}_{a} &= 
{\sum_{{\kappa}=0}^{N-1}} {\frac{ \sqrt{ {a_{\kappa}} } }{ \sqrt{ {\lambda} } }} {{\left| {\kappa} \right\rangle}}_{a} , 
\label{eq:LCU_Action_PREP}
 \\
  \lambda = {\sum_{{\kappa}=0}^{N-1}} {\sqrt{ {a_{\kappa}} }} \cdot { {( \sqrt{ {a_{\kappa}} } )}^{*} } &= {\sum_{{\kappa}=0}^{N-1}} {{| \sqrt{ {a_{\kappa}} } |}^2} 
 \label{eq:LCU_Action_PREP_lambda} 
\end{align}
The operation ({\hyperref[eq:LCU_Action_PREP]{\ref*{eq:LCU_Action_PREP}}}) means simply a procedure for amplitude encoding of the coefficients $ \sqrt{ {a_0} }, \dots, \sqrt{ {a_{N-1}} } $, where remaining amplitudes are set to zero if more than $ N $ basis states are formed by the ancilla qubits. 
In contrast to subsec. {\hyperref[subsec:Initialization]{\ref*{subsec:Initialization}}} however, this operation was not labeled $ [INIT] $ since the preparation procedure considered here has to be unitary, which does not necessarily have to apply for the initialization procedure in subsec. {\hyperref[subsec:Initialization]{\ref*{subsec:Initialization}}} as the {\emph{Qiskit}}-function {\texttt{initialize()}} can resort to qubit-reset operations in contrast to the function {\texttt{prepare\_state()}}. 
The selection gate is given by
\begin{align}
{\left[ SELECT ({U_0}, \dots , {U_{N-1}}) \right]} &= {\sum_{{\kappa}=0}^{N-1}}  {{\left| {\kappa} \right\rangle}}_{a} {{\left\langle {\kappa} \right|}}_{a} \otimes {U_{\kappa}} ,
\end{align}
i.e. the application of $ U_{\kappa} $ to the state $ \left| {\psi} \right\rangle $ conditioned on the basis state of the ancilla qubit register $ {\left| {\kappa} \right\rangle}_{a} $:
\begin{align}
 {\left[ SELECT ({U_0}, \dots , {U_{N-1}}) \right]} {{\left| {\kappa} \right\rangle}}_{a} \otimes {{\left| \psi \right\rangle}} &= {{\left| {\kappa} \right\rangle}}_{a} \otimes {U_{\kappa}} {{\left| \psi \right\rangle}} 
\end{align}
From this, it can be deduced that applying the sequence of gates shown in Fig. {\hyperref[Fig:LCU_q_circuit]{\ref*{Fig:LCU_q_circuit}}} and measuring the the ancilla qubits then in the $ \left| 0 \right\rangle $-state yields the application of $ {\frac{1}{\lambda}} A $ to $ \left| {\psi} \right\rangle $:
\begin{align}
 &  {\underbrace{ {{\left\langle 0 \right|}}_{a} {\left[ PREP( {( \sqrt{ {a_0} })}^{*}, \dots, {( \sqrt{ {a_{N-1}} }) }^{*} ) \right]}^{\dag} }_{ = ~ {\left( {\left[ PREP( {( \sqrt{ {a_0} } )}^{*}, \dots, {( \sqrt{ {a_{N-1}} } )}^{*} ) \right]} {{\left| 0 \right\rangle}}_{a} \right)}^{\dag} }} \nonumber \\[0.15cm] 
 & ~ \cdot {\left[ SELECT \right]} \cdot  {\left[ PREP( \sqrt{ {a_0} }, \dots, \sqrt{ {a_{N-1}} } ) \right]} {{\left| 0 \right\rangle}}_{a} \otimes {{\left| \psi \right\rangle}} \nonumber \\[0.3cm]
 & ~ = 
{\sum_{{\kappa}'=0}^{N-1}} {\frac{ \sqrt{{a_{{\kappa}'}}} }{ \sqrt{\lambda} }} {{\left\langle {\kappa}' \right|}}_{a} {\left[ SELECT \right]} {\sum_{{\kappa}=0}^{N-1}} {\frac{ \sqrt{ {a_{\kappa}} } }{ \sqrt{ \lambda } }} {{\left| {\kappa} \right\rangle}}_{a} \otimes {{\left| \psi \right\rangle}} \nonumber \\
 & ~ = {\sum_{{\kappa}'=0}^{N-1}} {\sum_{{\kappa}=0}^{N-1}} {\frac{ \sqrt{ a_{{\kappa}'} } \sqrt{ a_{{\kappa}} } }{ \lambda }} { \underbrace{ { \left\langle {\kappa}' \middle| {\kappa} \right\rangle}_{a} }_{={{\delta}_{{\kappa}, {\kappa}'}}} } {U_{\kappa}} {{\left| \psi \right\rangle}} \nonumber \\
 & ~ = {\frac{1}{\lambda}}  {\sum_{{\kappa}=0}^{N-1}} {a_{\kappa}} {U_{\kappa}} {{\left| \psi \right\rangle}} 
 = {\frac{1}{\lambda}}  A {{\left| \psi \right\rangle}} 
\end{align}
Accordingly, the probability for obtaining the desired state $ {\left| 0 \right\rangle}_{a} $ in a measurement is given by
\begin{align}
 {{{Pr}}( {| a \rangle} = {{| 0 \rangle}_{a}} )} &= 
{\sum_{i=0}^{{N_{\psi}}-1}} {\Bigl| {{\langle i |}} {\frac{1}{\lambda}}A | \psi \rangle \Bigr|}^{2} \nonumber \\
 &= 
{\frac{1}{{\lambda}^2}} {\sum_{i=0}^{{N_{\psi}}-1}} ~ {\left| {{\langle i |}} A {\sum_{j=0}^{{N_{\psi}}-1}} {{{\langle j |}} \psi \rangle} {\left| j \right\rangle} \right|}^{2} \nonumber \\
 &= {\frac{1}{{\lambda}^2}} {\sum_{i=0}^{{N_{\psi}}-1}} ~ { \left| {\sum_{j=0}^{{N_{\psi}}-1}}  {\left\langle i \middle| A \middle| j  \right\rangle}  {\langle j | \psi \rangle} \right| }^{2} ,
 \label{eq:Succes_probability_LCU_general}
\end{align}
which depends on the state $ {\left| \psi \right\rangle} $, resulting from the previous computation. 
If eq. ({\hyperref[eq:Succes_probability_LCU_general]{\ref*{eq:Succes_probability_LCU_general}}}) is referred to the considered problem, it is $ {N_{\psi}} = {\prod_{r=1}^{d}} {2^{n_r}} $, the $ {\langle j | \psi \rangle} $ are the normalized DFT components $ {{\hat{S}}_{\underline{q}}} $ of the source field ({\hyperref[eq:discretized_source_field_long_vector]{\ref*{eq:discretized_source_field_long_vector}}}) and $ A $ is a diagonal matrix with the DFT components of the Green function or the analytical values ({\hyperref[eq:sequence_sampled_analytical_values]{\ref*{eq:sequence_sampled_analytical_values}}}) as the entries. 
For the special case of a diagonal matrix, i.e. $ {\left\langle i \middle| A \middle| j  \right\rangle} = {A_{i,j}} = {{\delta}_{i,j}} {A_{i,i}} $, eq. ({\hyperref[eq:Succes_probability_LCU_general]{\ref*{eq:Succes_probability_LCU_general}}}) becomes
\begin{align}
 {{{Pr}}( {| a \rangle} = {{| 0 \rangle}_{a}} )} &= {\frac{1}{{\lambda}^2}} {\sum_{i=0}^{{N_{\psi}}-1}} {{| {A_{i,i}} |}^2} \cdot {{| \left\langle {i} \middle| {\psi} \right\rangle |}^2}
 \label{eq:Succes_probability_LCU_specific_problem_here}
\end{align}

Fig. {\hyperref[Fig:LCU_in_q_alg]{\ref*{Fig:LCU_in_q_alg}}} summarizes the circuit structure of the algorithm resulting with this method. 
For further details and an overview of related techniques, it is referred here to the works \cite{Over_et_al_Q_alg_for_AD_eq} and \cite{Bengoechea_et_al_Q_algs_BCs}.

\subsubsection{Sequence of Multi-controlled Rotation Gates}
\label{subsubsec:Sequence_of_multi-controlled_rotation_gates}

An implementation for realizing specifically the multiplication of the state vector by a diagonal matrix is given by the circuit pattern shown in Fig. {\hyperref[Fig:Sequence_multi-controlled_R_gates_q_circuit]{\ref*{Fig:Sequence_multi-controlled_R_gates_q_circuit}}} (a), which requires just one ancilla qubit.

For this, the diagonal entries $ {A_{i,i}} $ of the diagonal matrix $ A $ have to be given in terms of amplitude and complex phase, corresponding to the product
\begin{align}
A &= 
\begin{pmatrix}
{A_{0,0}} & 0 & \dots & 0  \\
0 & {A_{1,1}} & \ddots &  \\
\vdots & \ddots & \ddots & \\
0 &  & &  
\end{pmatrix}
 \nonumber \\[0.15cm]
 &= 
\begin{pmatrix}
 |{A_{0,0}}| & 0 & \dots & 0 \\
 0 & |{A_{1,1}}| & \ddots &   \\
 \vdots & \ddots & \ddots & \\
 0 &  & & 
\end{pmatrix}
 \cdot 
\begin{pmatrix}
 {e^{{\text{i}} {{\phi}_{0}} }} & 0 & \dots & 0  \\
0 & {e^{{\text{i}} {{\phi}_{1}} }} & \ddots &  \\
 \vdots & \ddots & \ddots & \\
 0 &  & &  
\end{pmatrix} 
 .
 \label{eq:Sequence_of_multi-controlled_R-gates_matrix_decomposition}
\end{align}
Furthermore, the norm of the vector $ \underline{A} $ that shall be given via the entries $ {A_{i,i}} $ has to be computed.

Here, the $ R_y $- and the $ R_z $-1-qubit rotation gates are considered in the forms
\begin{align}
{R_y}({\theta})&= 
\begin{pmatrix}
{\cos{\left( {\frac{\theta}{2}} \right)}} & {\sin{\left( {\frac{\theta}{2}} \right)}} 
 \\
-{\sin{\left( {\frac{\theta}{2}} \right)}} & {\cos{\left( {\frac{\theta}{2}} \right)}}
\end{pmatrix} , 
 \\
{R_z}({\theta})&= 
\begin{pmatrix}
{\exp{\left( -{\text{i}} {\frac{\theta}{2}} \right)}} & 0 \\
0 & {\exp{\left( {\text{i}} {\frac{\theta}{2}} \right)}}
\end{pmatrix}
 ,
\end{align}
respectively. 
W.r.t. the subspace given by the basis state $ {| {i} \rangle}_{\psi} $ of the $ \psi $-register, the multiplication factor $ {\frac{|{A_{i,i}}|}{|{\underline{A}}|}} $ is realized for the amplitude of the state $ {| {1} \rangle}_{a} \otimes {| {i} \rangle}_{\psi} $ if it is chosen
\begin{align}
 {{\theta}_{y,i}} &= -2 ~ \mbox{arcsin}\left( {\frac{ {|{A_{i,i}}|} }{ {|{\underline{A}}|} }} \right)
 ,
\end{align}
for the angle of an $ R_y $-gate that is applied to the ancilla qubit in the state $ {| {0} \rangle}_{a} $ and conditioned on $ {| {i} \rangle}_{\psi} $, whereas the factor $ \sqrt{ 1 - {{\left( {\frac{ {|{A_{i,i}}|} }{ {|{\underline{A}}|} }} \right)}^2} } $ appears for the state $ {| {0} \rangle}_{a} \otimes {| {i} \rangle}_{\psi} $. 
After that, the complex phase factors $ {e^{{\text{i}} {{\phi}_i} }} $ can be realized for the amplitudes of the states $ {| {1} \rangle}_{a} \otimes {| {i} \rangle}_{\psi} $ by applying an $ R_z $-gate to the ancilla qubit conditioned on the $ \psi $-register basis state $ {| {i} \rangle}_{\psi} $ and with the angle
\begin{align}
 {{\theta}_{z,i}} &= 2 {{\phi}_i}
 ,
\end{align}
whereas the factor $ {e^{-{\text{i}} {{\phi}_i} }} $ is implemented for $ {| {0} \rangle}_{a} \otimes {| {i} \rangle}_{\psi} $. 
For the setup for periodic BCs, it is to note that the complex phases of the regularized analytical values ({\hyperref[eq:sequence_sampled_analytical_values]{\ref*{eq:sequence_sampled_analytical_values}}}) are
\begin{align}
{{\phi}_0} = 0 , & \qquad {{\phi}_i} = {\pi} , ~ i \in \{ 1, \dots, {N_{\psi}} - 1 \}
\label{eq:sequence_sampled_analytical_values_complex_phases}
\end{align}
and since the DFT components of the Green function for periodic BCs should correspond to the analytical values, the values ({\hyperref[eq:sequence_sampled_analytical_values_complex_phases]{\ref*{eq:sequence_sampled_analytical_values_complex_phases}}}) should hold for them as well. 
Therefore, the gate sequence of $ {R_z} $-gates in the general circuit in Fig. {\hyperref[Fig:Sequence_multi-controlled_R_gates_q_circuit]{\ref*{Fig:Sequence_multi-controlled_R_gates_q_circuit}}} (a) can be shortened by applying $ {R_z}({\theta} = 2{\pi}) $ to the ancilla qubit, which causes a negative sign in all amplitudes, and correcting this sign then for the amplitude of the $ {| {0} \rangle}_{\psi} $-basis state by applying a corresponding multi-controlled $ R_z $-gate with the inverse angle as shown in Fig. {\hyperref[Fig:Sequence_multi-controlled_R_gates_q_circuit]{\ref*{Fig:Sequence_multi-controlled_R_gates_q_circuit}}} (b). 
However, it is of course to note that the uncontrolled $ R_z $-gate with angle $ 2{\pi} $ before causes a change of the global phase of the qubit system and is thus irrelevant for measurement operations after this quantum circuit.

For the implementation of the multiplication by a diagonal matrix considered in this subsection, the probability for measuring the ancilla qubit after the gate sequence in the desired state $ {{| 1 \rangle}_{a}} $ is
\begin{align}
 {{{Pr}}( {| a \rangle} = {{| 1 \rangle}} )} &= {\frac{1}{{|{\underline{A}}|}^2}} {\sum_{i=0}^{{N_{\psi}}-1}} {{| {A_{i,i}} |}^2} \cdot {{| \left\langle {i} \middle| {\psi} \right\rangle |}^2} .
 \label{eq:Success_probability_sequence_of_multi-controlled_R-gates}
\end{align}
Again, as for eq. ({\hyperref[eq:Succes_probability_LCU_specific_problem_here]{\ref*{eq:Succes_probability_LCU_specific_problem_here}}}), the $ {\langle j | \psi \rangle} $ are given here by the normalized DFT components $ {{\hat{S}}_{\underline{q}}} $ of the source field ({\hyperref[eq:discretized_source_field_long_vector]{\ref*{eq:discretized_source_field_long_vector}}}) and the $ {A_{i,i}} $ are the DFT components of the Green function or the analytical values ({\hyperref[eq:sequence_sampled_analytical_values]{\ref*{eq:sequence_sampled_analytical_values}}}).

\section{Functionality Tests}
\label{sec:Functionality_tests}

To verify the functionality of the quantum algorithm, it was simulated via IBM's software kit {\emph{Qiskit}} \cite{qiskit} in the version 1.2.2, where the state vector simulation was used. 
This was done for one- and two-dimensional test examples, for which the solution values for $ \varphi $ that are encoded in the resulting state vector were compared with the analytical solutions. 
The actual solution values are obtained from the corresponding state vector components via multiplying them by the norm of the vector of the values of the source field ({\hyperref[eq:discretized_source_field_long_vector]{\ref*{eq:discretized_source_field_long_vector}}}) and further by $ \lambda $ according to ({\hyperref[eq:LCU_Action_PREP_lambda]{\ref*{eq:LCU_Action_PREP_lambda}}}) or the norm of the vector of the diagonal entries of the multiplied diagonal matrix $ |{\underline{A}}| $ according to ({\hyperref[eq:Sequence_of_multi-controlled_R-gates_matrix_decomposition]{\ref*{eq:Sequence_of_multi-controlled_R-gates_matrix_decomposition}}}), respectively.

As stated in the paragraph below eq. ({\hyperref[eq:FT_integral_over_finite_interval_approximated_as_sum]{\ref*{eq:FT_integral_over_finite_interval_approximated_as_sum}}}), periodic BCs demand $ k = q \cdot {\Delta}k , q \in \mathbb{Z} $ and $ x = j \cdot {\Delta}x , j \in \mathbb{Z}  $ for the application of the DFT, which has to be realized in particular for the computation of the DFT of the Green function since for it, a spatial shift is not canceled in the procedure in contrast to a shift of the source field. 
In accordance with this, the $ N $ sample points for discretizing an interval of the representation $ [{x_0}, {x_0} + L] $, where this representation means here also the interval $ [-{L_{GF}}, {L_{GF}}] $ according to ({\hyperref[eq:Integral_over_GF_shall_be_0]{\ref*{eq:Integral_over_GF_shall_be_0}}}) considered for the DFT of the Green function, were set here as follows: 
It was computed $ \zeta $ according to $ {\zeta} \cdot {\Delta}x = {x_0} $ and determined whether $ \zeta $ is closer to its rounded-down value $ \lfloor \zeta \rfloor $ or its rounded-up value $ \lceil \zeta \rceil $. 
The numerical starting position $ x_{0, num} $ was then set as the multiple of $ {\Delta}x $ that is the closest to $ {\zeta} {\Delta}x $, i.e.
\begin{align}
x_{0, num} &=  \left\{ 
 \begin{array}{ll}
 {\lfloor \zeta \rfloor} \cdot {\Delta}x , & {\zeta} - { \lfloor \zeta \rfloor } \leq 0.5 \\
 {\lceil \zeta \rceil}  \cdot {\Delta}x, &  {\zeta} - { \lfloor \zeta \rfloor } > 0.5
 \end{array} 
  \right.
\end{align}
and for the Green function in spatial representation in particular, the data set of these sample values, referred to the ordering $ x_{0, num} + j \cdot {\Delta}x , j \in \{ 0, \dots , N - 1 \} $, was periodically shifted so that the data set starts with the function value at the position $ 0 \cdot {\Delta}x $, as the QFT implements the DFT for the convention that it is applied to a periodically continued data set that is referred to the ordering $ \{ 0 \cdot {\Delta}x, 1 \cdot {\Delta}x, \dots , (N-1) \cdot {\Delta}x  \} $ (and the same ordering w.r.t. multiples of $ {\Delta}k $ applies for the inverse DFT, cf. ({\hyperref[eq:sequence_sampled_analytical_values]{\ref*{eq:sequence_sampled_analytical_values}}})).

As stated in subsection {\hyperref[subsec:Numerical_realization]{\ref*{subsec:Numerical_realization}}}, the solution of the Poisson equation is determined up to a constant $ C \in \mathbb{C} $ that can be added, where this constant is fixed in the numerical treatment such that the average value of the computed field over the whole computation domain, which includes the padding domains in the case of the free field problem, is zero. 
However, this was not enforced for the calculated analytical solutions. 
To compare the numerical and the analytical results, a mismatching offset was aligned here via considering just the values of the real part for the first position in the source domain. 
For the one-dimensional examples, this hence means the position that is the most left in the possibly padded source domain, i.e. the position $ x_{0, num} $, and for the two-dimensional examples, it means the position in the lower left corner of the possibly padded source domain.

\begin{figure*}[t]
\begin{minipage}[c]{0.325\textwidth}
\centering
\begin{tikzpicture}
\draw (0,0) node[inner sep=0]{\includegraphics[height=1.0\linewidth]{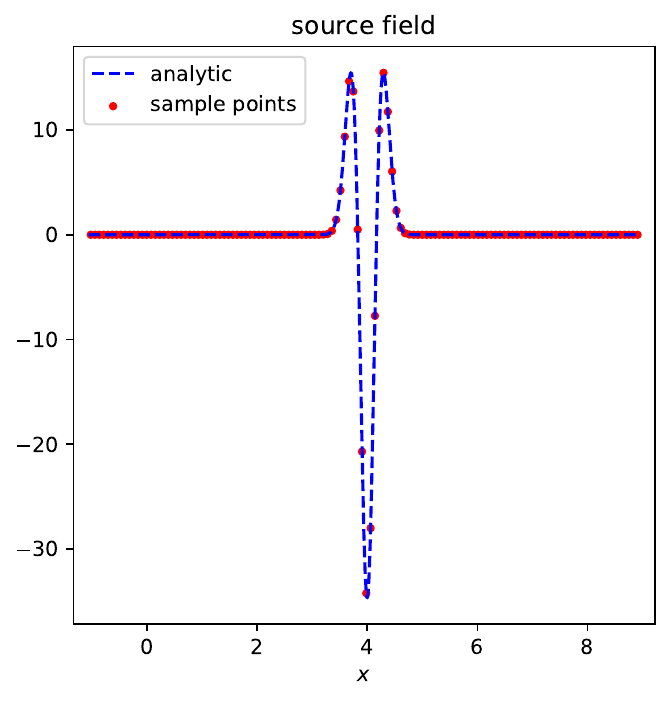}};
\end{tikzpicture}
\end{minipage}
\hfill
\begin{minipage}[c]{0.325\textwidth}
\centering
\begin{tikzpicture}
\draw (0,0) node[inner sep=0]{\includegraphics[height=1.0\linewidth]{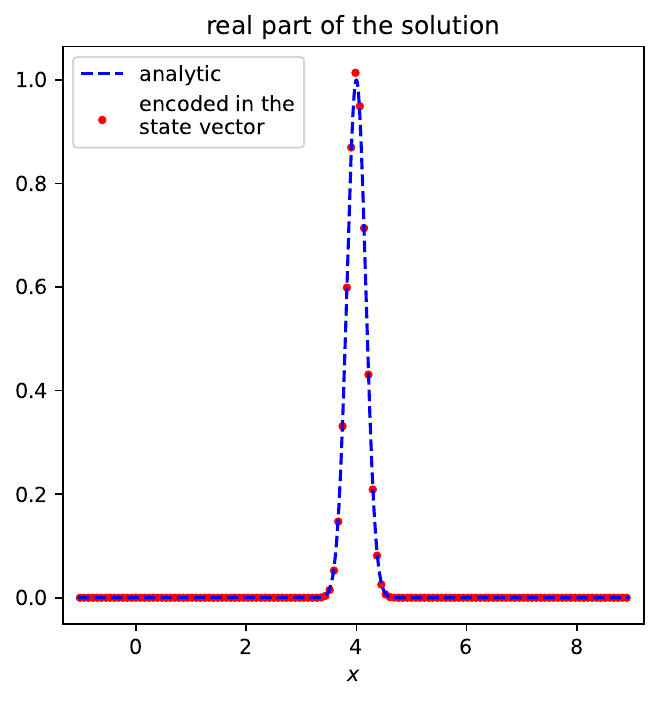}};
\end{tikzpicture}
\end{minipage}
\hfill
\begin{minipage}[c]{0.325\textwidth}
\centering
\begin{tikzpicture}
\draw (0,0) node[inner sep=0]{\includegraphics[height=1.0\linewidth]{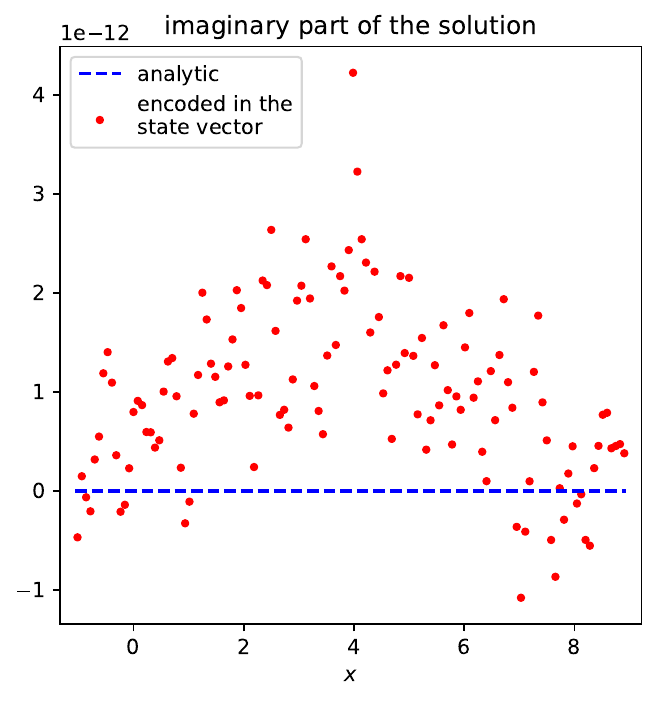}};
\end{tikzpicture}
\end{minipage}
\hfill
\begin{minipage}[c]{0.325\textwidth}
\centering
\begin{tikzpicture}
\draw (0,0) node[inner sep=0]{\includegraphics[height=1.0\linewidth]{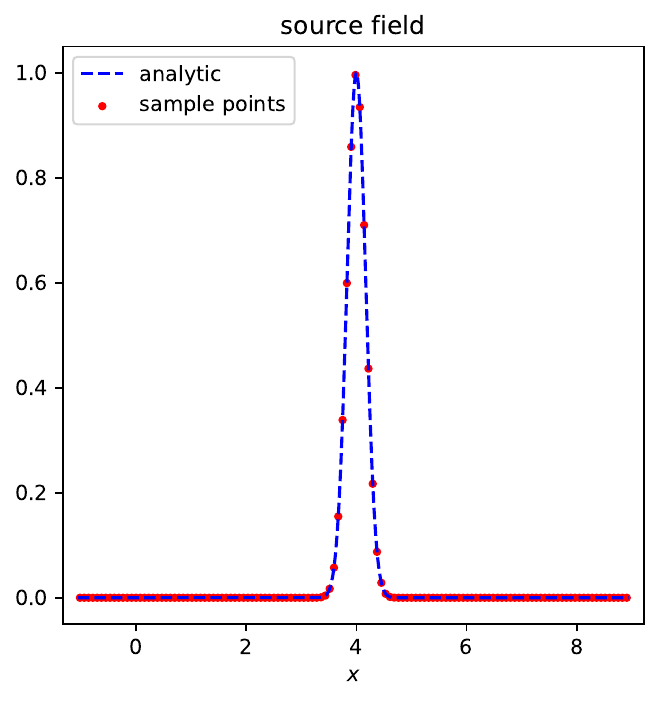}};
\end{tikzpicture}
\end{minipage}
\hfill
\begin{minipage}[c]{0.325\textwidth}
\centering
\begin{tikzpicture}
\draw (0,0) node[inner sep=0]{\includegraphics[height=1.0\linewidth]{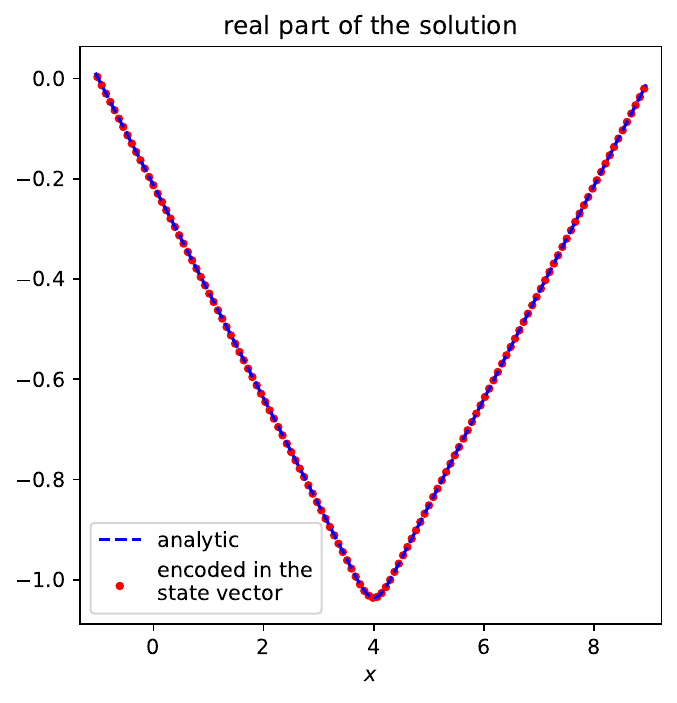}};
\end{tikzpicture}
\end{minipage}
\hfill
\begin{minipage}[c]{0.325\textwidth}
\centering
\begin{tikzpicture}
\draw (0,0) node[inner sep=0]{\includegraphics[height=1.0\linewidth]{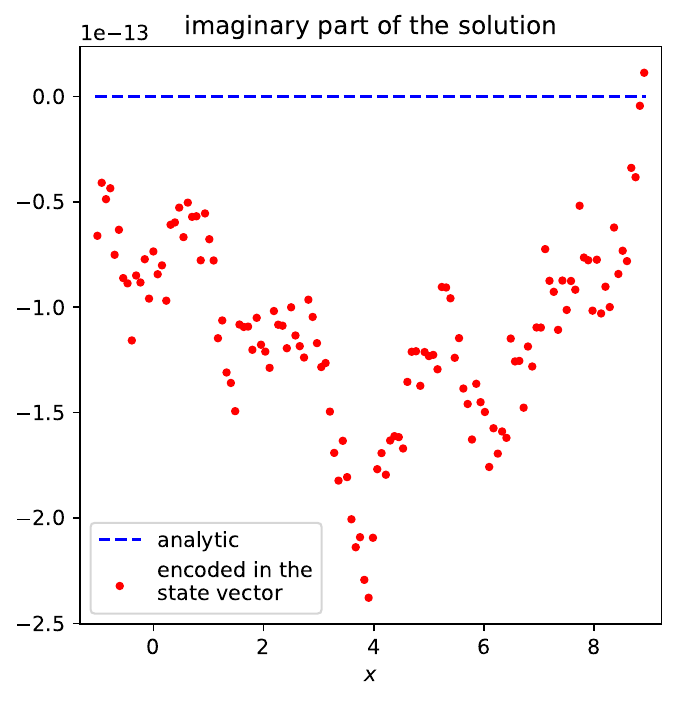}};
\end{tikzpicture}
\end{minipage}
\captionsetup{justification=raggedright, singlelinecheck=false}
\caption[]{Simulation results obtained via the first implementation variant for the test examples considered for free field conditions in one dimension, using $ 7 $ qubits to realize a resolution of $ 128 $ discretization points for the source domain of extension $ 10 $. 
In total $ 15 $ qubits were needed in these cases. 
The panels in the upper row are referred to the source ({\hyperref[eq:Test_example_1D_source_Gauss_2nd_deriv]{\ref*{eq:Test_example_1D_source_Gauss_2nd_deriv}}}) and those in the lower row to the source ({\hyperref[eq:Test_example_1D_source_Gauss]{\ref*{eq:Test_example_1D_source_Gauss}}}). 
The panels on the left show the sampled source distribution. 
The panels in the middle and on the right show the real and imaginary parts of the computed solution values $ \varphi $ in the computational subdomain of the source, respectively. 
The respective analytical solutions for these two cases are given by ({\hyperref[eq:Test_example_1D_source_Gauss_2nd_deriv_analytic_solution]{\ref*{eq:Test_example_1D_source_Gauss_2nd_deriv_analytic_solution}}}) and ({\hyperref[eq:Test_example_1D_source_Gauss_analytic_solution_free_field]{\ref*{eq:Test_example_1D_source_Gauss_analytic_solution_free_field}}}), where the additative constant was adjusted to match the real part of the computed solution value at the most left grid point of the source domain.}
\label{Fig:Test_examples_1D_free_field_variant_1_all}
\end{figure*}

For free field conditions in one dimension, two source functions were considered, for which the results of the implementation of the multiplication step via the LCU version are shown in Fig. {\hyperref[Fig:Test_examples_1D_free_field_variant_1_all]{\ref*{Fig:Test_examples_1D_free_field_variant_1_all}}}: 
As a first test example, the second derivative of a Gauss distribution according to
\begin{align}
S(x) &= 
 {\frac{{\partial}^2}{{{{\partial}x}^2}}} e^{ -{\frac{{(x - {x_s})}^2}{{\sigma}^2}} } = {\frac{2}{{\sigma}^2}} { e^{ -{\frac{{(x - {x_s})}^2}{{\sigma}^2}} } } { \left( 2{\frac{{(x - {x_s})}^2}{{\sigma}^2}} - 1 \right) }
 \label{eq:Test_example_1D_source_Gauss_2nd_deriv}
\end{align}
with $ {x_s} = 4 $, $ {\sigma} = {\frac{0.2}{\sqrt{ \ln(2) }}} $, $ [ {x_0}, {x_0} + L ] = [ {x_s} - {\frac{L}{2}}, {x_s} + {\frac{L}{2}} ] $, $ L = 10 $  was considered as the source, so that a solution of the Poisson equation is given by this Gauss distribution itself:
\begin{align} 
 {\varphi}(x) &=  { e^{- {\frac{ {{\left( x - {x_s} \right)}^2}  }{ {\sigma}^2 }}  } } + const.
 \label{eq:Test_example_1D_source_Gauss_2nd_deriv_analytic_solution}
\end{align}
As the second example, this Gauss distribution itself was considered as the source
\begin{align} 
{{{S}}({x})} 
 &= { e^{- {\frac{ {{\left( x - {x_s} \right)}^2}  }{ {\sigma}^2 }}  } } ,
 \label{eq:Test_example_1D_source_Gauss} 
\end{align}
for which
\begin{align} 
 {\varphi}(x) &= {\frac{{\sigma}^2}{2}}  e^{-{\frac{{(x-{x_s})}^2}{{\sigma}^2}}} + {\frac{{\sigma}{\sqrt{\pi}}}{2}}  (x-{x_s}) \cdot {\mbox{erf}{ \left( {\frac{ x-{x_s} }{\sigma}} \right) }} \nonumber \\
 & \quad + const.
 \label{eq:Test_example_1D_source_Gauss_analytic_solution_free_field}
\end{align}
was found as an analytical solution (see appendix {\hyperref[App_subsec:Calculation_Analytical_references]{\ref*{App_subsec:Calculation_Analytical_references}}} for details). 
Furthermore, the free field Green function of the Poisson equation for one dimension is given by \cite{Delfs_Lecture_notes}
\begin{align}
 G_{ent, 1D}(x - {\xi}) &= {\frac{| x - {\xi} |}{2}} .
\label{eq:GF_1D_free_field}
\end{align}

\begin{figure*}[t]
\begin{minipage}[c]{0.325\textwidth}
\centering
\begin{tikzpicture}
\draw (0,0) node[inner sep=0]{\includegraphics[height=1.0\linewidth]{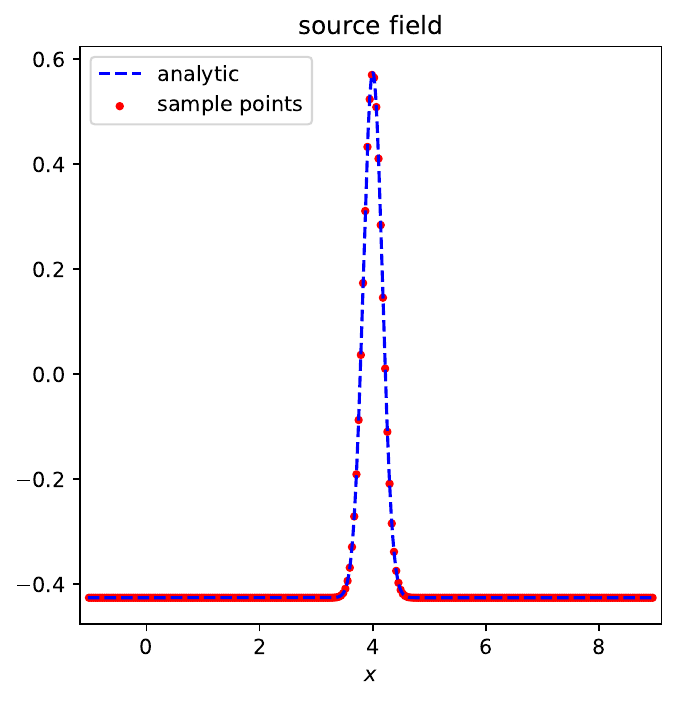}};
\end{tikzpicture}
\end{minipage}
\hfill
\begin{minipage}[c]{0.325\textwidth}
\centering
\begin{tikzpicture}
\draw (0,0) node[inner sep=0]{\includegraphics[height=1.0\linewidth]{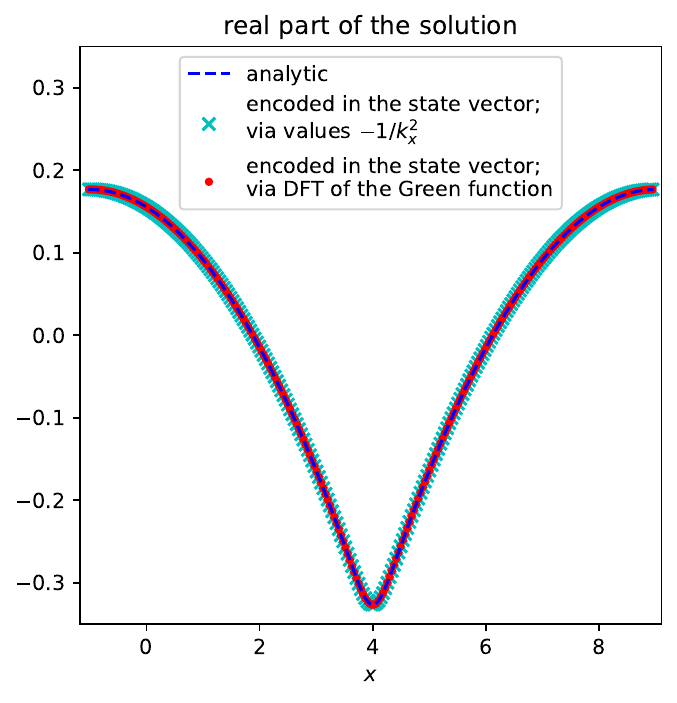}};
\end{tikzpicture}
\end{minipage}
\hfill
\begin{minipage}[c]{0.325\textwidth}
\centering
\begin{tikzpicture}
\draw (0,0) node[inner sep=0]{\includegraphics[height=1.0\linewidth]{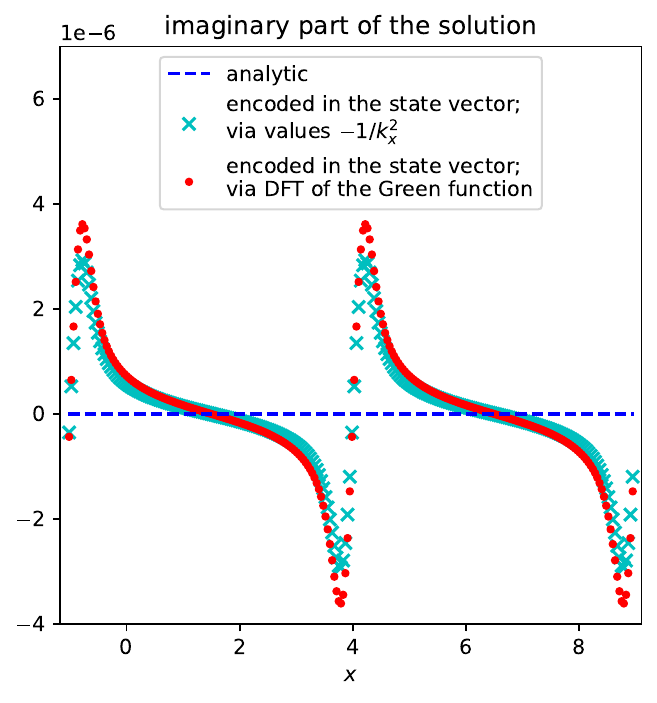}};
\end{tikzpicture}
\end{minipage}
\captionsetup{justification=raggedright, singlelinecheck=false}
\caption[]{Simulation results obtained via the first implementation variant for the test example considered for periodic BCs in one dimension, using $ 8 $ qubits to realize a resolution of $ 256 $ discretization points for the source domain of extension $ 10 $. 
In total $ 16 $ qubits were needed in this case.  
The panel on the left shows the sampled source distribution ({\hyperref[eq:Test_example_1D_source_Gauss_minus_offset]{\ref*{eq:Test_example_1D_source_Gauss_minus_offset}}}). 
The panels in the middle and on the right show the real and imaginary parts of the computed solution values $ \varphi $, respectively, where the computation of the multiplication step was done using the sampled analytical values ({\hyperref[eq:sequence_sampled_analytical_values]{\ref*{eq:sequence_sampled_analytical_values}}}) as well as the DFT of the corresponding Green function in position representation ({\hyperref[eq:GF_1D_per_BCs]{\ref*{eq:GF_1D_per_BCs}}}). 
The analytical solution for this case is given by  ({\hyperref[eq:Test_example_1D_source_Gauss_analytic_solution_per_BCs]{\ref*{eq:Test_example_1D_source_Gauss_analytic_solution_per_BCs}}}), where the additative constant was adjusted to match the real part of the computed solution value at the most left grid point of the source domain for the use of the DFT of the Green function for the multiplied matrix.}
\label{Fig:Test_example_Gauss_1D_per_BCs_variant_1_all}
\end{figure*}

It can be seen that the analytical solutions are relatively well approximated. 
The numerically resulting values of the imaginary parts, which should actually be zero, are of the order of magnitude $ 10^{-11} $ and $ 10^{-12} $ for the first and second test example, respectively and are therefore considered as neglectibly small. 
The corresponding plots for the second implementation variant of the multiplication step can be found in Fig. {\hyperref[Fig:Test_examples_1D_free_field_variant_2_all]{\ref*{Fig:Test_examples_1D_free_field_variant_2_all}}} in appendix {\hyperref[App_subsec:Further_simulation_results]{\ref*{App_subsec:Further_simulation_results}}}. 
In comparison to results from the LCU variant, the imaginary parts obtained from the simulation of this second variant are three orders of magnitude higher.

\begin{figure*}[t]
\begin{minipage}[c]{0.240\textwidth}
\centering
\begin{tikzpicture}
\draw (0,0) node[inner sep=0]{\includegraphics[height=1.0\linewidth]{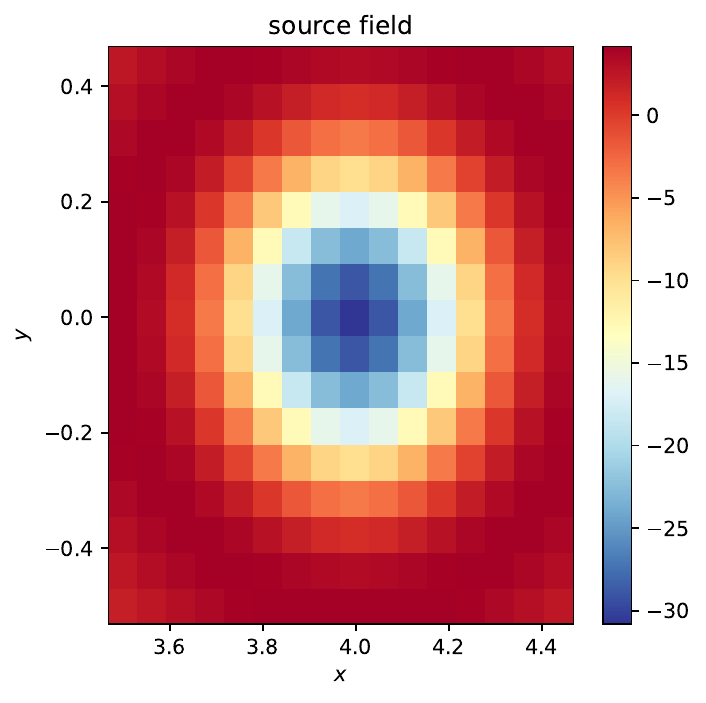}};
\end{tikzpicture}
\end{minipage}
\hfill
\begin{minipage}[c]{0.240\textwidth}
\centering
\begin{tikzpicture}
\draw (0,0) node[inner sep=0]{\includegraphics[height=1.0\linewidth]{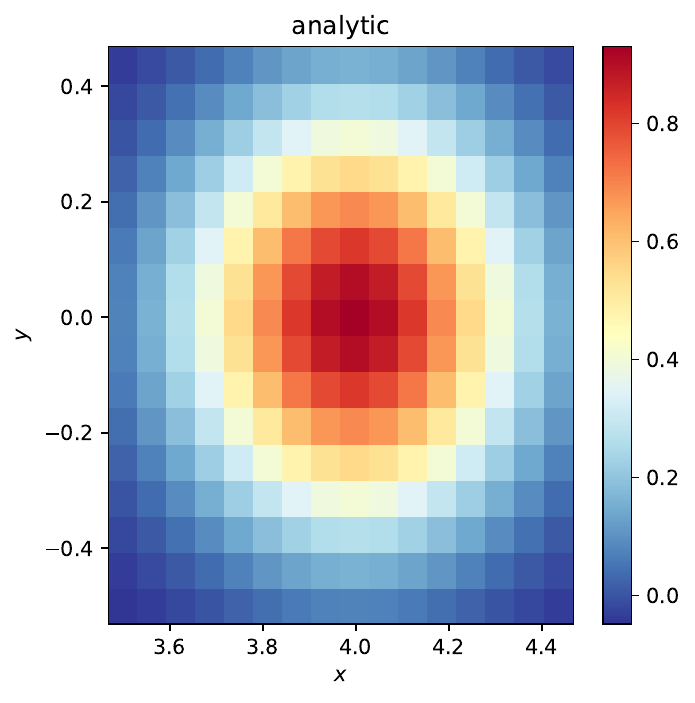}};
\end{tikzpicture}
\end{minipage}
\hfill
\begin{minipage}[c]{0.240\textwidth}
\centering
\begin{tikzpicture}
\draw (0,0) node[inner sep=0]{\includegraphics[height=1.0\linewidth]{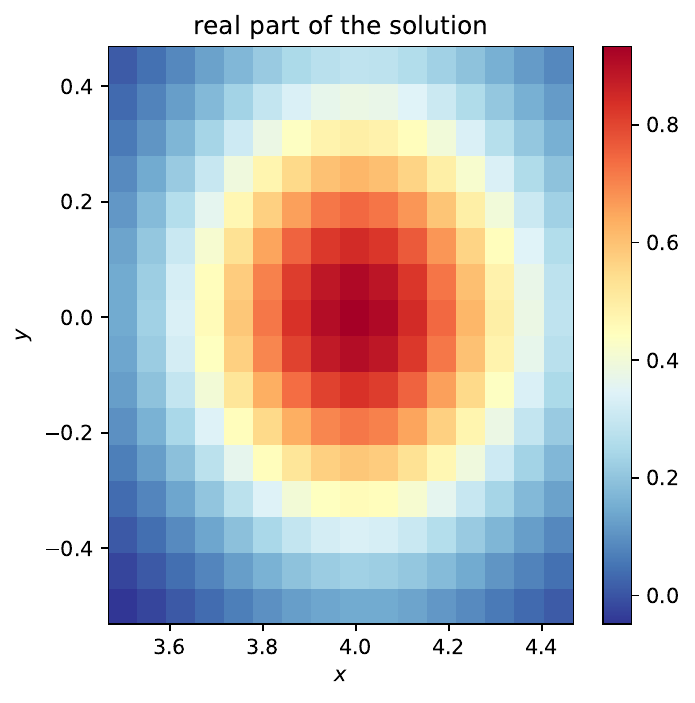}};
\end{tikzpicture}
\end{minipage}
\hfill
\begin{minipage}[c]{0.240\textwidth}
\centering
\begin{tikzpicture}
\draw (0,0) node[inner sep=0]{\includegraphics[height=1.0\linewidth]{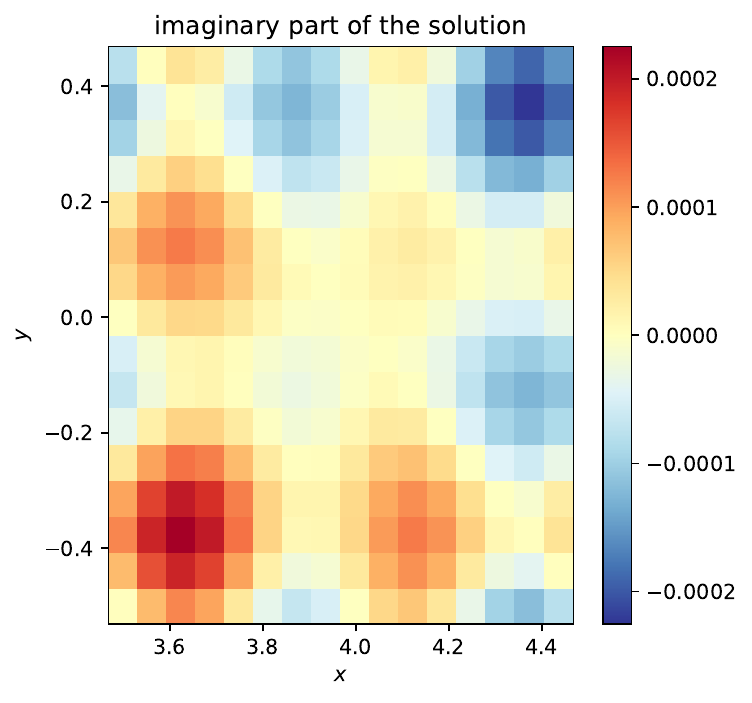}};
\end{tikzpicture}
\end{minipage}
\hfill
\begin{minipage}[c]{0.240\textwidth}
\centering
\begin{tikzpicture}
\draw (0,0) node[inner sep=0]{\includegraphics[height=1.0\linewidth]{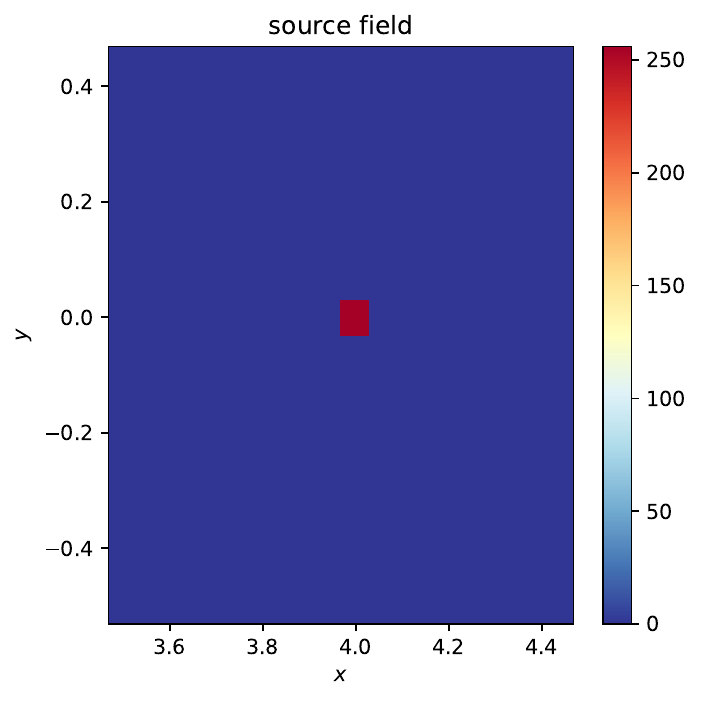}};
\end{tikzpicture}
\end{minipage}
\hfill
\begin{minipage}[c]{0.240\textwidth}
\centering
\begin{tikzpicture}
\draw (0,0) node[inner sep=0]{\includegraphics[height=1.0\linewidth]{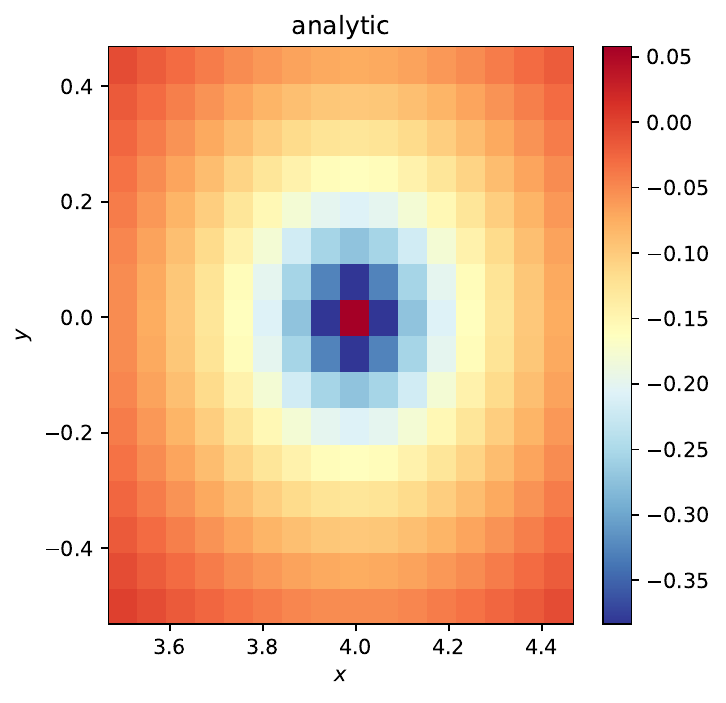}};
\end{tikzpicture}
\end{minipage}
\hfill
\begin{minipage}[c]{0.240\textwidth}
\centering
\begin{tikzpicture}
\draw (0,0) node[inner sep=0]{\includegraphics[height=1.0\linewidth]{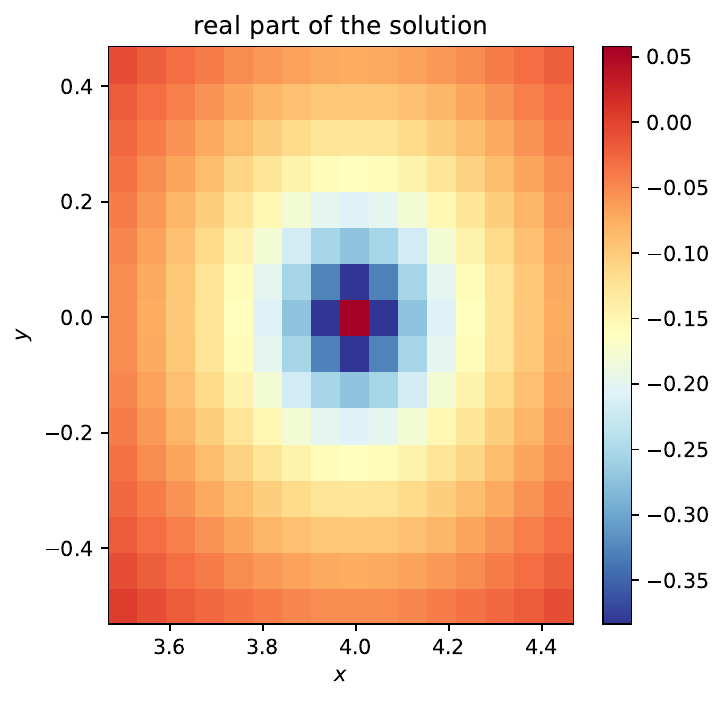}};
\end{tikzpicture}
\end{minipage}
\hfill
\begin{minipage}[c]{0.240\textwidth}
\centering
\begin{tikzpicture}
\draw (0,0) node[inner sep=0]{\includegraphics[height=1.0\linewidth]{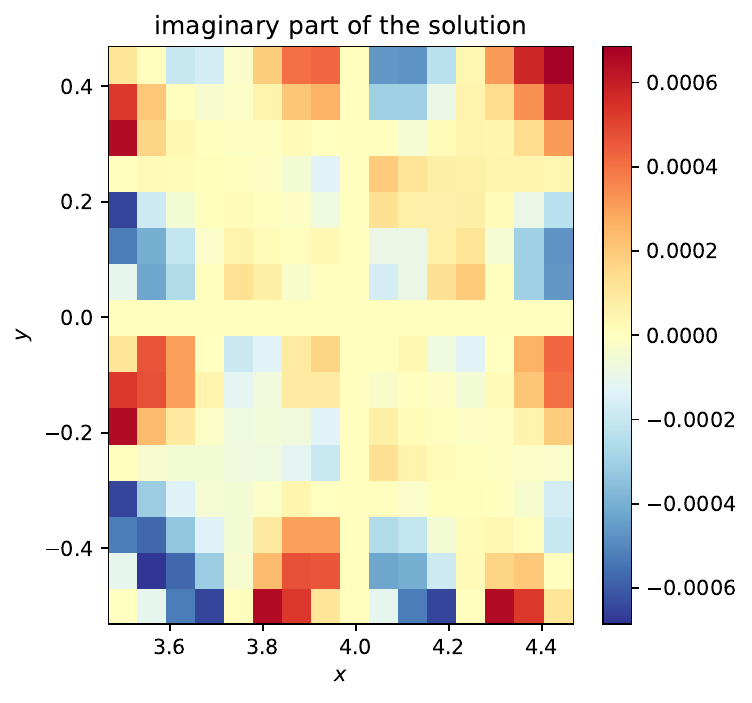}};
\end{tikzpicture}
\end{minipage}
\captionsetup{justification=raggedright, singlelinecheck=false}
\caption[]{Simulation results obtained via the first implementation variant for the test examples considered for free field conditions in two dimensions, using $ 8 $ qubits to realize a resolution of $ 16 $ discretization points per dimension for the source domain of extension $ 1 $ in each dimension. 
In total $ 20 $ qubits were needed in these cases. 
The panels in the upper row are referred to the source ({\hyperref[eq:Test_example_2D_source_Gauss_2nd_deriv]{\ref*{eq:Test_example_2D_source_Gauss_2nd_deriv}}}) and those in the lower row to the source ({\hyperref[eq:Test_example_2D_source_approximated_delta-distribution]{\ref*{eq:Test_example_2D_source_approximated_delta-distribution}}}). 
The panels most left and on the left in the middle show the sampled source distribution and sampled analytical solution, respectively. 
The panels on the right in the middle and most right show the real and imaginary parts of the computed solution values $ \varphi $ in the computational subdomain of the source, respectively. 
The respective analytical solutions for these two cases are given by ({\hyperref[eq:Test_example_2D_source_Gauss_2nd_deriv_analytic_solution]{\ref*{eq:Test_example_2D_source_Gauss_2nd_deriv_analytic_solution}}}) and ({\hyperref[eq:Test_example_2D_source_approximated_delta-distribution_analytic_solution]{\ref*{eq:Test_example_2D_source_approximated_delta-distribution_analytic_solution}}}), where the additative constant was adjusted to match the real part of the computed solution value at the grid point in the lower left corner of the source domain.}
\label{Fig:Test_examples_2D_free_field_variant_1_all}
\end{figure*}

For one dimension, it was also considered the case of periodic BCs for the test example of the Gauss distribution ({\hyperref[eq:Test_example_1D_source_Gauss]{\ref*{eq:Test_example_1D_source_Gauss}}}). 
This was done using the regularized analytical values ({\hyperref[eq:sequence_sampled_analytical_values]{\ref*{eq:sequence_sampled_analytical_values}}}) for the multiplied diagonal matrix as well as the DFT of the spatial representation of the Green function of the Poisson equation in one dimension for periodic BCs, which is given by \cite{Marshall_Periodic_Green_functions}
\begin{align}
 G_{per, 1D}(x - {\xi}) &=  -{\frac{1}{2L}} \cdot \Bigl( {{(x - {\xi})}^2} - {L{|x - {\xi}|}} + {\frac{{L^2}}{6}} \Bigr) .
\label{eq:GF_1D_per_BCs}
\end{align}
In accordance with the explanations given in subsection {\hyperref[subsec:Analytical_consideration]{\ref*{subsec:Analytical_consideration}}} for the requirement ({\hyperref[eq:Per_BCs_requirement_source_integral_over_domain_zero]{\ref*{eq:Per_BCs_requirement_source_integral_over_domain_zero}}}),  however, an offset was introduced for the source function in this case according to
\begin{align}
 S(x) &= {e^{ -{\frac{{( {x} - {x_s} )}^2}{{\sigma}^2}} }} - {\overline{S}} 
 \label{eq:Test_example_1D_source_Gauss_minus_offset}
\end{align}
where the specific expression for $ \overline{S} $ is given in ({\hyperref[eq:Test_example_1D_Gauss_per_BCs_expression_integral_over_S]{\ref*{eq:Test_example_1D_Gauss_per_BCs_expression_integral_over_S}}}). 
As an analytical solution, it was calculated
\begin{align} 
 {\varphi}(x) &= 
 {\frac{{\sigma}^2}{2}}  e^{-{\frac{{(x-{x_s})}^2}{{\sigma}^2}}} + {\frac{{\sigma}{\sqrt{\pi}}}{2}}  (x-{x_s}) \cdot {\mbox{erf}{ \left( {\frac{ x-{x_s} }{\sigma}} \right) }} \nonumber \\
 & \quad - {\frac{1}{2L}} \left( {x^2} + {\frac{L^2}{6}} \right) {\sqrt{\pi}} {\sigma} ~ {\mbox{erf}{\left( {\frac{L}{2{\sigma}}} \right)}} \nonumber \\
 & \quad  + {\frac{x}{L}} {\sqrt{\pi}} {\sigma} {x_s} ~ {\mbox{erf}{\left( {\frac{L}{2{\sigma}}} \right)}}
   + {\frac{{{\sigma}^2}}{4}} {e^{ -{\frac{L^2}{{(2{\sigma})}^2}} }} \nonumber \\
 & \quad  - {\frac{{{\sigma}^3}{\sqrt{\pi}}}{4L}} ~ {\mbox{erf}{\left( {\frac{L}{2{\sigma}}} \right)}}
   - {\frac{{{x_s^2}{\sigma}}{\sqrt{\pi}}}{2L}} ~ {\mbox{erf}{\left( {\frac{L}{2{\sigma}}} \right)}} \nonumber \\
 & \quad + {\frac{ {\overline{S}} }{2L}} \cdot \biggl( {\frac{1}{3}} \biggl( {{\left( {x_s} + {\frac{L}{2}} \right)}^3} - {{\left( {x_s} - {\frac{L}{2}} \right)}^3}  \biggr) \nonumber \\
 & \quad - x ~ \biggl( {{\left( {x_s} + {\frac{L}{2}} \right)}^2} - {{{\left( {x_s} - {\frac{L}{2}} \right)}}^2}  \biggr)  \nonumber \\
 & \quad + L ~ \left( {x^2} + {\frac{L^2}{6}} \right)  \biggr) \nonumber \\
 & \quad - {\frac{ {\overline{S}} }{4}} \cdot \biggl( \left({ {x_s} + {\frac{L}{2}} - x} \right) \cdot \left|{ {x_s} + {\frac{L}{2}} - x}\right| \nonumber \\
 & \quad - \left({ {x_s} - {\frac{L}{2}} - x}\right) \cdot \left|{ {x_s} - {\frac{L}{2}} - x}\right| \biggr) + const.
 \label{eq:Test_example_1D_source_Gauss_analytic_solution_per_BCs}
\end{align}
(see appendix {\hyperref[App_subsec:Calculation_Analytical_references]{\ref*{App_subsec:Calculation_Analytical_references}}} for details). 
The simulation results of the LCU variant are shown in Fig. {\hyperref[Fig:Test_example_Gauss_1D_per_BCs_variant_1_all]{\ref*{Fig:Test_example_Gauss_1D_per_BCs_variant_1_all}}} and the results of the second implementation variant are given by Fig. {\hyperref[Fig:Test_example_Gauss_1D_per_BCs_variant_2_all]{\ref*{Fig:Test_example_Gauss_1D_per_BCs_variant_2_all}}}, which is included again in appendix {\hyperref[App_subsec:Further_simulation_results]{\ref*{App_subsec:Further_simulation_results}}}.

In the simulation results for this case, it is to observe that the imaginary parts obtained for the LCU implementation of the multiplication step have a relatively high order of magnitude of $ 10^{-5} $, whereas in contrast to the previous test cases for free field conditions, now, the alternative implementation of the multiplication results in lower imaginary parts. 
For this second implementation variant, they have again an order of magnitude of $ 10^{-10} $.

As a first test example for a free field problem in two dimensions, the corresponding case of the second derivative of a Gauss distribution according to
\begin{align}
 S(x, y) &= \left[ {\frac{{\partial}^2}{{{{\partial}x}^2}}} + {\frac{{\partial}^2}{{{{\partial}y}^2}}} \right] {e^{ -{\frac{{(x - {x_s})}^2}{{\sigma}^2}} }} {e^{ -{\frac{{y}^2}{{\sigma}^2}} }} \nonumber \\
 &= {\frac{2}{{\sigma}^2}} {e^{ -{\frac{{(x - {x_s})}^2}{{\sigma}^2}} }} {e^{ -{\frac{{y}^2}{{\sigma}^2}} }} { \left( 2{\frac{{{(x - {x_s})}^2} + {y^2}}{{\sigma}^2}} - 2 \right) }
 \label{eq:Test_example_2D_source_Gauss_2nd_deriv}
\end{align}
with $ x_s = 4 $, $ [ {x_0}, {x_0} + {L_x} ] = [ {x_s} - {\frac{{L_x}}{2}}, {x_s} + {\frac{{L_x}}{2}} ] $, $ [ {y_0}, {y_0} + {L_y} ] = [ -{\frac{{L_y}}{2}}, +{\frac{{L_y}}{2}} ] $, $ {L_x} = {L_y} = 1 $ 
was considered. 
However, in order to obtain a simulation result that looks already similar to the analytical solution
\begin{align}
 {\varphi}(x, y) &= {e^{ -{\frac{{(x - {x_s})}^2}{{\sigma}^2}} }} {e^{ -{\frac{{y}^2}{{\sigma}^2}} }} + const.
 \label{eq:Test_example_2D_source_Gauss_2nd_deriv_analytic_solution}
\end{align}
for the quite low resolution of the source domain via $ 2^{4} = 16 $ grid points for each dimension, it was used $ {\sigma} = {\frac{0.3}{\sqrt{ \ln(2) }}} $ in this case, for which the results are depicted in the upper row of Fig. {\hyperref[Fig:Test_examples_2D_free_field_variant_1_all]{\ref*{Fig:Test_examples_2D_free_field_variant_1_all}}} for the LCU implementation and in the upper row of Fig. {\hyperref[Fig:Test_examples_2D_free_field_variant_2_all]{\ref*{Fig:Test_examples_2D_free_field_variant_2_all}}} in appendix {\hyperref[App_subsec:Further_simulation_results]{\ref*{App_subsec:Further_simulation_results}}} for the alternative implementation.

\begin{figure*}[t]
\begin{minipage}[c]{0.240\textwidth}
\centering
\begin{tikzpicture}
\draw (0,0) node[inner sep=0]{\includegraphics[height=1.0\linewidth]{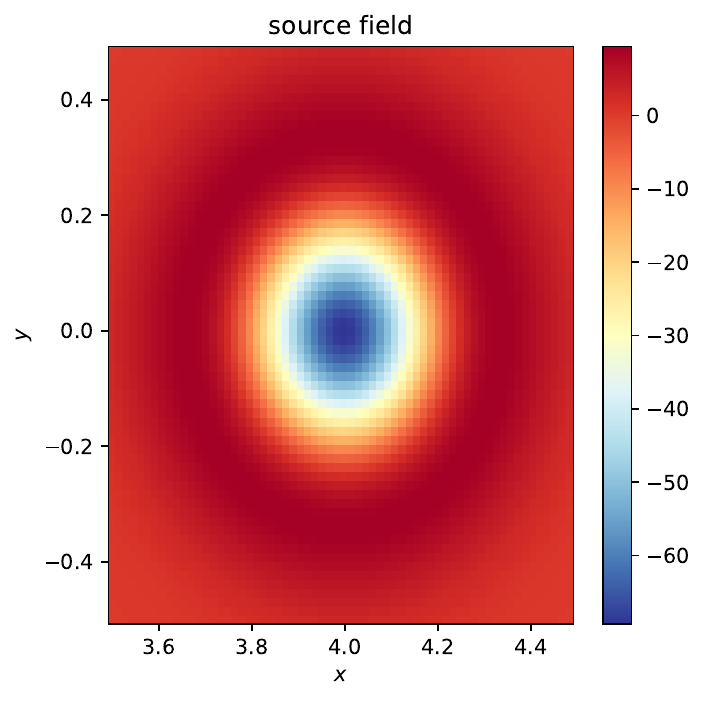}};
\end{tikzpicture}
\end{minipage}
\hfill
\begin{minipage}[c]{0.240\textwidth}
\centering
\begin{tikzpicture}
\draw (0,0) node[inner sep=0]{\includegraphics[height=1.0\linewidth]{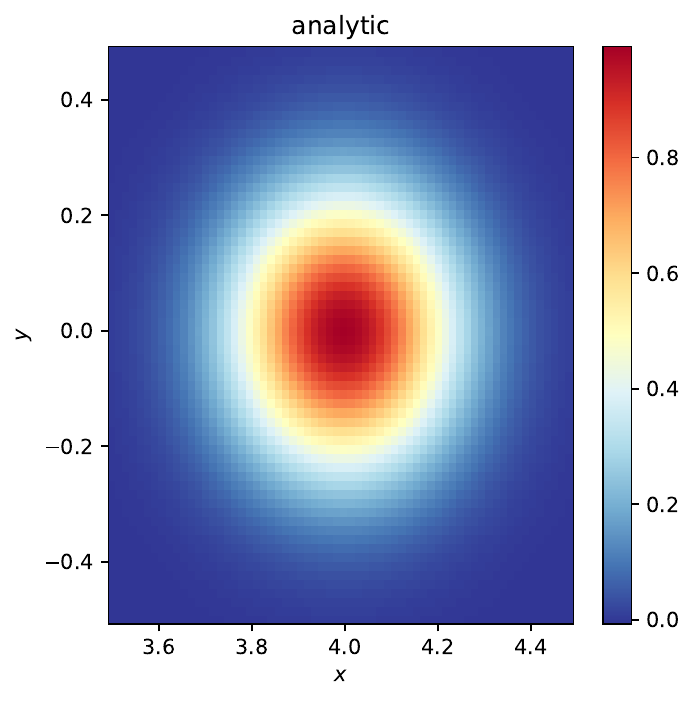}};
\end{tikzpicture}
\end{minipage}
\hfill
\begin{minipage}[c]{0.240\textwidth}
\centering
\begin{tikzpicture}
\draw (0,0) node[inner sep=0]{\includegraphics[height=1.0\linewidth]{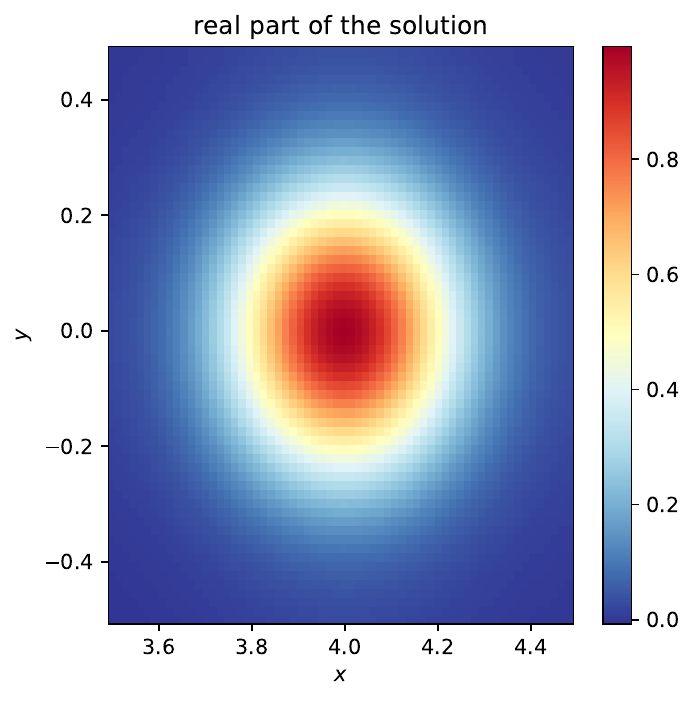}};
\end{tikzpicture}
\end{minipage}
\hfill
\begin{minipage}[c]{0.240\textwidth}
\centering
\begin{tikzpicture}
\draw (0,0) node[inner sep=0]{\includegraphics[height=1.0\linewidth]{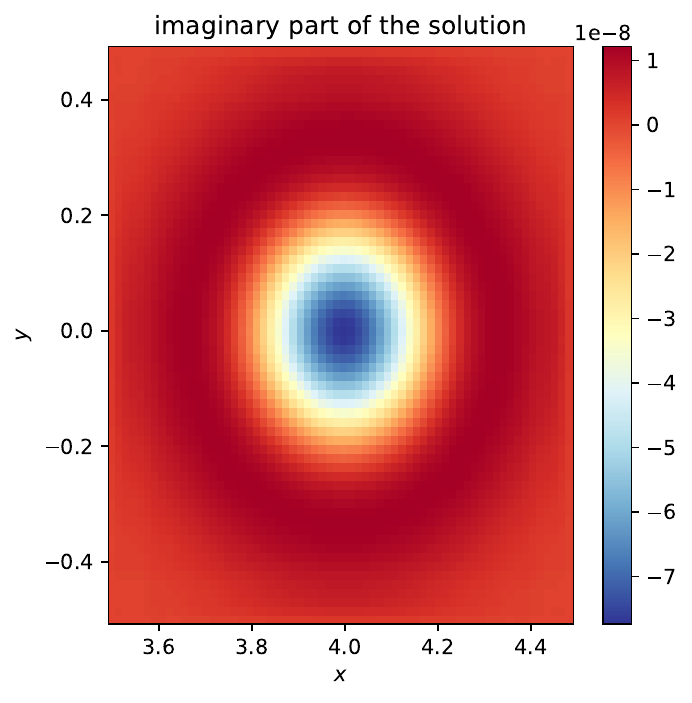}};
\end{tikzpicture}
\end{minipage}
\hfill
\begin{minipage}[c]{0.240\textwidth}
\centering
\begin{tikzpicture}
\draw (0,0) node[inner sep=0]{\includegraphics[height=1.0\linewidth]{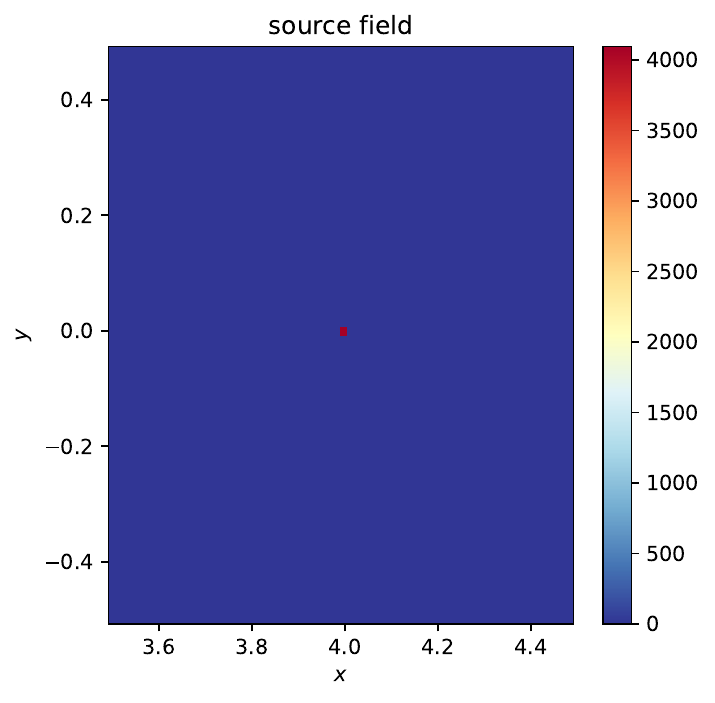}};
\end{tikzpicture}
\end{minipage}
\hfill
\begin{minipage}[c]{0.240\textwidth}
\centering
\begin{tikzpicture}
\draw (0,0) node[inner sep=0]{\includegraphics[height=1.0\linewidth]{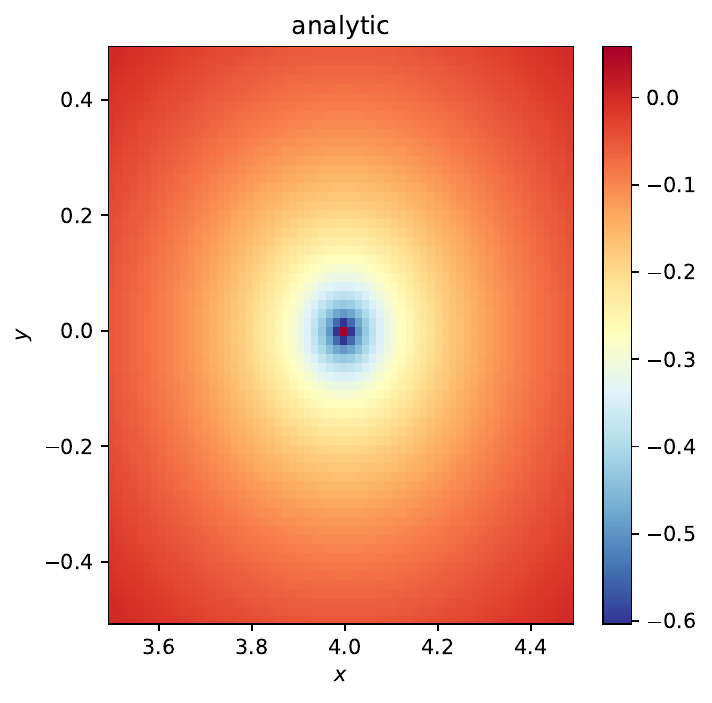}};
\end{tikzpicture}
\end{minipage}
\hfill
\begin{minipage}[c]{0.240\textwidth}
\centering
\begin{tikzpicture}
\draw (0,0) node[inner sep=0]{\includegraphics[height=1.0\linewidth]{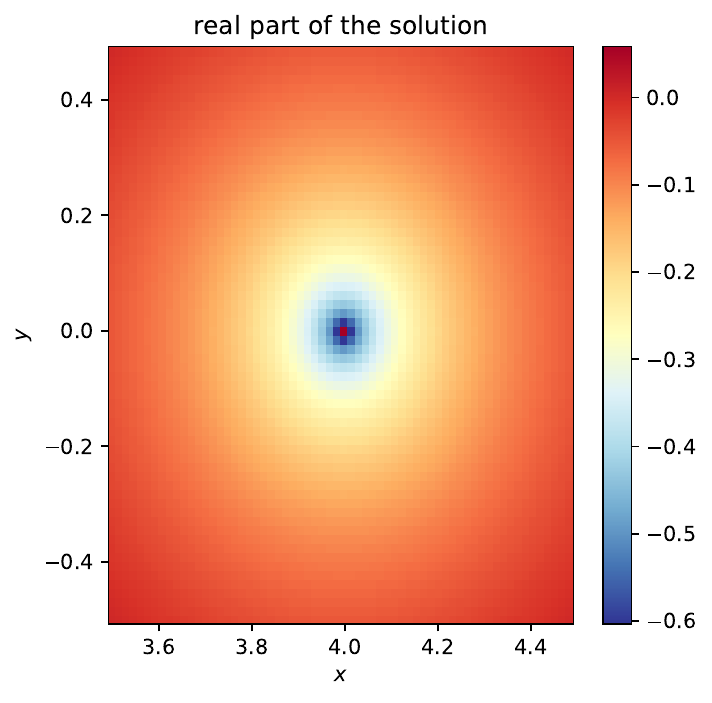}};
\end{tikzpicture}
\end{minipage}
\hfill
\begin{minipage}[c]{0.240\textwidth}
\centering
\begin{tikzpicture}
\draw (0,0) node[inner sep=0]{\includegraphics[height=1.0\linewidth]{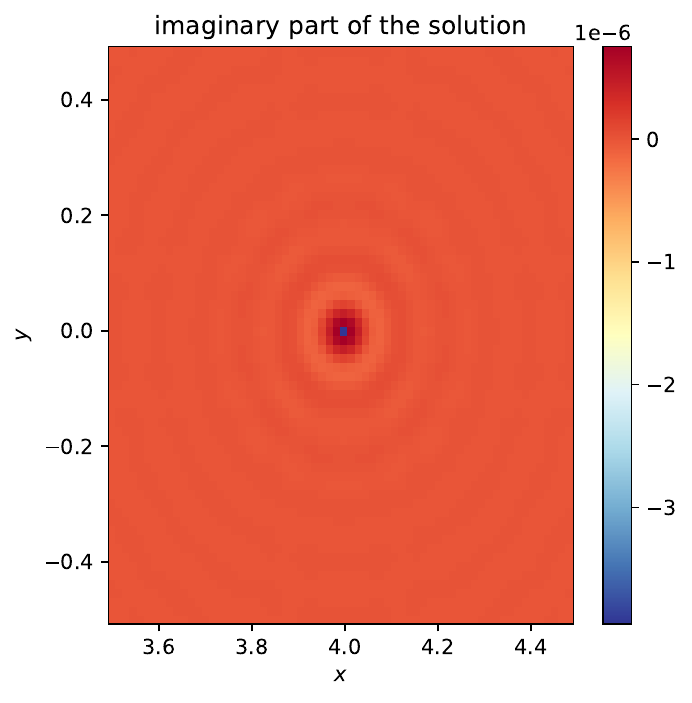}};
\end{tikzpicture}
\end{minipage}
\captionsetup{justification=raggedright, singlelinecheck=false}
\caption[]{Simulation results obtained via the second implementation variant for the test examples considered for free field conditions in two dimensions, using $ 12 $ qubits to realize a resolution of $ 64 $ discretization points per dimension for the source domain of extension $ 1 $ in each dimension. 
In total $ 15 $ qubits were needed in these cases. 
The panels in the upper row are referred to the source ({\hyperref[eq:Test_example_2D_source_Gauss_2nd_deriv]{\ref*{eq:Test_example_2D_source_Gauss_2nd_deriv}}}) and those in the lower row to the source ({\hyperref[eq:Test_example_2D_source_approximated_delta-distribution]{\ref*{eq:Test_example_2D_source_approximated_delta-distribution}}}). 
The panels most left and on the left in the middle show the sampled source distribution and sampled analytical solution, respectively. 
The panels on the right in the middle and most right show the real and imaginary parts of the computed solution values $ \varphi $ in the computational subdomain of the source, respectively. 
The respective analytical solutions for these two cases are given by ({\hyperref[eq:Test_example_2D_source_Gauss_2nd_deriv_analytic_solution]{\ref*{eq:Test_example_2D_source_Gauss_2nd_deriv_analytic_solution}}}) and ({\hyperref[eq:Test_example_2D_source_approximated_delta-distribution_analytic_solution]{\ref*{eq:Test_example_2D_source_approximated_delta-distribution_analytic_solution}}}), where the additative constant was adjusted to match the real part of the computed solution value at the grid point in the lower left corner of the source domain.}
\label{Fig:Test_examples_2D_free_field_variant_2_higher_resolution_all}
\end{figure*}

As a second test example for two dimensions, the asymptotic case of a Gauss distribution according to a $ \delta $-distribution was considered:
\begin{align}
 {{{S}}({x}, {y})} &= {{\delta}(x - {x_s})} {{\delta}(y)} 
  \label{eq:Test_example_2D_source_approximated_delta-distribution}
\end{align}
Accordingly, an analytical solution is given just by the free field Green function of the Poisson equation in two dimensions ({\hyperref[eq:GF_2D_free_field]{\ref*{eq:GF_2D_free_field}}}) itself \cite{Delfs_Lecture_notes}:
\begin{align} 
 {\varphi}(x, y) &= {\frac{1}{4{\pi}}} { \ln( { { {{\left( x - {x_s} \right)}^2} + {{y}^2}  } }  ) } + const.
 \label{eq:Test_example_2D_source_approximated_delta-distribution_analytic_solution}
\end{align}
For the numerical consideration of this case, the source distribution was approximated as a rectangular function via setting the value to $ \frac{1}{ {\Delta}x ~ {\Delta}y } $ for the grid point that is the closest to the point $ (x = {x_s}, y = 0) $ and to zero otherwise. 
Furthermore, it is to note again that the free field Green function was regularized, i.e., the value for it was set to zero if an argument of zero was encountered in the logarithm. 
The results for this test case from the first and second variant of implementing the multiplication step are shown in the lower row of Fig. {\hyperref[Fig:Test_examples_2D_free_field_variant_1_all]{\ref*{Fig:Test_examples_2D_free_field_variant_1_all}}} and Fig. {\hyperref[Fig:Test_examples_2D_free_field_variant_2_all]{\ref*{Fig:Test_examples_2D_free_field_variant_2_all}}}, respectively.

Regarding the results for these test cases of the Hockney method for two dimensions, it is to note that the imaginary parts obtained via the LCU variant are relatively high since they have an order of magnitude of $ 10^{-3} $, while the values of the real part reach an order of magnitude of $ 1 $. 
In contrast, the imaginary parts obtained from the simulation of the circuit using the specific sequence of multi-controlled rotation gates for the matrix multiplication are again lower for these examples. 
They have an order of magnitude of $ 10^{-9} $ and $ 10^{-8} $ for the first and second test example, respectively.

That there is always just one ancilla qubit for the implementation of the multiplication step via the specific sequence of multi-controlled rotation gates described in subsection {\hyperref[subsubsec:Sequence_of_multi-controlled_rotation_gates]{\ref*{subsubsec:Sequence_of_multi-controlled_rotation_gates}}} allows to simulate higher resolutions. 
For this, Fig. {\hyperref[Fig:Test_examples_2D_free_field_variant_2_higher_resolution_all]{\ref*{Fig:Test_examples_2D_free_field_variant_2_higher_resolution_all}}} shows the results for the test examples for two dimensions at a resolution of the source domain via $ 2^{6} = 64 $ grid points for each dimension, where it was however taken again $ {\sigma} = {\frac{0.2}{\sqrt{ \ln(2) }}} $ in the case of the second derivative of the Gauss distribution as the source.

\begin{table*}[t]
\begin{center}
\begin{tabular}[c]{c|c|c|c|c|c|c}
 \multicolumn{1}{c|}{ $~$dimensionality$~$ } & \multicolumn{1}{c|}{ BCs } & \multicolumn{1}{c|}{ source } & \multicolumn{1}{c|}{ implementation } & \multicolumn{1}{c|}{ resolution of } & \multicolumn{1}{c|}{ number of } &
  \multicolumn{1}{c}{ success }  
  \\
  \multicolumn{1}{c|}{  } & \multicolumn{1}{c|}{  } & \multicolumn{1}{c|}{ distribution } & \multicolumn{1}{c|}{ variant } & \multicolumn{1}{c|}{ $~$source domain$~$ } & \multicolumn{1}{c|}{ $~$needed qubits$~$ } & \multicolumn{1}{c}{ $~$probability$~$ }    \\[0.1cm] \hline
 \multicolumn{1}{c|}{  } & \multicolumn{1}{c|}{  } &  \multicolumn{1}{c|}{ $ ~ {{\underline{\nabla}}^2} $ of Gaussian$~$ }  & \multicolumn{1}{l|}{ $ ~ 1 $ } & \multicolumn{1}{l|}{  } & $ 15 $ & 
 \multicolumn{1}{l}{ $ ~ \approx 0.000006 $ }  
   \\[0.1cm] \cline{4-4} \cline{6-7}
 \multicolumn{1}{c|}{  } & \multicolumn{1}{c|}{ $~$free field$~$ } &  \multicolumn{1}{c|}{  }  & \multicolumn{1}{l|}{ $ ~ 2 $ } & \multicolumn{1}{l|}{ $ ~ {n_x} = 7 $ } & $ 9 $ & 
 \multicolumn{1}{l}{ $ ~ \approx 0.000003 $ } 
  \\[0.1cm] \cline{3-4} \cline{6-7}
 \multicolumn{1}{c|}{  } & \multicolumn{1}{c|}{  } &  \multicolumn{1}{c|}{ Gaussian }  & \multicolumn{1}{l|}{ $ ~ 1 $ } & \multicolumn{1}{l|}{  } & $ 15 $ & 
 \multicolumn{1}{l}{ $ ~ \approx 0.060894 $ }  \\[0.1cm] \cline{4-4} \cline{6-7}
 \multicolumn{1}{c|}{ 1D }  &  \multicolumn{1}{c|}{  }  &  \multicolumn{1}{c|}{  } &  \multicolumn{1}{l|}{ $ ~ 2 $ } & \multicolumn{1}{l|}{  } & $ 9 $ & 
 \multicolumn{1}{l}{ $ ~ \approx 0.030006 $ } \\[0.1cm] \cline{2-7}
  \multicolumn{1}{c|}{  }  & \multicolumn{1}{c|}{  }  &  \multicolumn{1}{c|}{  }  & \multicolumn{1}{l|}{ $ ~ 1 $; analytical values } & \multicolumn{1}{l|}{  } & $ 16 $ & 
 \multicolumn{1}{l}{ $ ~ \approx 0.017421 $ }  \\[0.1cm] \cline{4-4} \cline{6-7}
 \multicolumn{1}{c|}{  }  & \multicolumn{1}{c|}{ periodic }  &  \multicolumn{1}{c|}{ Gaussian }  & \multicolumn{1}{l|}{ $ ~ 2 $; analytical values } & \multicolumn{1}{l|}{ $  ~{n_x} = 8 $ } & $ 9 $ &  
 \multicolumn{1}{l}{ $ ~ \approx 0.010185 $ }  \\[0.1cm] \cline{4-4} \cline{6-7}
   \multicolumn{1}{c|}{  }  & \multicolumn{1}{c|}{  }  &  \multicolumn{1}{c|}{  }  & \multicolumn{1}{l|}{ $ ~ 1 $; DFT of Green function$~$ } & \multicolumn{1}{l|}{  } & $ 16 $ & 
 \multicolumn{1}{l}{ $ ~ \approx 0.017421 $ }  \\[0.1cm] \cline{4-4} \cline{6-7}
 \multicolumn{1}{c|}{  }  & \multicolumn{1}{c|}{  }  &  \multicolumn{1}{c|}{  }  & \multicolumn{1}{l|}{ $ ~ 2 $; DFT of Green function } & \multicolumn{1}{l|}{  } & $ 9 $ & 
 \multicolumn{1}{l}{ $ ~ \approx 0.010185 $ }  \\[0.1cm] \hline
  \multicolumn{1}{c|}{  } & \multicolumn{1}{c|}{  } &  \multicolumn{1}{c|}{ $ {{\underline{\nabla}}^2} $ of Gaussian }  & \multicolumn{1}{l|}{ $ ~ 1 $ } & \multicolumn{1}{l|}{  } & $ 20 $ & 
\multicolumn{1}{l}{ $ ~ \approx 0.057006 $ }  \\[0.1cm] \cline{4-4} \cline{6-7}
 \multicolumn{1}{c|}{  }  &  \multicolumn{1}{c|}{  }  &  \multicolumn{1}{c|}{  } &  \multicolumn{1}{l|}{ $ ~ 2 $ } & \multicolumn{1}{l|}{ $ ~ {n_x} = {n_y} = 4 ~ $ } & $ 11 $ & 
 \multicolumn{1}{l}{ $ ~ \approx 0.030392 $ }  \\[0.1cm] \cline{3-4} \cline{6-7}
   \multicolumn{1}{c|}{ 2D } & \multicolumn{1}{c|}{ free field } &  \multicolumn{1}{c|}{ $~$approximated $ \delta ~ $ }  & \multicolumn{1}{l|}{ $ ~ 1 $ } & \multicolumn{1}{l|}{  } & $ 20 $ & 
 \multicolumn{1}{l}{ $ ~ \approx 0.001832 $ }  
 \\[0.1cm] \cline{4-4} \cline{6-7}
 \multicolumn{1}{c|}{  }  &  \multicolumn{1}{c|}{  }  &  \multicolumn{1}{c|}{  } &  \multicolumn{1}{l|}{ $ ~ 2 $ } & \multicolumn{1}{l|}{  } & $ 11 $ & 
 \multicolumn{1}{l}{ $ ~ \approx 0.000977 $ } 
  \\[0.1cm] \cline{3-7}
  \multicolumn{1}{c|}{  } & \multicolumn{1}{c|}{  } &  \multicolumn{1}{c|}{ $ {{\underline{\nabla}}^2} $ of Gaussian }  & \multicolumn{1}{l|}{ $ ~ 2 $ } & \multicolumn{1}{l|}{ $ ~ {n_x} = {n_y} = 6 ~ $ } & $ 15 $ & \multicolumn{1}{l}{ $ ~ \approx 0.004925 $ }   \\[0.1cm] \cline{3-3} \cline{6-7}
 \multicolumn{1}{c|}{  }  &  \multicolumn{1}{c|}{  }  &  \multicolumn{1}{c|}{ approximated $ \delta $ } &  \multicolumn{1}{l|}{  } & \multicolumn{1}{l|}{  } & $ 15 $ & \multicolumn{1}{l}{ $ ~ \approx 0.000061 $ } 
\end{tabular}
\end{center}
\captionsetup{justification=raggedright, singlelinecheck=false}
\caption[]{Required total number of qubits and success probability for measuring the required configuration of the ancilla qubits for the test examples for which plots of the simulation results were included in this work. 
The values stated for the success probabilities were obtained from the simulated state vector. 
Implementation variant 1 and 2 refer to the used method for implementing the multiplication of the state vector by a diagonal matrix, i.e. the method of subsection {\hyperref[subsubsec:Lin_combination_of_unitary_matrices]{\ref*{subsubsec:Lin_combination_of_unitary_matrices}}} and the method of subsection {\hyperref[subsubsec:Sequence_of_multi-controlled_rotation_gates]{\ref*{subsubsec:Sequence_of_multi-controlled_rotation_gates}}}, respectively, and for the test examples for periodic BCs, it is moreover stated which values were taken as the entries in the multiplied diagonal matrix. 
The stated values for the resolution of the source mean that the source domain was discretized via $ {2^{n_i}} , i \in \{ x, y \} $ points in the corresponding dimension, realized via $ {n_i} $ qubits.}
\label{Fig_Table:Test_examples_with_plots_Parameters_of_q_alg}
\end{table*}

The observations from the plots of the imaginary parts that different orders of magnitude and also different patterns result for the two variants of the quantum circuit implementation are attributed to the respective numerical simulation processes on the classical computer and in general, since our work presents no new numerical method but just the implementation of an established method as a quantum algorithm, the plots of the real and imaginary parts encoded in the state vector were taken here just as a qualitative evidence of the correct implementation. 
Of course, these results, obtained via an ideal simulation, do not give any insight into the quantitative performance of a real quantum computer.

\begin{table*}[t]
\begin{center}
\begin{tabular}[c]{l|c|c|c|c}
 \multicolumn{1}{c|}{ resolution of } & \multicolumn{2}{c|}{ $ ~ {{\underline{\nabla}}^2} $ of Gaussian;$~$ } & \multicolumn{2}{c}{ Gaussian; }   
  \\
  \multicolumn{1}{c|}{ $~$source domain$~$ } & \multicolumn{1}{c|}{ $~$1st variant$~$ } & \multicolumn{1}{c|}{ $~$2nd variant$~$ } & \multicolumn{1}{c|}{ $~$1st variant$~$ } & \multicolumn{1}{c}{ $~$2nd variant$~$ }   
  \\[0.1cm] \hline
 $ ~ \quad ~~ n_x = 4 ~ $ & $ ~ \approx 0.12482331 ~ $ & $ ~ \approx 0.06142557 ~ $ & $ ~ \approx 0.09353838 ~ $ & $ ~ \approx 0.04603025 ~ $  \\[0.1cm] \hline
 $ ~ \quad ~~ n_x = 5 ~ $ & $ ~ \approx 0.00005394 ~ $ & $ ~ \approx 0.00002657 ~ $ & $ ~ \approx 0.05915747 ~ $ & $ ~ \approx 0.02914081 ~ $  \\[0.1cm] \hline
  $ ~ \quad ~~ n_x = 6 ~ $ & $ ~ \approx 0.00000585 ~ $ & $ ~ \approx 0.00000288 ~ $ & $ ~ \approx 0.06089817 ~ $ & $ ~ \approx 0.03000602 ~ $  \\[0.1cm] \hline
  $ ~ \quad ~~ n_x = 7 ~ $ & $ ~ \approx 0.00000550 ~ $ & $ ~ \approx 0.00000271 ~ $ & $ ~ \approx 0.06089410 ~ $ & $ ~ \approx 0.03000596 ~ $  \\[0.1cm] \hline 
  $ ~ \quad ~~ n_x = 8 ~ $ & $ ~ \approx 0.00000543 ~ $ & $ ~ \approx 0.00000268 ~ $ & $ ~ \approx 0.06089331 ~ $ & $ ~ \approx 0.03000606 ~ $  \\[0.1cm] \hline 
  $ ~ \quad ~~ n_x = 9 ~ $ & $ ~ \approx 0.00000541 ~ $ & $ ~ \approx 0.00000267 ~ $ & $ ~ \approx 0.06089311 ~ $ & $ ~ \approx 0.03000608 ~ $  \\[0.1cm] \hline
  $ ~ \quad ~~ n_x = 10 ~ $ & $ ~ \approx 0.00000541 ~ $ & $ ~ \approx 0.00000267 ~ $ & $ ~ \approx 0.06089306 ~ $ & $ ~ \approx 0.03000609 ~ $ 
\end{tabular}
\end{center}
\captionsetup{justification=raggedright, singlelinecheck=false}
\caption[]{Computed success probabilities for the source fields considered for free field conditions in one dimension according to the formulas ({\hyperref[eq:Succes_probability_LCU_specific_problem_here]{\ref*{eq:Succes_probability_LCU_specific_problem_here}}}) and ({\hyperref[eq:Success_probability_sequence_of_multi-controlled_R-gates]{\ref*{eq:Success_probability_sequence_of_multi-controlled_R-gates}}}) for the first and second implementation variant, respectively. 
The table states the success probabilities for increasing the resolution of the source distribution via $ 2^{n_x} $ discretization points for its domain. 
It can be observed from these values that the success probabilities converge.}
\label{Fig_Table:Test_examples_free_field_1D_convergence_study}
\end{table*}

\begin{table*}[t]
\begin{center}
\begin{tabular}[c]{c|c|c|c|c|c}
 \multicolumn{2}{c|}{  } & \multicolumn{2}{c|}{ number of native gates } & \multicolumn{2}{c}{ circuit depth for native gates }   
  \\[0.1cm] 
  \multicolumn{2}{c|}{  } & \multicolumn{1}{c|}{ $~$1-qubit rotation gates$~$ } & \multicolumn{1}{c|}{ $~$CX-gates$~$ } & \multicolumn{1}{c|}{ $~$1-qubit rotation gates$~$ } & \multicolumn{1}{c}{ $~$CX-gates$~$ }   
  \\[0.1cm] \hline
 \multicolumn{2}{l|}{ $~$state preparation w.r.t.$~$ } & $ ~ {2^{n+1}} - 2 ~ $ & $ ~ {2^{n+1}} - 2(n+1) ~ $ & $ ~ {2^{n+1}} - (n+1) ~ $ & $ ~ {2^{n+1}} - 2(n+1) ~ $  \\
  \multicolumn{2}{l|}{ $~$amplitude encoding$~$ } &  &  &  &  \\[0.1cm] \hline
 \multicolumn{1}{l|}{ $~$1D-QFT$~$ } & \multicolumn{1}{l|}{ $~$H- and CP-gates$~$ } & $ ~ 2n + {\frac{3}{2}}n(n-1) ~ $ & $ ~ n(n-1) ~ $ & $ ~ 4n - 2 ~ $ & \multicolumn{1}{l}{ $ ~ \text{for } n > 1: ~ 4n - 6 ~ $ }  \\[0.1cm] \cline{2-6}
 \multicolumn{1}{l|}{  } & \multicolumn{1}{l|}{ $~$multi-qubit-SWAP gate$~$ } & $ ~ 0 ~ $ & $ ~ 3 \cdot {\left\lfloor {\frac{n}{2}} \right\rfloor} ~ $ & $ ~ 0 ~ $ & \multicolumn{1}{l}{ $ ~ \text{for } n > 1: ~ 3 ~ $ }  \\[0.1cm] \hline
 \multicolumn{2}{l|}{ $~$SELECT-gate block of$~$ } & $ ~ {2^{n-1}} n - {2^{{\frac{n}{2}}-2}} n ~ $ & $ ~ {2^{n-1}} n - {2^{{\frac{n}{2}}-1}} n ~ $ & $ ~ {2^{n-1}} n - {2^{{\frac{n}{2}}-2}} n \left( {\frac{n}{2}} + 1 \right) ~ $ & $ ~ {2^{n-1}} n - {2^{{\frac{n}{2}}-1}}  n ~ $  \\ 
  \multicolumn{2}{l|}{ $~$LCU circuit (for $ n = 2r , r \in \mathbb{N} $)$~$ } &  &  &  &  \\[0.1cm] \hline 
 \multicolumn{2}{l|}{ $~$sequence of multi-controlled$~$ } & $ ~ {2^{2n-1}} - {2^{n}} ~ $ & $ ~ {2^{2n-1}} - {2^{n+1}} + 2 ~ $ & $ ~ {2^{2n-1}} - {2^{n}} ~ $ & $ ~ {2^{2n-1}} - {2^{n+1}} + 2 ~ $ \\
  \multicolumn{2}{l|}{ $~$rotation gates$~$ } &  &  &  &  
\end{tabular}
\end{center}
\captionsetup{justification=raggedright, singlelinecheck=false}
\caption[]{Estimated contributions to the number of gates and the circuit depth for the individual routines in the presented quantum algorithm w.r.t. their application to $ n $ qubits. 
The contributions are referred to native 1- and 2-qubit gates, for which the 1-qubit rotation gates and the CX-gate were considered here, respectively as in \cite{Shende_et_al_Synthesis_of_q_circuits}. 
The formulas entered here for 'sequence of multi-controlled rotation gates' are referred to the second implementation variant of the multiplication step for the free field case.}
\label{Fig_Table:Estimated_computational_resources_scaling_formulas}
\end{table*}

Concerning parameters of the quantum algorithm, Table {\hyperref[Fig_Table:Test_examples_with_plots_Parameters_of_q_alg]{\ref*{Fig_Table:Test_examples_with_plots_Parameters_of_q_alg}}} documents the needed qubit numbers and the success probabilities for ending up in the desired ancilla qubit subspace for the stated test examples. 
The stated values of the success probabilities were obtained from the state vector simulation (but they can also be computed via the formulas ({\hyperref[eq:Succes_probability_LCU_specific_problem_here]{\ref*{eq:Succes_probability_LCU_specific_problem_here}}}) and ({\hyperref[eq:Success_probability_sequence_of_multi-controlled_R-gates]{\ref*{eq:Success_probability_sequence_of_multi-controlled_R-gates}}}) for the first and second implementation variant, respectively). 
It can be noted from Table {\hyperref[Fig_Table:Test_examples_with_plots_Parameters_of_q_alg]{\ref*{Fig_Table:Test_examples_with_plots_Parameters_of_q_alg}}} that although the desired ancilla subspace in the LCU variant is smaller relative to the full state space of the system, the success probability for ending up in it is higher than that in the variant via the specific sequence of multi-controlled rotation gates for all considered test examples. 
The success probability of the used LCU implementation is always a factor of around two higher than the probability of the alternative implementation, where however the exact value of this factor varies from case to case. 
For the specific version of the LCU method with tensor products of $ {{\bf{1}}_2}, {{\sigma}_x}, {{\sigma}_y}, {{\sigma}_z} $ as basis matrices used here (cf. subsection {\hyperref[subsubsec:Lin_combination_of_unitary_matrices]{\ref*{subsubsec:Lin_combination_of_unitary_matrices}}}), there are $ 4^n $ basis matrices for the representation of a general matrix of dimension $ {2^n} $\texttimes $ {2^n} $. 
However, for the diagonal matrices treated here, it was observed that non-vanishing expansion coefficients occur only for tensor products of $ {{\bf{1}}_2} $ and $ {{\sigma}_z} $. 
In accordance with this, it was observed for the considered test cases that the number of required qubits in the LCU version just doubles due to the needed ancilla qubits except for the free field test examples for one dimension, for which one ancilla qubit less is enough.

In general, it can be seen from Table {\hyperref[Fig_Table:Test_examples_with_plots_Parameters_of_q_alg]{\ref*{Fig_Table:Test_examples_with_plots_Parameters_of_q_alg}}} that the success probability is highly dependent on the specific problem that is considered. 
However, it can be seen that the success probabilities of both implementation variants converge with increasing resolution for a fixed source distribution (which does accordingly not hold for the used procedure to approximate a $ \delta $-distribution since for it, the shape of the numerically represented source distribution depends on the grid spacing). 
For this, Table {\hyperref[Fig_Table:Test_examples_free_field_1D_convergence_study]{\ref*{Fig_Table:Test_examples_free_field_1D_convergence_study}}} documents the respective probability values for the one-dimensional free field test examples, where these values were obtained by computing them via the formulas ({\hyperref[eq:Succes_probability_LCU_specific_problem_here]{\ref*{eq:Succes_probability_LCU_specific_problem_here}}}) and ({\hyperref[eq:Success_probability_sequence_of_multi-controlled_R-gates]{\ref*{eq:Success_probability_sequence_of_multi-controlled_R-gates}}}). 
For this, it is however to note that the situation is not sufficiently resolved before $ N_x = 6 $.

\section{Estimations of Required Computational Resources}
\label{sec:Estimations_of_required_computational_resources}

Since the used procedure for numerically solving the Poisson equation starts and ends with the conventional QFT, it requires an input state according to the conventional amplitude encoding w.r.t. multiple qubits and also produces an output state of this encoding format. 
If this procedure is considered as a stand-alone quantum circuit, i.e. in the form of the presented quantum algorithm according to Fig. {\hyperref[Fig:Q_alg_modules]{\ref*{Fig:Q_alg_modules}}} and not as a building block in a larger quantum circuit, therefore also the computational efforts of the initialization and the read-out w.r.t. this encoding format have to be taken into account.

\begin{figure*}[t]
\begin{adjustbox}{width=1.0\textwidth}
\tikzset{invisible/.style={fill=none,draw=none,line width=0pt,inner xsep=0pt,inner ysep=0pt}}
\begin{quantikz}
\lstick{${q_0}$} & \gategroup[1,steps=1, style={invisible}, label style={label position=above,anchor=mid,yshift=-0.3cm, xshift=-0.2cm}]{LSB} &[-0.2cm]  &[-0.2cm]  &[-0.2cm] \ \ldots\  &[-0.2cm]  &[-0.2cm]  &[-0.2cm] \ctrl{5} &[-0.2cm] \ \ldots\  &[-0.2cm] &[-0.2cm]  &[-0.2cm] \ctrl{2} &[-0.2cm] &[-0.2cm] \ctrl{1} &[-0.2cm]  \gate{H} & 
 \swap{5}\gategroup[6, steps=3, style={dashed, rounded corners}, background, label style={label position=below, anchor=north,yshift=-0.2cm}]{{\text{${n}$-qubit-SWAP}}} & & 
 & \\
 \lstick{${q_1}$} &  &  &   & \ \ldots\ &  & \ctrl{4} &  &  \ \ldots\  &  & \ctrl{1} &  &  \gate{H}  & \gate{P({\frac{2{\pi}}{2^{2}}})} &  &  & \swap{3} & &  \\
 \lstick{${q_2}$} &  &  &  & \ \ldots\ & \ctrl{3} &  &  &  \ \ldots\  & \gate{H}  & \gate{P({\frac{2{\pi}}{2^{2}}})} &  \gate{P({\frac{2{\pi}}{2^{3}}})} & &  &  & & & \swap{1} & \\
\lstick{$ \vdots ~~~ $} & \wave&&&&&&&&&&&&&&&& &\\
 \lstick{${q_{{n}-2}}$} &  &  & \ctrl{1} & \ \ldots\ &  &  &  & \ \ldots\  & &  &  &  &  &  & & \targX{} & &  \\
 \lstick{${q_{{n}-1}}$} & \gategroup[1,steps=1, style={invisible}, label style={label position=below,anchor=mid,yshift=-0.05cm, xshift=-0.2cm}]{MSB} & \gate{H} & \gate{P({\frac{2{\pi}}{2^{2}}})} & \ \ldots\ & \gate{P({\frac{2{\pi}}{2^{{n}-2}}})} & \gate{P({\frac{2{\pi}}{2^{{n}-1}}})} & \gate{P({\frac{2{\pi}}{2^{{n}}}})} & \ \ldots\  &  & & &  &  &  & \targX{} &    & & 
\end{quantikz}
\end{adjustbox}
\caption[]{Quantum circuit pattern of the one-dimensional QFT.}
\label{Fig:1-dim_QFT}
\end{figure*}
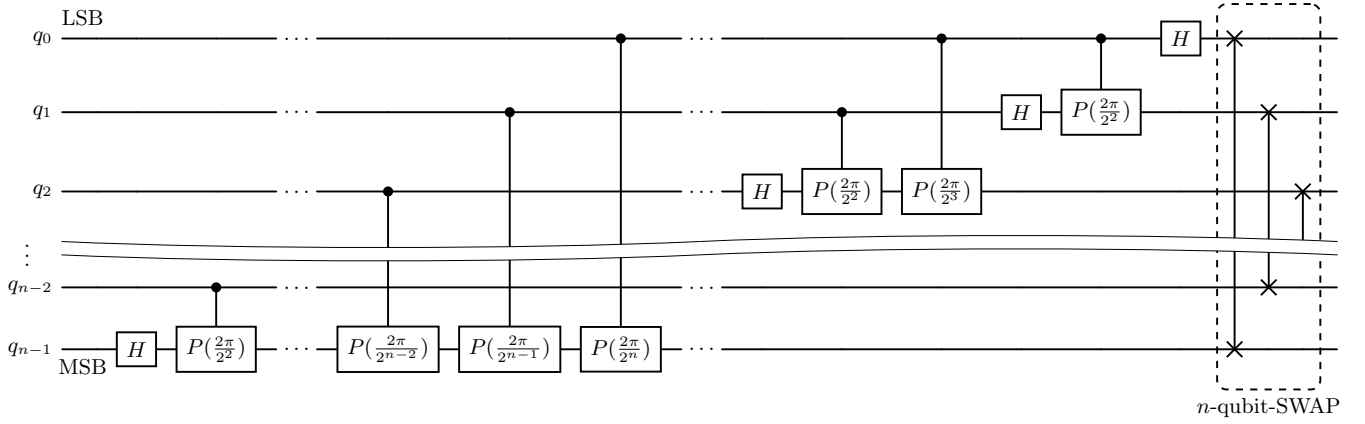

For this setup, the number of required qubits is directly apparent: 
Storing the possibly padded source field that is given via $ {N_{\psi}} = {2^{n_{\psi}}} , {n_{\psi}} \in \mathbb{N} $ values in amplitude encoding, demands $ n_{\psi} $ qubits and as documented in the previous two sections, the multiplication step requires up to the same amount of ancilla qubits for the use of the LCU variant and one ancilla qubit for the use of the specific sequence of multi-controlled rotation gates. 
This section gives now estimations for the computational efforts w.r.t. the number of gates and the runtime needed for the individual circuit routines of the algorithm. 
As a measure for the runtime, the circuit depth is considered here, which means the minimum number of {\emph{gate layers}} that have to be performed in one run of the quantum algorithm. 
A gate layer is a group of gates in a quantum circuit diagram that can be executed in parallel, defining a processing cycle, for which the runtime is given by the execution time of the gate in this layer that demands the most. 
Table {\hyperref[Fig_Table:Estimated_computational_resources_scaling_formulas]{\ref*{Fig_Table:Estimated_computational_resources_scaling_formulas}}} summarizes the scaling formulas found for the application of the listed circuit routines to $ n $ qubits. 
They were obtained following mainly the rules given in V. V. Shende et al.  \cite{Shende_et_al_Synthesis_of_q_circuits} for decomposing depicted gates in terms of native gates, i.e. gates that are in fact physically implemented. 
In \cite{Shende_et_al_Synthesis_of_q_circuits}, the three 1-qubit rotation gates $ R_i, i \in \{ x, y, z \} $ in the form
\begin{align}
{R_x}({\theta})&= 
\begin{pmatrix}
{\cos{\left( {\frac{\theta}{2}} \right)}} & {\text{i}} ~ {\sin{\left( {\frac{\theta}{2}} \right)}} \\
{\text{i}} ~ {\sin{\left( {\frac{\theta}{2}} \right)}} & {\cos{\left( {\frac{\theta}{2}} \right)}}
\end{pmatrix} , \label{eq:Rx_matrix_form} \\
{R_y}({\theta})&= 
\begin{pmatrix}
{\cos{\left( {\frac{\theta}{2}} \right)}} & {\sin{\left( {\frac{\theta}{2}} \right)}} 
 \\
-{\sin{\left( {\frac{\theta}{2}} \right)}} & {\cos{\left( {\frac{\theta}{2}} \right)}}
\end{pmatrix} , \label{eq:Ry_matrix_form} \\
{R_z}({\theta})&= 
\begin{pmatrix}
{\exp{\left( -{\text{i}} {\frac{\theta}{2}} \right)}} & 0 \\
0 & {\exp{\left( {\text{i}} {\frac{\theta}{2}} \right)}}
\end{pmatrix}
 \label{eq:Rz_matrix_form}
\end{align}
and the controlled NOT-gate (denoted here as CX-gate) are used as native gates. 
Furthermore, it was assumed here all-to-all connectivity for the estimations in Table {\hyperref[Fig_Table:Estimated_computational_resources_scaling_formulas]{\ref*{Fig_Table:Estimated_computational_resources_scaling_formulas}}} and that gates can be executed in parallel if they do not involve the same qubits (which means e.g. that if the control and target qubit of a CX-gate span over one qubit in between of them, a gate involving this qubit in between can be executed in parallel). 
To obtain the actual runtime of a routine from the individual circuit depth contributions, the stated numbers of layers of 1-qubit rotation gates and CX-gates would have to be multiplied by the execution time of the respective gate type and summed up then.

In the following three subsections, the derivations of the found scaling formulas are explained and further implications of the used routines in the quantum algorithm for its performance w.r.t. resources are discussed. 
Finally, it is commented in the fourth subsection on the number of runs of the quantum algorithm that should be necessary to infer the probability distribution of the state vector prepared by the setup of Fig. {\hyperref[Fig:Q_alg_modules]{\ref*{Fig:Q_alg_modules}}}.

\subsection{State Preparation}
\label{subsec:Resource_estimations_state_preparation}

In the quantum algorithm, a procedure for preparing an arbitrary state in amplitude encoding, starting from the state in which all qubits are in the $ | 0 \rangle $-state, has to be performed for loading the source field as the initial state and also in terms of the two $ [PREP] $-gate blocks if the LCU implementation is used. 
For the latter, it is to note that $ {[PREP]}^{\dag} $ can be realized just by the application of a corresponding gate sequence for $ [PREP] $ in reversed ordering and with negative angles for the 1-qubit rotation gates. 
Here, the procedure according to Theorem 9 in \cite{Shende_et_al_Synthesis_of_q_circuits} is considered for the task of state preparation. 
From the decomposition described in  \cite{Shende_et_al_Synthesis_of_q_circuits} for this procedure, the formulas given in Table {\hyperref[Fig_Table:Estimated_computational_resources_scaling_formulas]{\ref*{Fig_Table:Estimated_computational_resources_scaling_formulas}}} for 'state preparation w.r.t. amplitude encoding' can be deduced (see  \cite{Koesel_et_al_Resource_implications_QCFD_arxiv_v1} for an explicit documentation of the derivations, in particular the formulas that are not explicitly stated in \cite{Shende_et_al_Synthesis_of_q_circuits}, like e.g. the circuit depth contribution of the 1-qubit rotation gates).

As explained in subsection {\hyperref[subsec:Initialization]{\ref*{subsec:Initialization}}}, the state preparation in the initialization gate block involves also for free field conditions effectively just the qubits of the $ \psi $-register that are set to resolve the domain of the source. 
For the use of the LCU implementation however, as observed in section {\hyperref[sec:Functionality_tests]{\ref*{sec:Functionality_tests}}}, the number of qubits that has to be involved in the $ [PREP] $-operations is in general as large as the number of qubits in the full $ \psi $-register, i.e. including the qubits for a padding of the source domain.

\subsection{Quantum Fourier Transform}

There are two multi-dimensional QFT procedures in the quantum algorithm according to Fig. {\hyperref[Fig:Q_alg_modules]{\ref*{Fig:Q_alg_modules}}} (since '$ {[ \text{multi-dimensional QFT} ]}^{\dag} $' is given by '$ {[ \text{multi-dimensional QFT} ]} $' with negative angles in the CP-gates). 
In subsection {\hyperref[subsec:QFT]{\ref*{subsec:QFT}}}, it was explained that the multi-dimensional QFT results just via the parallel application of one-dimensional QFT circuits to subregisters of the register $ \psi $ that are set for a specific dimension $ r $ of the computational domain, for which they form $ 2^{n_r} $ basis states to represent this amount of discretization points, respectively. 
Accordingly, the number of gates in a multi-dimensional QFT is given by summing up the numbers of gates of the individual one-dimensional QFTs and its circuit depth is given by the circuit depth of the one-dimensional QFT that involves the most qubits, i.e. the one that is associated to the largest $ n_r $.

The circuit pattern of the one-dimensional QFT for the application to $ n $ qubits is depicted again in Fig. {\hyperref[Fig:1-dim_QFT]{\ref*{Fig:1-dim_QFT}}}. 
It is considered at first the gate sequence before the $ n $-qubit-SWAP gate block:

For each qubit, there is an H-gate, which has the matrix representation $ {\frac{1}{\sqrt{2}}}  \left( \begin{smallmatrix} 1 & 1 \\ 1 & -1  \end{smallmatrix} \right) $. 
For the use of the conventions ({\hyperref[eq:Rx_matrix_form]{\ref*{eq:Rx_matrix_form}}}) and ({\hyperref[eq:Ry_matrix_form]{\ref*{eq:Ry_matrix_form}}}), it results that an H-gate can be decomposed as an $ R_x $-gate followed by an $ R_y $-gate according to
\begin{align}
{R_x}({\pi}) \cdot {R_y}{\left( -{\frac{\pi}{2}} \right)} &= {\text{i}}
\begin{pmatrix}
 0 & 1 \\
 1 & 0
\end{pmatrix} 
 \cdot {\frac{1}{\sqrt{2}}}
\begin{pmatrix}
 1 & -1 \\
 1 & 1
\end{pmatrix} \nonumber \\
 &= {\frac{{\text{i}}}{\sqrt{2}}}
\begin{pmatrix}
 1 & 1 \\
 1 & -1
\end{pmatrix} 
 = {\text{i}} H 
\label{eq:H_decomposition_matrix_form}
\end{align}
since a global complex phase factor like $ \text{i} $ is physically undetectable. 
This gives a contribution of $ 2n $ to the number of needed 1-qubit rotation gates.

For the decomposition of the CP-gates, given in matrix representation via ({\hyperref[eq:CP_matrix_form]{\ref*{eq:CP_matrix_form}}}), it is to note at first
\begin{align}
{U_{CP}}({\phi}) &= 
\begin{pmatrix}
1 & 0 & 0 & 0 \\
0 & 1 & 0 & 0 \\
0 & 0 & 1 & 0 \\
0 & 0 & 0 & {e^{{\text{i}} {\phi} }}
\end{pmatrix} \nonumber \\ 
 &= 
\begin{pmatrix}
1 & 0 & 0 & 0 \\
0 & 1 & 0 & 0 \\
0 & 0 & {e^{{\text{i}} {\frac{\phi}{2}} }} & 0 \\
0 & 0 & 0 & {e^{{\text{i}} {\frac{\phi}{2}} }}
\end{pmatrix} 
 \cdot 
\begin{pmatrix}
1 & 0 & 0 & 0 \\
0 & 1 & 0 & 0 \\
0 & 0 & {e^{-{\text{i}} {\frac{\phi}{2}} }} & 0 \\
0 & 0 & 0 & {e^{{\text{i}} {\frac{\phi}{2}} }}
\end{pmatrix} .
\end{align}
This means the application of a controlled gate that has the effect of a multiplication by $ {e^{{\text{i}} {\frac{\phi}{2}} }} $ for the states $ | 0 \rangle $ and $ | 1 \rangle $ of the target qubit and a controlled $ {R_z} $-gate ({\hyperref[eq:Rz_matrix_form]{\ref*{eq:Rz_matrix_form}}}) with argument $ \theta = \phi $ (cf. Theorem 7 in \cite{Shende_et_al_Synthesis_of_q_circuits}), where the order in which these operations are applied does not matter, i.e.:
\begin{align}
\begin{quantikz}[align equals at=1.5]
 & \ctrl{1} &  \\ 
 & \gate{P({\phi})} & 
\end{quantikz}
 ~ & \widehat{=} ~
 \begin{quantikz}[align equals at=1.5]
 & \ctrl{1} & \ctrl{1} &  \\ 
 & \gate{ {e^{{\text{i}} {\frac{\phi}{2}} }}  } & \gate{{R_z}({\phi})} & 
\end{quantikz} \nonumber \\
 ~ & \widehat{=} ~
  \begin{quantikz}[align equals at=1.5]
 & \ctrl{1} & \ctrl{1} &  \\ 
 & \gate{{R_z}({\phi})} & \gate{ {e^{{\text{i}} {\frac{\phi}{2}} }}  } &
\end{quantikz}
 \label{eq:circuit_diagram_Decomposition_CP}
\end{align}
Further, Theorem 3 in \cite{Shende_et_al_Synthesis_of_q_circuits} states that the controlled gate mediating a scalar multiplication for the target qubit is equivalent to the application of a diagonal gate to the control qubit according to
\begin{align}
\begin{quantikz}[align equals at=1.5]
 & \ctrl{1} &  \\ 
  & \gate{ {e^{{\text{i}} {\frac{\phi}{2}} }}  }  & 
\end{quantikz}
 ~ & \widehat{=} ~
 \begin{quantikz}[align equals at=1.5]
 & \gate{P({\frac{\phi}{2}})} &   \\ 
  & &  
\end{quantikz} 
\end{align}
or in matrix representation
\begin{align}
\begin{pmatrix}
1 & 0 & 0 & 0 \\
0 & 1 & 0 & 0 \\
0 & 0 & {e^{{\text{i}} {\frac{\phi}{2}} }} & 0 \\
0 & 0 & 0 & {e^{{\text{i}} {\frac{\phi}{2}} }}
\end{pmatrix} 
 &= 
 {\underbrace{ \begin{pmatrix} 1 & 0 \\ 0 & {e^{{\text{i}} {\frac{\phi}{2}} }} \end{pmatrix} }_{ = P({\frac{\phi}{2}}) }}
 \otimes 
 \begin{pmatrix} 1 & 0 \\ 0 & 1 \end{pmatrix}
 ,
\end{align}
where $ P $ is the phase shift gate (abbreviated as P-gate). 
Since global phase factors are undetectable, a P-gate corresponds just to an $ R_z $-gate. 
For the controlled $ R_z $-gate in ({\hyperref[eq:circuit_diagram_Decomposition_CP]{\ref*{eq:circuit_diagram_Decomposition_CP}}}), it follows according to Theorem 4 in \cite{Shende_et_al_Synthesis_of_q_circuits} the decomposition into two $ R_z $-gates and two CX-gates, where the ordering of the gates can be reversed:
\begin{align}
\begin{quantikz}[align equals at=1.5]
 & \ctrl{1} &  \\ 
 & \gate{{R_z}({\phi})} & 
\end{quantikz}
 ~ & \widehat{=} ~
 \begin{quantikz}[align equals at=1.5]
 & & \ctrl{1} & & \ctrl{1} &  \\ 
 & \gate{ {R_z}({\frac{\phi}{2}}) } & \targ{} &  \gate{ {R_z}(-{\frac{\phi}{2}}) } & \targ{} & 
\end{quantikz} \nonumber \\
 ~ & \widehat{=} ~
  \begin{quantikz}[align equals at=1.5]
 & \ctrl{1} & & \ctrl{1} & &  \\ 
 & \targ{} & \gate{ {R_z}(-{\frac{\phi}{2}}) } & \targ{} & \gate{ {R_z}({\frac{\phi}{2}})  } &
\end{quantikz}
 \label{eq:circuit_diagram_Decomposition_C-Rz}
\end{align}

\begin{figure*}[t]
\sbox0{\begin{adjustbox}{width=1.0\textwidth}
\parbox{0.75\textwidth}{
\tikzset{ invisible/.style={fill=none,draw=none,line width=0pt,inner xsep=0pt,inner ysep=0pt}}
\begin{align*}
 & 
\begin{quantikz}[row sep=0.4cm]
\lstick{${q_0}$} & \gategroup[1,steps=1, style={invisible}, label style={label position=above,anchor=mid,yshift=-0.3cm, xshift=-0.2cm}]{LSB} &[-0.2cm]  &[-0.2cm] \ghost{R_z}\gategroup[7, steps=21, style={dashed, blue, rounded corners, inner xsep=2pt, inner ysep = 0.65cm}, background, label style={label position=above, anchor=north,yshift=0.3cm}]{ $ {[{\text{block}}]}_{1} $ }
 &[-0.2cm] &[-0.2cm]  &[-0.2cm]  &[-0.2cm] &[-0.2cm] &[-0.2cm]  &[-0.2cm]  &[-0.2cm] \ \ldots\  &[-0.2cm] &[-0.2cm] &[-0.2cm] &[-0.2cm]   &[-0.2cm]  &[-0.2cm] &[-0.2cm] &[-0.2cm]  &[-0.2cm] \gate{{R_z}} \gategroup[7, steps=4, style={dashed, red, rounded corners, inner sep = 0pt}, background, label style={label position=below, anchor=north,yshift=-0.2cm}]{ $ {[{\text{subblock}}]}_{1, n-1} $ } &[-0.2cm] \ctrl{6} &[-0.2cm] &[-0.2cm] \ctrl{6} &[-0.2cm] &[-0.2cm] &[-0.2cm] &[-0.2cm] &[-0.2cm] &[-0.2cm] &[-0.2cm] &[-0.2cm] &[-0.2cm] &[-0.2cm]  \ \ldots\   \\
 \lstick{${q_1}$} &  &  &  &  &  &  &  &  &  &  & \ \ldots\ & & & &  & \gate{{R_z}}\gategroup[6, steps=4, style={dashed, red, rounded corners, inner sep = 0pt}, background, label style={label position=below, anchor=north,yshift=-0.2cm}]{ $ {[{\text{subblock}}]}_{1, n-2} $ }  & \ctrl{5} & & \ctrl{5} & & &   &  & & & & & & & & & &  \ \ldots\  \\
 \lstick{${q_2}$} &  &  &  &  &  &  &  &  &  & &  \ \ldots\ &  \gate{{R_z}}\gategroup[5, steps=4, style={dashed, red, rounded corners, inner sep = 0pt}, background, label style={label position=below, anchor=north,yshift=-0.2cm}]{ $ {[{\text{subblock}}]}_{1, n-3} $ }  &  \ctrl{4} &  & \ctrl{4} & & & & & & &  &  & & & & & & & & & &   \ \ldots\  \\[0.5cm]
\lstick{$ \vdots ~~~ $} & \wave&&&&&&&&&&&&&&&&&&&&&&&&&&&&&&&& \\
 \lstick{${q_{{n}-3}}$} &  &  &   &  & &  & \gate{R_z}\gategroup[3, steps=4, style={dashed, red, rounded corners, inner sep = 0pt}, background, label style={label position=below, anchor=north,yshift=-0.2cm}]{ $ {[{\text{subblock}}]}_{1,2} $ } & \ctrl{2}  &  & \ctrl{2} &  \ \ldots\ & & & &  &  &  &  &  &  &  &  &  &  &  
  & \ctrl{1}\gategroup[2, steps=4, style={dashed, red, rounded corners, inner sep = 0pt}, background, label style={label position=below, anchor=north,yshift=-0.2cm}]{ $ {[{\text{subblock}}]}_{2,1} $ } & & \ctrl{1} & \gate{{R_z}} &  & & &  \ \ldots\   \\
 \lstick{${q_{{n}-2}}$} &  &  &  \gate{{R_z}}\gategroup[2, steps=4, style={dashed, red, rounded corners, inner sep = 0pt}, background, label style={label position=below, anchor=north,yshift=-0.2cm}]{ $ {[{\text{subblock}}]}_{1,1} $ }  & \ctrl{1} & & \ctrl{1}  &  &  &  &  & \ \ldots\ &  &  & & & &  &  &  &  &  &  &  & \gate{H} & \gate{{R_z}} & \targ{} & \gate{{R_z}} & \targ{} & \gate{{R_z}} & \targ{}\wire[u][2]{q} & \gate{{R_z}} & \targ{}\wire[u][2]{q} &  \ \ldots\   \\
 \lstick{${q_{{n}-1}}$} & \gategroup[1,steps=1, style={invisible}, label style={label position=below,anchor=mid,yshift=-0.075cm, xshift=-0.2cm}]{MSB} & \gate{H} & \gate{ {R_z} } & \targ{} & \gate{ {R_z} } & \targ{}  & \gate{ {R_z} } & \targ{} & \gate{ {R_z} } & \targ{} & \ \ldots\ & \gate{{R_z}}  & \targ{} & \gate{{R_z}}  & \targ{} & \gate{{R_z}}  & \targ{} & \gate{{R_z}}  & \targ{} & \gate{{R_z}}   & \targ{} & \gate{{R_z}}  & \targ{} & & & & & & & & & &    \ \ldots\   
\end{quantikz} \\[1.0cm]
 & 
\begin{quantikz}[row sep=0.4cm]
\lstick{${q_0}$}   \ \ldots\  &[-0.2cm] \gate{R_z} &[-0.2cm] \ctrl{3} &[-0.2cm] \hphantomgate{wide} &[-0.2cm] \ctrl{3} &[-0.2cm] &[-0.2cm] &[-0.2cm] \ghost{R_z}\gategroup[7, steps=8, style={dashed, blue, rounded corners, inner xsep=2pt, inner ysep = 0.65cm}, background, label style={label position=above, anchor=north,yshift=0.3cm}]{ $ {[{\text{block}}]}_{n-2} $ } &[-0.2cm]   &[-0.2cm] &[-0.2cm]  &[-0.2cm]  \gate{R_z}\gategroup[3, steps=4, style={dashed, red, rounded corners, inner sep = 0pt}, background, label style={label position=below, anchor=north,yshift=-0.2cm}]{ $ {[{\text{subblock}}]}_{n-2, 2} $ }  &[-0.2cm] \ctrl{2}  &[-0.2cm]  &[-0.2cm] \ctrl{2} &[-0.2cm]   &[-0.2cm] &[-0.2cm]  \gate{R_z}\gategroup[2, steps=4, style={dashed, red, rounded corners, inner sep = 0pt}, background, label style={label position=below, anchor=north,yshift=-0.2cm}]{ $ {[{\text{subblock}}]}_{n-1, 1} $ } \gategroup[7, steps=4, style={dashed, blue, rounded corners, inner xsep=2pt, inner ysep = 0.65cm}, background, label style={label position=above, anchor=north,yshift=0.3cm}]{ $ {[{\text{block}}]}_{n-1} $ } &[-0.2cm] \ctrl{1} &[-0.2cm]  &[-0.2cm] \ctrl{1} &[-0.2cm]  \gate{H} &   \\
 \lstick{${q_1}$}   \ \ldots\    &  &  &  &  &  &  & \ctrl{1}\gategroup[2, steps=4, style={dashed, red, rounded corners, inner sep = 0pt}, background, label style={label position=below, anchor=north,yshift=-0.2cm}]{ $ {[{\text{subblock}}]}_{n-2, 1} $ } &  & \ctrl{1} & \gate{R_z}  &  &  &  &  &  \gate{H} & \gate{R_z}  &  & \targ{} & \gate{R_z}  &  \targ{} &  &  \\
 \lstick{${q_2}$}  \ \ldots\  &  &  &  &  & \gate{H} & \gate{R_z} &  \targ{} &  \gate{R_z}  &  \targ{} &  \gate{R_z} &  & \targ{} & \gate{R_z} &  \targ{}  &  &  &  &  &  &  &  &   \\[0.5cm]
\lstick{$ \vdots ~~~ $}  \wave&&&&&&&&&&&&&&&&&&&&&&&&\\
 \lstick{${q_{{n}-3}}$}   \ \ldots\  & & & & & & & & &  &  &  & & & &  &  &  &  &  &  & \ghost{R_z} &  \\
 \lstick{${q_{{n}-2}}$}   \ \ldots\  & & & & & & & &  &  &  &  &  &  & & & &  &  &  &  & \ghost{R_z}  &   \\
 \lstick{${q_{{n}-1}}$}  \ \ldots\  & & & &  & & & & & &  &  &  &  &  &  &  & & &  &  & \ghost{R_z} &  
\end{quantikz}
\end{align*}
}
\end{adjustbox}}\usebox0
\captionsetup{justification=raggedright, singlelinecheck=false}
\caption[]{Quantum circuit pattern of the decomposition of the one-dimensional QFT considered here. 
The lower row continues the upper row.}
\label{Fig:1-dim_QFT_decomposition_pattern}
\end{figure*}
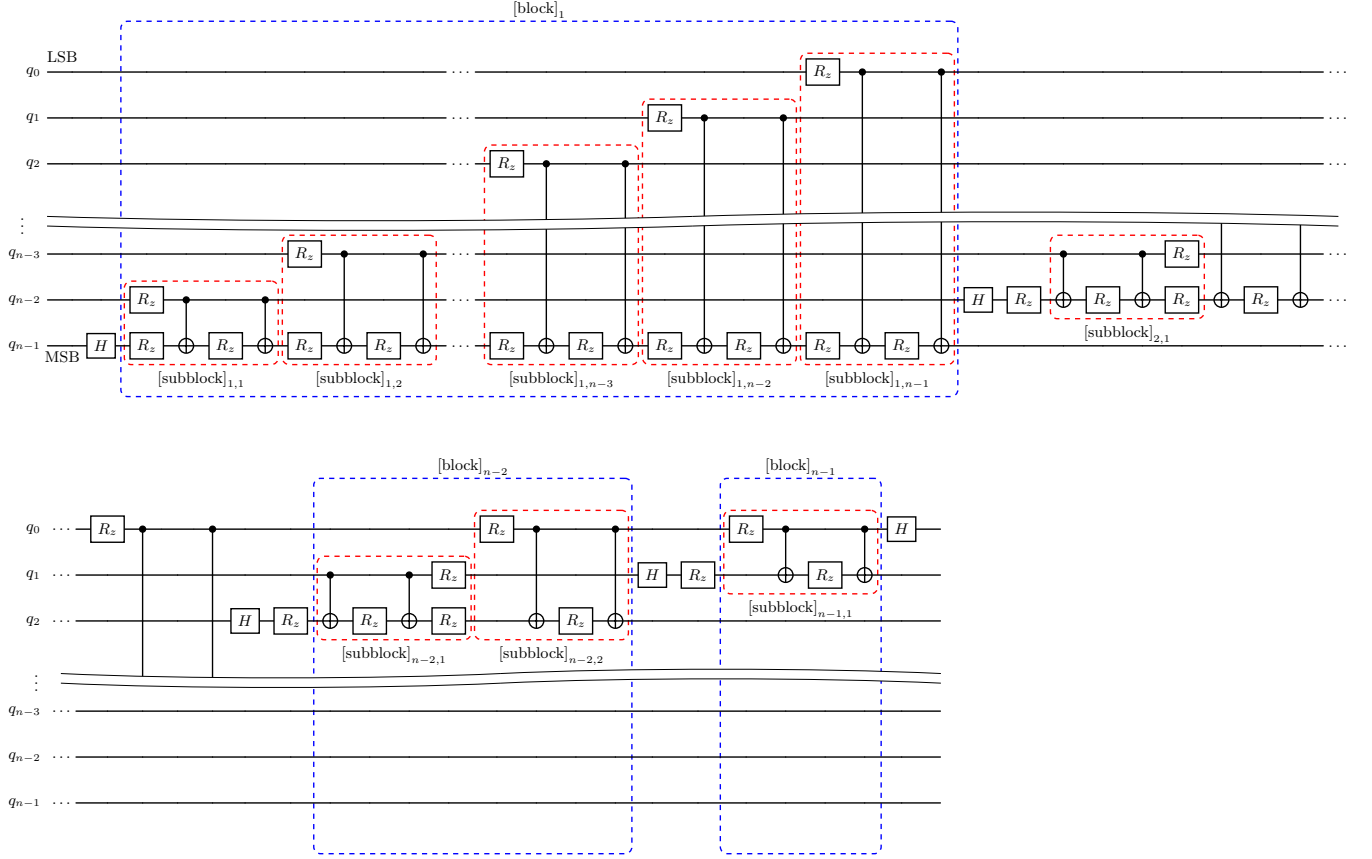

At this point, it shall be remarked that the decomposition rules in \cite{Shende_et_al_Synthesis_of_q_circuits} are given for the case that the control qubits are the MSBs and there are no further qubits between the control and the target qubits. 
The decomposition rules should apply however also if there are further uninvolved qubits and the bit ordering of the qubits is different since further qubits that are isolated from the system at this point of the circuit by definition do not influence the physical processes that the gates represent and the assigned bit ordering is irrelevant for the physically performed operations but is just defined w.r.t. the encoding for the computation. 
Thus, the decomposition of a controlled gate
\begin{align*}
\begin{quantikz}
 & \ctrl{4} &  \\ 
 &  &  \\
 \wave&& \\
 &  &  \\
 & \gate{U} & 
\end{quantikz}
\end{align*}
should always be the same w.r.t. control and target qubits and for the QFT without the $ n $-qubit-SWAP block, it can be obtained the pattern of Fig. {\hyperref[Fig:1-dim_QFT_decomposition_pattern]{\ref*{Fig:1-dim_QFT_decomposition_pattern}}}, where the H-gates are shown not decomposed.

The structure of this pattern is as follows: 
The first gate block $ {[{\text{block}}]}_{1} $ is given via the CP-gates between the first and the second H-gate in the pattern of Fig. {\hyperref[Fig:1-dim_QFT]{\ref*{Fig:1-dim_QFT}}}, i.e. the sequence of CP-gates between the H-gate acting on qubit $ {q_{n-1}} $ and the H-gate acting on qubit $ {q_{n-2}} $. 
For these CP-gates, the decomposition according to the upper row of the relation ({\hyperref[eq:circuit_diagram_Decomposition_C-Rz]{\ref*{eq:circuit_diagram_Decomposition_C-Rz}}}) was inserted, with the $ R_z $-gate resulting for the control qubit in the decomposition ({\hyperref[eq:circuit_diagram_Decomposition_CP]{\ref*{eq:circuit_diagram_Decomposition_CP}}}) added on the left. 
For the $ k $th CP-gate in this first block, where $ k \in \{ 1, \dots, n-1 \} $, this decomposition is defined as $ {[{\text{subblock}}]}_{1, k} $. 
In general, the {\emph{length}} of a controlled gate is defined here as the number of qubits spanned between the target and the control qubit plus one. 
E.g., for the CX-gates in $ {[{\text{subblock}}]}_{1, k} $, the length is $ k $. 
The CP-gates that come after $ {[{\text{block}}]}_{1} $ in Fig. {\hyperref[Fig:1-dim_QFT]{\ref*{Fig:1-dim_QFT}}} were also decomposed by applying the upper row of the relation ({\hyperref[eq:circuit_diagram_Decomposition_C-Rz]{\ref*{eq:circuit_diagram_Decomposition_C-Rz}}}) for the resulting controlled $ R_z $-gate. 
However, for them, the $ R_z $-gate for the control qubit was added on the left only for the CP-gates that have the qubit $ q_0 $ as the control qubit. 
For all other CP-gates, the $ R_z $-gate resulting from the decomposition for the control qubit was added on the right. 
The blocks $ {[{\text{block}}]}_{k} $ for $ k \in \{ 2, \dots, n-1 \} $ are then set respectively as these decompositions of the CP-gates between two H-gates in Fig. {\hyperref[Fig:1-dim_QFT]{\ref*{Fig:1-dim_QFT}}}, where however the first $ R_z $-gate that occurs for the target qubit of the considered CP-gate sequence is excluded. 
In these blocks for $ k > 1 $, the last subblock, given by $ {[{\text{subblock}}]}_{k, n-k} $, is defined in turn as the stated decomposition of the associated CP-gate without the first $ R_z $-gate that occurs for the target qubit in this decomposition. 
The other subblocks $ {[{\text{subblock}}]}_{k, r} , r \in \{ 1, \dots, n-k-1 \} $ in the blocks for $ k > 1 $ are defined then via the stated decomposition of the CP-gate of length $ r $ in the considered CP-gate sequence of Fig. {\hyperref[Fig:1-dim_QFT]{\ref*{Fig:1-dim_QFT}}}, where the first $ R_z $-gate that occurs for the target qubit is excluded but the first $ R_z $-gate that occurs for the target qubit in the decomposition of the next CP-gate is included in turn.

It is to note that under the assumption of all-to-all connectivity, the QFT circuit pattern has a telescopic structure. 
The circuit depth contributions given in Table {\hyperref[Fig_Table:Estimated_computational_resources_scaling_formulas]{\ref*{Fig_Table:Estimated_computational_resources_scaling_formulas}}} are based on the following specific considerations: 
Assuming that the execution of one $ R_z $-gate and two CX-gates takes longer than the execution of one H-gate, the H-gate and the $ R_z $-gate in front of $ {[{\text{block}}]}_k  $ for $ k \in \{ 2, \dots, n-1 \} $ can be moved into $ {[{\text{subblock}}]}_{{k-1}, 2} $, i.e., they can be executed in parallel with the gates of this subblock. 
Further, $ {[{\text{subblock}}]}_{k, r} $ can be absorbed by $ {[{\text{subblock}}]}_{k-1, r+2} $ for $ k \in \{ 2, \dots, n-2 \} $, $ r \in \{ 1, \dots, n-k-1 \} $, since all subblocks consist of two $ R_z $-gate layers and two CX-gate layers. 
Regarding this inclusion of one subblock into another one, it is to note that due to the reversed ordering of $ R_z $-gates and CX-gates in the subblocks of the first block $ {[{\text{block}}]}_{1} $ compared to the subblocks of the other blocks, there is then always the situation of four gate layers, where all have a CX-gate in it. 
However, it is assumed here that the system does not have to wait for the termination of the gate with the highest execution time in a layer until further gates can be applied for qubits for which the respective gates of the layer are already executed. 
For the described incorporation of a subblock into another one, therefore, a count of two $ R_z $-gate layers and two CX-gate layers is taken here and not a count of four CX-gate layers.

For $ n > 3 $ qubits, the circuit depth of this decomposition is then given via the gate layers from the H-gate at the beginning, the H-gate at the end, $ {[{\text{block}}]}_{1} $ and the last subblock from each of the $ n-2 $ remaining blocks $ {[{\text{block}}]}_{k} , k \in \{ 2, \dots, n-1 \} $. 
The latter means $ {[{\text{subblock}}]}_{k, n-k} , k \in \{ 2, \dots, n-1 \} $, consisting of two $ R_z $-gate layers and two CX-gate layers. 
As deduced above, an H-gate layer means a circuit depth contribution of two layers of 1-qubit rotation gates and the $ (n-1) $ subblocks of $ {[{\text{block}}]}_{1} $, which have absorbed all other subblocks, also consist of two $ R_z $-gate layers and  two CX-gate layers. 
In total, this gives
\begin{align}
 4 + (n-1) \cdot 2 + (n-2) \cdot 2 &= 4n - 2
\end{align}
layers of 1-qubit rotation gates and
\begin{align}
 (n-1) \cdot 2 + (n-2) \cdot 2 &= 4n - 6
\end{align}
layers of CX-gates. 
For $ n = 3 $, where just the H-gate and the $ R_z $-gate in front of $ {[{\text{block}}]}_{2} $ can be incorporated in $ {[{\text{subblock}}]}_{1, 2} $, it results $ 6 $ layers of CX-gates and $ 10 $ layers of 1-qubit rotation gates, since the $ R_z $-gates after the second CX-gate of $ {[{\text{subblock}}]}_{1, 1} $ can be executed in parallel to the $ R_z $-gate for the control qubit on the left of the first CX-gate of $ {[{\text{subblock}}]}_{1, 2} $. 
For $ n = 2 $, it results $ 6 $ layers of 1-qubit rotation gates and $ 2 $ layers of CX-gates, whereas for $ n = 1 $, there is just one H-gate, i.e. two layers of 1-qubit rotation gates. 
Hence, the formulas for the circuit depth contributions derived for $ n > 3 $ are also applicable for $ n > 1 $ and moreover, the formula for the circuit depth contribution due to the 1-qubit rotation gates holds also for $ n = 1 $.

The total numbers of 1-qubit rotation gates and CX-gates resulting from the CP-gates can be deduced already via inserting the decomposition of the CP-gate into the standard quantum circuit diagram of the QFT according to Fig. {\hyperref[Fig:1-dim_QFT]{\ref*{Fig:1-dim_QFT}}}, which yields gate sequences of the structure $ {[{\text{block}}]}_{1} $, where there is one such structure applied to $ k $ qubits for $ k \in \{ 2, \dots, n \} $. 
Such a block applied to $ k $ qubits has $ k-1 $ CP-gates, which are decomposed into three $ R_z $-gates and two CX-gates, respectively. 
Using the summation formula
\begin{align}
 {\sum_{k=1}^{n}} ~ k &= {\frac{ n (n+1) }{2}} ,
 \label{eq:Gauss_summation_formula}
\end{align}
this gives a number of
\begin{align}
 {\sum_{k=2}^{n}} (k-1) = {\sum_{k=1}^{n-1}} k &= {\frac{ n (n-1) }{2}}
\end{align}
CP-gates in the QFT and thus a contribution of $ {\frac{3}{2}} n (n-1) $ $ R_z $-gates and $ n (n-1) $ CX-gates to the numbers of native gates.

Finally, there is the optional $n$-qubit-SWAP gate block, consisting of $ \left\lfloor {\frac{n}{2}} \right\rfloor $ SWAP-gates. 
According to \cite{Nielsen_and_Chuang_QC_Book}, a SWAP-gate can be expressed via three CX-gates, yielding a gate count of $ 3 \cdot \left\lfloor {\frac{n}{2}} \right\rfloor $ w.r.t. CX-gates. 
Since these SWAP-gates act on different qubits, they should be executable in parallel, resulting in a circuit depth of $ 3 $ layers of CX-gates.

In the considered setup however, these SWAP-gates can be left away in the QFTs, since they just reorder the qubits in the circuit diagram and thereby the entries of the resulting output state vectors but the setup can be adapted according to another ordering. 
Specifically, the entries of the multiplied diagonal matrix would have to be rearranged and the one-dimensional QFT circuits of the second multi-dimensional QFT in the circuit would have to be inserted upside down in order to account for the changed bit ordering. 
If the algorithm is in fact considered as a stand-alone circuit, as it could be desired for testing, saving these SWAP-gates in particular for the second QFT offers a further advantage: 
The second QFT is directly followed by the measurement operations at the end, which means that the implementation of the measured QFT (MQFT) \cite{Nielsen_and_Chuang_QC_Book, Griffiths_and_Niu_MQFT, Baeumer_et_al_QFT_using_dynamic_circuits, Baeumer_et_al_QFT_using_dynamic_circuits_Sup_Mat} can be applied. 
In the MQFT, the CP-gates can be realized as classically controlled 1-qubit gates. 
In this implementation, the necessary measurement operations are however performed sequentially instead of parallelized but it could allow to complete all required operations for some qubits in shorter time, which could be advantageous if some qubits have shorter coherence times than others. 
Furthermore, it shall be noted that always the approximate QFT (AQFT) \cite{Steijl_Q_algorithms_for_fluid_simulations} can be considered for the QFT, which trades circuit depth for accuracy of the output and is realized by gradually omitting CP-gates w.r.t. the magnitude of their angle arguments.

With regard to the implementation of the DFT, the following comparison can be made: 
Referred to $ 2^{n} $ processed values with $ n \in \mathbb{N} $, the complexity, i.e. the needed number of processing operations, scales like $ 2^{n} ~ n $ for the classical FFT, whereas the QFT has a complexity that scales with $ n^2 $ in the leading order term \cite{Nielsen_and_Chuang_QC_Book, Pfeffer_Multi-dim_QFT_arxiv_v1}. 
This scaling was also derived here in terms of native gates (cf. Table {\hyperref[Fig_Table:Estimated_computational_resources_scaling_formulas]{\ref*{Fig_Table:Estimated_computational_resources_scaling_formulas}}}). 
According to \cite{Koopman_and_Bisseling_Minimizing_communication_in_FFT_arxiv_v2}, the scaling of $ 2^{n} ~ n $ applies also for the runtime of the FFT for a serial execution of it. 
Concerning runtimes, however, the comparison has to made of course for a parallelized execution of the FFT. 
According to \cite{Koopman_and_Bisseling_Minimizing_communication_in_FFT_arxiv_v2}, the FFT can be distributed across several processing units, which results in a scaling like $ \frac{2^{n} ~ n}{P}  $, where the number of processors $ P $ is in general a function of the number of treated values $ 2^n $, i.e. $ P = P(n) $. 
Furthermore, it is to note that in general some effort for the communication between the processors has to be taken into account. 
In \cite{Koopman_and_Bisseling_Minimizing_communication_in_FFT_arxiv_v2}, it was found an optimal $ P $ according to $ P(n) = {\sqrt{ {2^n} }} $, resulting in a runtime scaling of $ {\sqrt{ {2^n} }} ~ n $ for the classical reference. 
Referred to the assumption that all necessary runs of a quantum algorithm that features the QFT are parallelized, the conclusion that can be drawn from this consideration is thus that the computation of the DFT via the QFT should then scale better also w.r.t. runtime.

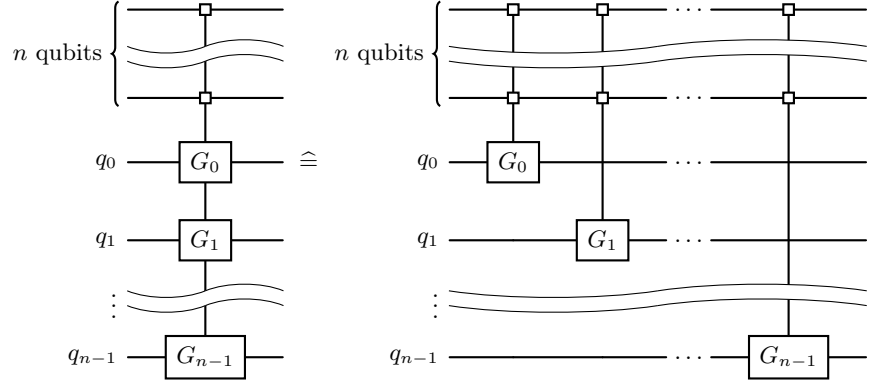
\begin{figure*}[t]
\begin{center}
\tikzset{invisible/.style={fill=none,draw=none,line width=0pt,inner xsep=0pt,inner ysep=0pt}}
\tikzset{transparent/.style={fill=none}}
\begin{align*}
\begin{quantikz}[align equals at=4]
\lstick[3]{$ n $ qubits} & |[operator]| &  \\ 
 \wave&& \\
 & |[operator]| &  \\ 
\lstick[1]{$ {q_{0}} $} & \gate{G_0} &  \\
\lstick[1]{$ {q_{1}} $} & \gate{G_1} &  \\
\lstick[1]{$ \vdots $} \wave&& \\
\lstick[1]{$ {q_{n-1}} $} & \gate{G_{n-1}}\wire[u][6]{q} & 
\end{quantikz}
 ~ &  \widehat{=} ~
\begin{quantikz}[align equals at=4]
\lstick[3]{$ n $ qubits} 
 & |[operator]| & |[operator]| & \ \ldots\ & |[operator]| & \\ 
 \wave&&&&& \\
 & |[operator]| & |[operator]| & \ \ldots\ & |[operator]| & \\ 
 \lstick[1]{$ {q_0} $} & \gate{G_0}\wire[u][3]{q} & & \ \ldots\ & &  \\
 \lstick[1]{$ {q_1} $} &  & \gate{G_1}\wire[u][4]{q} & \ \ldots\ &  & \\
\lstick[1]{$ \vdots $}  \wave&&&&& \\
 \lstick[1]{$ {q_{n-1}} $} &  &  & \ \ldots\ & \gate{G_{n-1}}\wire[u][6]{q} &
\end{quantikz} 
\end{align*}
\end{center}
\captionsetup{justification=raggedright, singlelinecheck=false}
\caption[]{Gate type considered for the resource estimation of the $ [SELECT] $-gate block. 
It is $ {G_k} \in \{ {{\bf{1}}_2}, {{\sigma}_{z}} \} $ and the $ \square $-symbols can be here open or filled circles, i.e. a control conditioned on the 1-qubit state $ | 0 \rangle $ or the 1-qubit state $ | 1 \rangle $, respectively.}
\label{Fig:Circuit_diagram_LCU_variant_considered_gate_for_estimation_of_resources}
\end{figure*}

\subsection{Multiplication by a Diagonal Matrix}

Based on the observations in section {\hyperref[sec:Functionality_tests]{\ref*{sec:Functionality_tests}}}, the number of gates and circuit depth for the implementations of the multiplication step were estimated as follows:

Regarding the LCU variant, the computational costs of the $ [PREP] $-gate blocks were derived already in subsection {\hyperref[subsec:Resource_estimations_state_preparation]{\ref*{subsec:Resource_estimations_state_preparation}}}, so that just the $ [SELECT] $-gate block is considered now. 
For this, it is assumed that the number of qubits doubles due to the needed ancilla qubits, so that a number of qubits according to $ 2n $ is considered for it here. 
For this situation, it is considered a gate of the form shown in Fig. {\hyperref[Fig:Circuit_diagram_LCU_variant_considered_gate_for_estimation_of_resources]{\ref*{Fig:Circuit_diagram_LCU_variant_considered_gate_for_estimation_of_resources}}} where a gate $ G_k $ can be the Z-gate or the $ {\bf{1}} $-gate, which means no applied operation. 
Here, the squares '$ \square $' mean control signs, where these can be open circles according to a conditioning w.r.t. the 1-qubit state $ | 0 \rangle $ or filled circles according to a conditioning w.r.t. the 1-qubit state $ | 1 \rangle $, corresponding to the concept of 'multiplexors' in \cite{Shende_et_al_Synthesis_of_q_circuits}, which means generalized controlled gates. 
The target operations realize the tensor product of $ {\bf{1}}_{2} $- and Z-gates w.r.t. the $ n $ target qubits, where the Z-gate is given in matrix representation by $ {{\sigma}_{z}} = \left( \begin{smallmatrix} 1 & 0 \\ 0 & -1  \end{smallmatrix} \right) = {\text{i}} {R_z}({\theta} = {\pi} ) $. 
Such gate combinations are considered here since it was observed in section {\hyperref[sec:Functionality_tests]{\ref*{sec:Functionality_tests}}} that only these give contributions to the expansion ({\hyperref[eq:LCU_decomposition]{\ref*{eq:LCU_decomposition}}}).

For two matrices, $ 2^n $ such combinations can be formed. 
The number of combinations in which one of the matrices occurs $ k \in \{ 0, \dots, n \} $ times is given by the binomial coefficient
\begin{align}
  \begin{pmatrix}
 n \\ k
 \end{pmatrix} 
 &= \frac{n!}{k! \cdot (n-k)!} ,
\end{align}
where $ ! $ is the factorial. 
If $ k $ is set as the number of Z-gates in such a combination, this means that $ k $ multi-controlled Z-gates with $ n $ control qubits have to be decomposed into native gates. 
Using 
\begin{align}
 \sum_{k=1}^{n} 
 \begin{pmatrix}
 n \\ k
 \end{pmatrix}
 \cdot k 
  &= {2^{n-1}} ~ n ,
\end{align}
this means a total number of
\begin{align}
 \sum_{k=0}^{n} 
 \begin{pmatrix}
 n \\ k
 \end{pmatrix}
 \cdot k 
  &= {2^{n-1}} ~ n ,
\end{align}
such multi-controlled Z-gates for the full $ [SELECT] $-block. 
Since the aim here is merely to provide estimations for what is to expect for transpiling the circuit routines, it was just considered a sequence of decomposed multi-controlled Z-gates and not looked for further possibilities for optimization, like e.g. canceling CX-gates for other arrangements. 
Therefore, the numbers of needed gates and the circuit depth contributions are determined here just by multiplying the corresponding quantities for one multi-controlled Z-gate by this number.

\begin{figure*}[t]
\sbox0{\begin{adjustbox}{width=1.0\textwidth}
\parbox{0.75\textwidth}{
\tikzset{invisible/.style={fill=none,draw=none,line width=0pt,inner xsep=0pt,inner ysep=0pt}}
\tikzset{transparent/.style={fill=none}}
\begin{align*}
\begin{quantikz}[align equals at=4]
\lstick[6]{$ n $ qubits} &[-0.2cm] |[operator]| &[-0.2cm] \ghost{R_z} \\ 
 & |[operator]| & \ghost{R_z} \\ 
 & |[operator]| & \ghost{R_z} \\ 
 & |[operator]| & \ghost{R_z} \\ 
  \wave&& \\
 & |[operator]| & \ghost{R_z} \\ 
 & \gate{Z}\wire[u][6]{q} & \ghost{R_z}
\end{quantikz} 
 ~ & \widehat{=} ~
\begin{quantikz}[align equals at=4]
\lstick[6]{$ n $ qubits} &[-0.2cm] \gate{ R_z } &[-0.2cm] |[operator]| &[-0.2cm] |[operator]| &[-0.2cm] |[operator]| &[-0.2cm] \ \ldots\ &[-0.2cm] |[operator]| &[-0.2cm]  \\ 
 &  & \gate{ R_z }\wire[u][1]{q} & |[operator]| & |[operator]| &  \ \ldots\ & |[operator]| &  \\ 
 &  &  & \gate{ R_z }\wire[u][2]{q} & |[operator]| & \ \ldots\ & |[operator]| &  \\ 
 &  &  &  & \gate{ R_z }\wire[u][3]{q}  & \ \ldots\ & |[operator]| &  \\ 
  \wave&&&&&&& \\
 &  &  &  &   & \ \ldots\ & |[operator]| & \ghost{R_z} \\ 
 &  &  &  &  & \ \ldots\ & \gate{ R_z }\wire[u][6]{q} &
\end{quantikz}
 \\[1.5cm]
 ~ &  \widehat{=} ~
\begin{quantikz}[align equals at=4]
\lstick[6]{$ n $ qubits} &[-0.2cm] \gate{ R_z } &[-0.2cm]  &[-0.2cm]  \ctrl{1} &[-0.2cm]  &[-0.2cm] \ctrl{1} &[-0.2cm]   &[-0.2cm]  &[-0.2cm]  &[-0.2cm] \ctrl{2} &[-0.2cm]  &[-0.2cm]  &[-0.2cm]  &[-0.2cm] \ctrl{2} &[-0.2cm]  &[-0.2cm]  &[-0.2cm]  \ \ldots\ &[-0.2cm]  &[-0.2cm]  \ctrl{6} &[-0.2cm]  \\ 
 &  & \gate{ R_z } & \targ{} & \gate{ R_z } & \targ{}  &  & \ctrl{1} &  &  &  &  \ctrl{1} &  &  &  &  &  \ \ldots\ &  &  &  \\ 
 &  &  &  &  &  & \gate{ R_z } & \targ{} & \gate{ R_z } & \targ{} & \gate{ R_z } &  \targ{}  &  \gate{ R_z }  &  \targ{}  &  & \ctrl{1} & \ \ldots\ &  &  &  \\ 
 &  &  &  &  &  &  &  &  &  &  &  &  &  & \gate{ R_z }  & \targ{} & \ \ldots\ &  &  &  \\ 
  \wave&&&&&&&&&&&&&&&&&&& \\
 &  &  &  &   &  &  &  &  &  &  &  &  &  &  &  &  \ \ldots\ &  &  & \ghost{R_z} \\ 
 &  &  &  &  &  &  &  &  &  &  &  &  &  &  &  &  \ \ldots\ & \gate{ R_z } & \targ{} &
\end{quantikz}
\end{align*}
}
\end{adjustbox}}\usebox0
\captionsetup{justification=raggedright, singlelinecheck=false}
\caption[]{Quantum circuit pattern of the decomposition of a multi-controlled $ Z $-gate considered here. 
The $ \square $-symbols can be here open or filled circles, i.e. a control conditioned on the 1-qubit state $ | 0 \rangle $ or the 1-qubit state $ | 1 \rangle $, respectively.}
\label{Fig:Circuit_diagram_LCU_variant_decomposition_CZ}
\end{figure*}
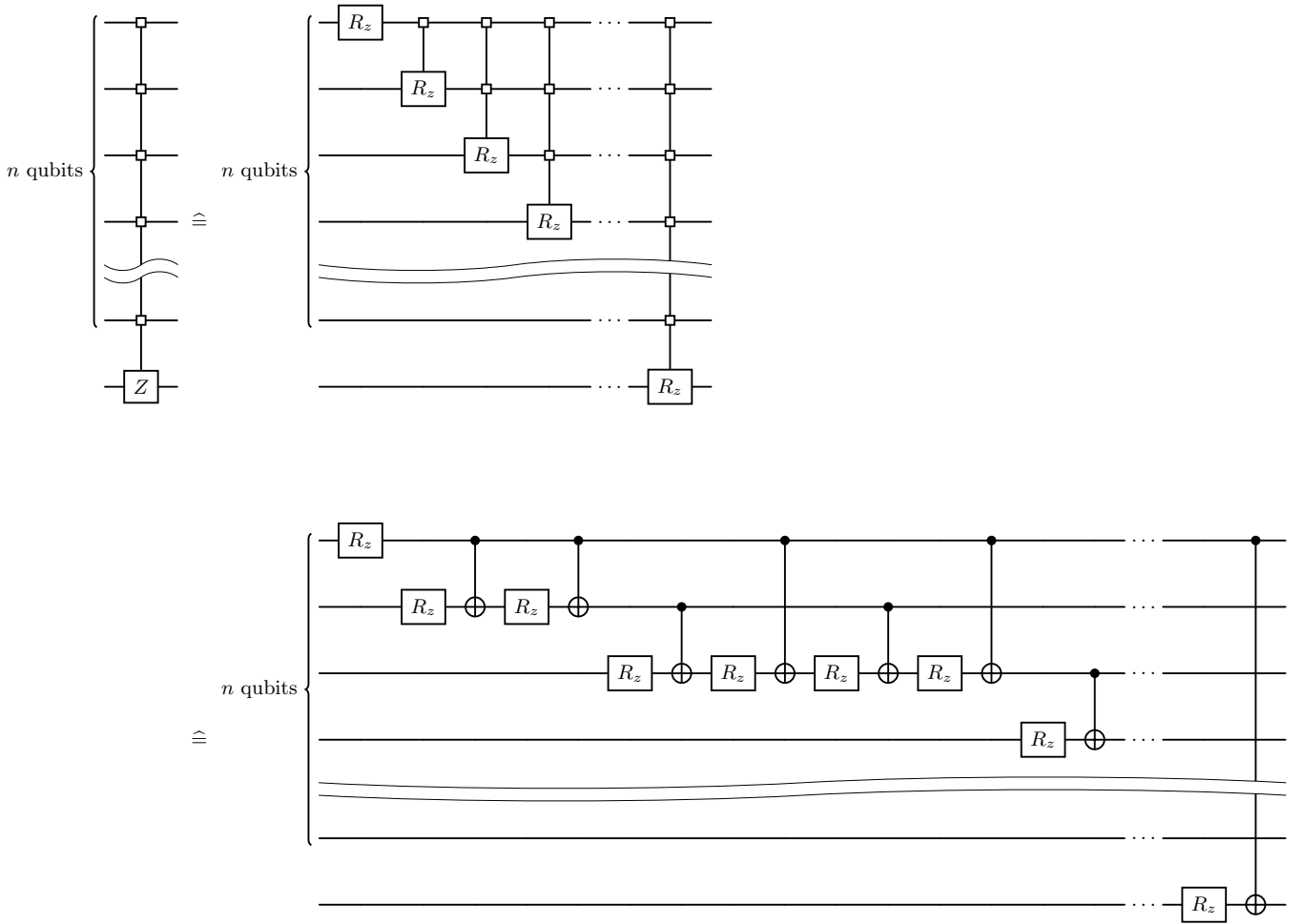

Since the Z-gate corresponds to an $ R_z $-Gate that is multiplied by the scalar $ \text{i} $, the situation of a multi-controlled Z-gate should correspond to Theorem 7 in \cite{Shende_et_al_Synthesis_of_q_circuits}: 
Since the Z-gate is a special case of the P-gate (see previous subsection), a multi-controlled Z-gate corresponds to the multiplication of the state vector formed by all considered qubits by a unitary diagonal matrix, denoted here as a gate with the specification $ \Delta $ in it. 
Based on the relation $ {{\sigma}_{z}} = {\text{i}} {R_z}({\theta} = {\pi} ) $, such a multi-controlled Z-gate is also equivalent to a corresponding multi-controlled $ R_z $-gate and the multiplication of the state vector formed by the control qubits by a unitary diagonal matrix:
\begin{align}
\begin{quantikz}[align equals at=1.5]
& \qwbundle{ } & |[operator]| & \ghost{R_z} \\ 
 &  & \gate{ Z }\wire[u][1]{q} &  \ghost{R_z}
\end{quantikz} 
 ~ &  \widehat{=} ~
\begin{quantikz}[align equals at=1.5]
& \qwbundle{ }  & \gate[2]{ \Delta } & \ghost{R_z}  \\ 
 &  &  & \ghost{R_z}
\end{quantikz}
 \nonumber \\
 ~ & \widehat{=} ~
\begin{quantikz}[align equals at=1.5]
& \qwbundle{ } & |[operator]| & |[operator]| & \ghost{R_z}  \\ 
 &  & \gate{ \text{i} }\wire[u][1]{q} &  \gate{ R_z }\wire[u][1]{q} & \ghost{R_z}
\end{quantikz}
 \nonumber \\
 ~ & \widehat{=} ~ 
\begin{quantikz}[align equals at=1.5]
& \qwbundle{ } & \gate{ \Delta } & |[operator]| & \ghost{R_z} \\ 
 &  &  &  \gate{ R_z }\wire[u][1]{q} & \ghost{R_z}
\end{quantikz}
\label{eq:Circuit_diagram_LCU_variant_Decomposition_multi-controlled_Z-gate_Theorem_7_in_Shende_et_al}
\end{align}
Thus, the decomposition of a multi-controlled Z-gate is obtained by applying ({\hyperref[eq:Circuit_diagram_LCU_variant_Decomposition_multi-controlled_Z-gate_Theorem_7_in_Shende_et_al]{\ref*{eq:Circuit_diagram_LCU_variant_Decomposition_multi-controlled_Z-gate_Theorem_7_in_Shende_et_al}}}) recursively. 
As shown in Fig. {\hyperref[Fig:Circuit_diagram_LCU_variant_decomposition_CZ]{\ref*{Fig:Circuit_diagram_LCU_variant_decomposition_CZ}}}, this yields one $ R_z $-gate with $ k $ control qubits for each $ k \in \{ 0, \dots , n \} $. 
By following the explanations for Theorem 8 in \cite{Shende_et_al_Synthesis_of_q_circuits}, it can be seen that a multi-controlled $ R_z $-gate with $ k > 0 $ control qubits can be decomposed into an alternating sequence of $ R_z $- and CX-gates, where the number of $ R_z $- and CX-gates is $ 2^k $, respectively, which gives accordingly also the respective circuit depth contributions of the decomposition of a single multi-controlled $ R_z $-gate. 
For the pattern resulting for the decomposition of a diagonal gate, it is however to note that the one $ R_z $-gate without control qubits that results in the decomposition of the multi-controlled Z-gate can be executed simultaneously with the $ R_z $-gate resulting at the beginning of the decomposition of the first multi-controlled $ R_z $-gate and the $ R_z $-gates at the beginning of the decompositions of the multi-controlled $ R_z $-gates with $ k > 1 $ control qubits can be moved into this first $ R_z $-gate layer as well. 
Using the finite geometric series
\begin{align}
 \sum_{k=0}^{n} {z^{k}} &= \frac{1 - {z^{n + 1}} }{1-z} , \quad z \neq 1 ,
\end{align}
the number of $ R_z $-gates for the decomposition of a $ Z $-gate with $ n $ control qubits is thus found to be
\begin{align}
 1 + \sum_{k=1}^{n} {2^{k}} &= \sum_{k=0}^{n} {2^{k}} = \frac{1 - {2^{n + 1}} }{1-2}  = {2^{n+1}} - 1
\end{align}
and the number of CX-gates is
\begin{align}
  \sum_{k=0}^{n} {2^{k}} - 1 &= {2^{n+1}} - 2 ,
\end{align}
which is also the circuit depth contribution of the CX-gates. 
The circuit depth contribution due to $ R_z $-gate layers is then given by subtracting the $ n $ absorbed $ R_z $-gate layers from the required number of $ R_z $-gates, i.e., it is $ {2^{n+1}} - 1 - n $. 
Accordingly, for the $ [SELECT] $-block in the presented quantum algorithm, the total number of 1-qubit rotation gates is estimated as
\begin{align}
  {2^{n-1}} ~ n ~ ({2^{n+1}} - 1) &= {2^{2n}} ~ n - {2^{n-1}} ~ n
 \label{eq:Resource_estimations_variant_1_number_1-qubit_rotation_gates}
\end{align}
and their circuit depth contribution as
\begin{align}
  {2^{n-1}} ~ n ~ ({2^{n+1}} - (n+1)) &= {2^{2n}} ~ n - {2^{n-1}} ~ n (n+1) .
 \label{eq:Resource_estimations_variant_1_circuit_depth_1-qubit_rotation_gates}
\end{align}
The total number of CX-gates as well as their circuit depth contribution is
\begin{align}
  {2^{n-1}} ~ n ~ ({2^{n+1}} - 2) &= {2^{2n}} ~ n - {2^{n}} ~ n  .
 \label{eq:Resource_estimations_variant_1_number_and_circuit_depth_CX-gates}
\end{align}

Regarding the implementation variant of the multiplication step via the specific sequence of multi-controlled $ R_y $- and $ R_z $-gates described in subsection {\hyperref[subsubsec:Sequence_of_multi-controlled_rotation_gates]{\ref*{subsubsec:Sequence_of_multi-controlled_rotation_gates}}}, the decomposition of such a gate is given by Theorem 8 of \cite{Shende_et_al_Synthesis_of_q_circuits}, used already before in this subsection w.r.t. $ R_z $. 
However, due to the fact that all of the multi-controlled gates in this implementation variant have the same target qubit, it can be exploited that there are pairs of these gates. 
As described in Theorem 9 of \cite{Shende_et_al_Synthesis_of_q_circuits}, two CX-gates can be canceled in this situation of a pair. 
So, in the decomposition of such a pair w.r.t. $ n $ control qubits, the number of 1-qubit rotation gates is $ 2 \cdot {2^n} $ and the number of CX-gates is $ 2 \cdot {2^n} - 2 $, where these numbers also give the respective circuit depth contributions. 
Specifically, $ {2^n} - 1  $ pairs can be formed for the implementation of the free field situation. 
The subtraction of $ 1 $ results here from the fact that the diagonal entry $ A_{0,0} $ was set to zero. 
For this value, no gate is needed to realize the multiplication by it. 
For the implementation of a periodic system, there is just the gate block of multi-controlled $ R_y $-gates for which such pairs can be formed, which is why here, the number of pairs is $ {2^{n-1}} - 1 $. 
However, in this case, it has to be taken into account that there is then one multi-controlled $ R_y $-gate is this gate block that has no partner and after the gate block of multi-controlled $ R_y $-gates, there is also an additional $ R_z $-gate and an additional multi-controlled $ R_z $-gate (cf. Fig. {\hyperref[Fig:Sequence_multi-controlled_R_gates_q_circuit]{\ref*{Fig:Sequence_multi-controlled_R_gates_q_circuit}}} (b)). 
It is to note that this additional $ R_z $-gate can be merged with the $ R_z $-gate at one end of the decomposition of the additional multi-controlled $ R_z $-gate, so that it yields no contribution to the circuit depth. 
However also this implementation variant, it was not looked for further optimization possibilities due to specific arrangements of the decompositions of the multi-controlled gates.

Referred to the application to $ n + 1 $ qubits, i.e. $ n $ control qubits, this estimation results then in total in an number of
\begin{align}
 {2^{n+1}} \cdot ( {2^n} - 1 ) &= {2^{2n+1}} - {2^{n+1}}
 \label{eq:Resource_estimations_variant_2_number_and_circuit_depth_1-qubit_rotation_gates_free_field}
\end{align}
1-qubit rotation gates and a number of
\begin{align}
 ( {2^{n+1}} - 2 ) \cdot ( {2^n} - 1 ) &= {2^{2n+1}} - {2^{n+2}} + 2
 \label{eq:Resource_estimations_variant_2_number_and_circuit_depth_CX-gates_free_field}
\end{align}
CX-gates for the free field case. 
For the case of periodic BCs, it is obtained
\begin{align}
 {2^{n+1}} \cdot ( {2^{n-1}} - 1 ) + 2 \cdot {2^n} &= {2^{2n}} 
 \label{eq:Resource_estimations_variant_2_number_and_circuit_depth_1-qubit_rotation_gates_per_BCs}
\end{align}
for the number of 1-qubit rotation gates and
\begin{align}
 ({2^{n+1}} - 2) \cdot ( {2^{n-1}} - 1 ) + 2 \cdot {2^n} &= {2^{2n}} - {2^n} + 2 
 \label{eq:Resource_estimations_variant_2_number_and_circuit_depth_CX-gates_per_BCs}
\end{align}
for the number of CX-gates. 
These formulas for the numbers of gates also represent the corresponding circuit depth contributions.

Since the derivations in this subsection are referred to $ n $ control qubits for both implementation variants, the formulas ({\hyperref[eq:Resource_estimations_variant_1_number_1-qubit_rotation_gates]{\ref*{eq:Resource_estimations_variant_1_number_1-qubit_rotation_gates}}}) -  ({\hyperref[eq:Resource_estimations_variant_1_number_and_circuit_depth_CX-gates]{\ref*{eq:Resource_estimations_variant_1_number_and_circuit_depth_CX-gates}}}) and ({\hyperref[eq:Resource_estimations_variant_2_number_and_circuit_depth_1-qubit_rotation_gates_free_field]{\ref*{eq:Resource_estimations_variant_2_number_and_circuit_depth_1-qubit_rotation_gates_free_field}}}) - ({\hyperref[eq:Resource_estimations_variant_2_number_and_circuit_depth_CX-gates_per_BCs]{\ref*{eq:Resource_estimations_variant_2_number_and_circuit_depth_CX-gates_per_BCs}}}) are directly comparable for a given number of grid points $ {N_{\psi}} $ w.r.t. the full computation domain, which is given by $ {N_{\psi}} = {2^n} $. 
For the first variant, the scaling of the leading order term is $ {{N^2_{\psi}}} ~ {\log_{2}({N_{\psi}})} $, i.e. worse than quadratically in the number of grid points, while the scaling of the leading order term of the second variant is quadratically for free field as well as periodic setups. 
Therefore, it can be concluded that the second variant is preferable, which was expectable since the decomposition of a multi-controlled gate scales exponentially with the number of qubits and the second variant involves only $ n + 1 $ qubits in contrast to the first variant, which required $ 2n $ qubits. 
However, a quadratic scaling for performing a multiplication of $ {N_{\psi}} $ values is unacceptable, so that the implementation of the multiplication step has to be significantly improved in future work.

\subsection{Required Number of Runs of the Quantum Algorithm}

Without the incorporation of techniques to infer the relative phases, the considered setup according to Fig. {\hyperref[Fig:Q_alg_modules]{\ref*{Fig:Q_alg_modules}}}, where the modules for implementing the solving procedure for the Poisson equation are directly followed by the final measurement operations, allows only to read out the absolute values of the probability amplitudes. 
To infer this probability distribution with a desired accuracy, a certain number of runs of the quantum algorithm has to be conducted.

Since the presented quantum algorithm prepares a state that encodes the result in amplitude encoding, it is to expect the effort examined for such a situation in \cite{Koesel_et_al_Resource_implications_QCFD_arxiv_v1}, in which an empirical study for a corresponding situation indicates a scaling like $ {2^n} ~ \ln({2^n}) $ w.r.t. the number of qubits $ n $ involved in the circuit. 
However, here, it is to take also into account that there is only a certain probability $ {Pr} $ that the desired matrix multiplication is performed in one run of the quantum algorithm. 
Thus, the necessary number of runs for the consideration of a successful runs has to be multiplied by $ \frac{1}{ {Pr} } $. 
The success probabilities for the first and second algorithm variant considered here were given via the formulas ({\hyperref[eq:Succes_probability_LCU_specific_problem_here]{\ref*{eq:Succes_probability_LCU_specific_problem_here}}}) and ({\hyperref[eq:Success_probability_sequence_of_multi-controlled_R-gates]{\ref*{eq:Success_probability_sequence_of_multi-controlled_R-gates}}}), respectively.

In principle, it can be regarded as a feature that QC allows to parallelize the computation by that many independent runs. 
However, regarding the high values resulting for the number of runs, it seems questionable whether this could in fact be a reasonable strategy concerning resource demands. 
In particular, if the presented implementation of the considered solving procedure for the Poisson equation would be a building block in a time-marching quantum algorithm, such low success probabilities as observed in section {\hyperref[sec:Functionality_tests]{\ref*{sec:Functionality_tests}}} 
would be highly problematic.

\section{Conclusion}
\label{sec:Conclusion}

In this work, we presented a quantum algorithm for solving the Poisson equation for free field conditions in multiple dimensions via the Hockney method. 
In accordance to this, the quantum algorithm is based on a corresponding procedure for numerically solving the situation for periodic BCs via the QFT. 
Besides the QFT, it requires a quantum circuit for multiplying the state vector by a diagonal matrix, for which two variants were compared based on simulated one- and two-dimensional test examples. 
The first variant is the LCU method using tensor products of the Pauli matrices and the second variant is a specific sequence of multi-controlled $ R_y $- and $ R_z $-gates.
While the latter variant demands only one ancilla qubit, it was observed that the LCU variant demanded a number of ancilla qubits up to the number of qubits set for the grid points. 
The lower number of qubits involved for the sequence of multi-controlled rotation gates leads also to lower numbers in the estimations for the numbers of required native gates and the circuit depth, which is why this variant is considered preferable. 
However, it was observed that the success probability of the LCU variant was a factor of around two higher than that for the sequence of multi-controlled rotation gates, where the specific value of this factor varied. 
In general, the success probability depends on the specific problem and converges with increasing resolution. 
W.r.t. one run of the circuit, the DFT steps should be implemented more efficiently via the QFT, compared to a parallelized FFT but the implementations considered for the other necessary steps in the quantum algorithm, i.e. the state preparation and the multiplication by the diagonal matrix, prevent a quantum advantage w.r.t. computation speed.

Concerning the implementation of the algorithm, the most important further work is therefore the search for techniques for lowering the circuit depth of the multiplication step and the state preparation step, e.g. via techniques trading circuit depth for an increased number of qubits. 
Regarding this task, it has to be also inspected for the LCU approach whether there are other matrix expansions that result in less ancilla qubits or a higher success probability. 
Concerning the numerical method, a next step would be to apply the quantum algorithm in the form presented here to other equations, like e.g. the Helmholtz equation. 
However conceptually, the most interesting further step would be to augment this quantum algorithm to realize a QC procedure for the extended solving procedures mentioned in the introduction. 
This means the use of the presented quantum algorithm as a basis for a procedure based on Picard iteration to numerically solve the Poisson equation in the presence of obstacles and the coupling of the presented quantum circuit to a quantum algorithm that requires the solution of the Poisson equation as an input state vector for solving a transport equation.

\acknowledgements

This project was made possible by the DLR Quantum Computing Initiative and the Federal Ministry for Research, Technology and Space; \url{qci.dlr.de/projects/toquaflics}.

H. A. Kösel acknowledges helpful exchange with Johannes Löwe.

Furthermore, we acknowledge the provision of the {\emph{quantikz}}-package \cite{Kay_quantikz_arxiv_v7} for \LaTeX , which was used to create the quantum circuit diagrams in this work, and the software development kit {\emph{Qiskit}} from IBM \cite{qiskit}, which was used for this work in the version 1.2.2 for simulating quantum circuits.

\appendix

\section{Additional Material for Section {\hyperref[sec:Functionality_tests]{\ref*{sec:Functionality_tests}}}}
\label{App_sec:Additional_material_for_functionality_tests}

\subsection{Calculation of Analytical References}
\label{App_subsec:Calculation_Analytical_references}

In this appendix, calculations of analytical reference solutions of the test examples used for one dimension in section {\hyperref[sec:Functionality_tests]{\ref*{sec:Functionality_tests}}} are documented. 
The solutions were obtained based on the relation ({\hyperref[eq:Def_solution_varphi_via_GF_in_general]{\ref*{eq:Def_solution_varphi_via_GF_in_general}}}).

The solution ({\hyperref[eq:Test_example_1D_source_Gauss_analytic_solution_free_field]{\ref*{eq:Test_example_1D_source_Gauss_analytic_solution_free_field}}}) is obtained via evaluating the integral
\begin{align}
 {\int_{-{\infty}}^{+{\infty}}} {e^{ -{\frac{{( {\xi} - {x_s} )}^2}{{\sigma}^2}} }} \cdot {\frac{|{\xi} - x|}{2}} ~ {d{\xi}} 
\end{align}
Using
\begin{align}
 & {\int_{a}^{b}} {e^{ -{\frac{{( {\xi} - {x_s} )}^2}{{\sigma}^2}} }} ~ {\xi} ~ {d{\xi}} \nonumber \\
 & ~ = {\int_{a}^{b}} {e^{ -{\frac{{( {\xi} - {x_s} )}^2}{{\sigma}^2}} }} ~ ({\xi} - {x_s}) ~ {d{\xi}}  + {x_s} {\int_{a}^{b}} {e^{ -{\frac{{( {\xi} - {x_s} )}^2}{{\sigma}^2}} }} ~ {d{\xi}} \nonumber \\
 & ~ = {\int_{ -{{\left( {\frac{a - {x_s}}{\sigma}} \right)}^2} }^{ -{{\left( {\frac{b - {x_s}}{\sigma}} \right)}^2} }} {e^{w}} ~ {\left( -{\frac{{\sigma}^2}{2}} \right)} ~ {d{w}} \nonumber \\
 & ~ \quad + {x_s} {\int_{a}^{b}} {e^{ -{\frac{{( {\xi} - {x_s} )}^2}{{\sigma}^2}} }} ~ {d{\xi}} \nonumber \\
 & ~ = { -{\frac{{\sigma}^2}{2}} } \cdot { \Bigl( {e^{ -{\frac{{( {b} - {x_s} )}^2}{{\sigma}^2}} }} - {e^{ -{\frac{{( {a} - {x_s} )}^2}{{\sigma}^2}} }} \Bigr) } \nonumber \\
 & ~ \quad + {x_s} {\int_{a}^{b}} {e^{ -{\frac{{( {\xi} - {x_s} )}^2}{{\sigma}^2}} }} ~ {d{\xi}} 
\end{align}
and
\begin{align}
 & {\int_{a}^{b}} {e^{ -{\frac{{( {\xi} - {x_s} )}^2}{{\sigma}^2}} }} ~ {d{\xi}} \nonumber \\
 & ~ = {\int_{ \frac{{a} - {x_s}}{\sigma} }^{ \frac{{b} - {x_s}}{\sigma} }} {e^{ -{y^2} }} {\sigma} ~ {d{y}} \nonumber \\
 & ~ = {\frac{{\sigma}{\sqrt{\pi}}}{2}} \cdot \left( {\mbox{erf}\left( { \frac{{b} - {x_s}}{\sigma} } \right)} - {\mbox{erf}\left( { \frac{{a} - {x_s}}{\sigma} } \right)} \right) ,
\end{align}
where
\begin{align}
 \mbox{erf}(x) := {\frac{2}{{\sqrt{\pi}}}} {\int_{0}^{x}} {e^{-{t^2}}} ~ dt
 \label{eq:Def_error_function}
\end{align}
is the error function and $ a, b \in \mathbb{R} $, it is found for \mbox{$ a \leq x \leq b $}:
\begin{align}
 & {\int_{a}^{b}} {e^{ -{\frac{{( {\xi} - {x_s} )}^2}{{\sigma}^2}} }} \cdot {\frac{|{\xi} - x|}{2}} ~ {d{\xi}} \nonumber \\
 & ~ = {\frac{1}{2}} \cdot \Bigl(  -{\int_{a}^{x}} {e^{ -{\frac{{( {\xi} - {x_s} )}^2}{{\sigma}^2}} }} \cdot ({\xi} - x) ~ d{\xi} \nonumber \\ 
 & ~ \quad +  {\int_{x}^{b}} {e^{ -{\frac{{( {\xi} - {x_s} )}^2}{{\sigma}^2}} }} \cdot ({\xi} - x) ~ d{\xi} \Bigr)  \nonumber \\
 & ~ = {\frac{1}{2}} \cdot  \Bigl(  -{\int_{a}^{x}} {e^{ -{\frac{{( {\xi} - {x_s} )}^2}{{\sigma}^2}} }} \cdot {\xi} ~ d{\xi}  + x {\int_{a}^{x}} {e^{ -{\frac{{( {\xi} - {x_s} )}^2}{{\sigma}^2}} }} ~ d{\xi}  \nonumber \\
 & ~ \quad +  {\int_{x}^{b}} {e^{ -{\frac{{( {\xi} - {x_s} )}^2}{{\sigma}^2}} }} \cdot {\xi} ~ d{\xi}  - x {\int_{x}^{b}} {e^{ -{\frac{{( {\xi} - {x_s} )}^2}{{\sigma}^2}} }} ~ d{\xi}  \Bigr) \nonumber \\
 & ~ = {\frac{1}{2}} \cdot  \biggl(  {\frac{{\sigma}^2}{2}} \cdot \left( {e^{ -{\frac{{( {x} - {x_s} )}^2}{{\sigma}^2}} }} - {e^{ -{\frac{{( {a} - {x_s} )}^2}{{\sigma}^2}} }}  \right) \nonumber \\
 & ~ \quad - {\frac{{\sigma}^2}{2}} \cdot \Bigl( {e^{ -{\frac{{( {b} - {x_s} )}^2}{{\sigma}^2}} }} - {e^{ -{\frac{{( {x} - {x_s} )}^2}{{\sigma}^2}} }}  \Bigr) \nonumber \\
 & ~ \quad  + x{\sigma}{\frac{\sqrt{\pi}}{2}} \cdot \left( \mbox{erf}\left( {\frac{x - {x_s}}{\sigma}} \right) - \mbox{erf}\left( {\frac{a - {x_s}}{\sigma}} \right)  \right) \nonumber \\
 & ~ \quad - x{\sigma}{\frac{\sqrt{\pi}}{2}} \cdot \left( \mbox{erf}\left( {\frac{b - {x_s}}{\sigma}} \right) - \mbox{erf}\left( {\frac{x - {x_s}}{\sigma}} \right)  \right) \nonumber \\
 & ~ \quad - {x_s}{\sigma}{\frac{\sqrt{\pi}}{2}} \cdot \left( \mbox{erf}\left( {\frac{x - {x_s}}{\sigma}} \right) - \mbox{erf}\left( {\frac{a - {x_s}}{\sigma}} \right)  \right) \nonumber \\
 & ~ \quad + {x_s}{\sigma}{\frac{\sqrt{\pi}}{2}} \cdot \left( \mbox{erf}\left( {\frac{b - {x_s}}{\sigma}} \right) - \mbox{erf}\left( {\frac{x - {x_s}}{\sigma}} \right)  \right)    \biggr) 
 \label{eq:Analytical_references_integral_Gauss_times_free_field_GF_general}
\end{align}
For free space, it has to be considered $ a \longrightarrow -{\infty} $, $ b \longrightarrow +{\infty} $, for which it results
\begin{align}
 \mbox{erf}\left( {\frac{a - {x_s}}{\sigma}} \right) & \longrightarrow -1 ,  \\
 \mbox{erf}\left( {\frac{b - {x_s}}{\sigma}} \right) & \longrightarrow 1 ,  \\
 {e^{ -{\frac{{( {b} - {x_s} )}^2}{{\sigma}^2}} }}  & \longrightarrow 0 ,  \\
 {e^{ -{\frac{{( {a} - {x_s} )}^2}{{\sigma}^2}} }} & \longrightarrow 0 
\end{align}
and hence
\begin{align}
 & {\int_{-{\infty}}^{+{\infty}}} {e^{ -{\frac{{( {\xi} - {x_s} )}^2}{{\sigma}^2}} }} \cdot {\frac{|{\xi} - x|}{2}} ~ {d{\xi}} \nonumber \\
 & ~ = {\frac{{\sigma}^2}{2}} {e^{ -{\frac{{( {x} - {x_s} )}^2}{{\sigma}^2}} }} + {\frac{{\sigma}{\sqrt{\pi}}}{2}} (x - {x_s}) \cdot \mbox{erf}\left( {\frac{x - {x_s}}{\sigma}} \right) .
 \label{eq:Analytical_references_integral_free_field}
\end{align}

The solution ({\hyperref[eq:Test_example_1D_source_Gauss_analytic_solution_per_BCs]{\ref*{eq:Test_example_1D_source_Gauss_analytic_solution_per_BCs}}}) follows from evaluating the integral
\begin{align}
 & {\int_{a}^{b}} \Bigl( {e^{ -{\frac{{( {\xi} - {x_s} )}^2}{{\sigma}^2}} }} - {\overline{S}} \Bigr) \nonumber \\
 & \quad \cdot \left( -{\frac{1}{2L}} \right) \cdot \left( {{({\xi} - x)}^2} - {L{|{\xi} - x|}} + {\frac{{L^2}}{6}} \right) ~ {d{\xi}} .
 \label{eq:Analytical_references_integral_per_BCs_general}
\end{align}
One part of this integral was already determined in ({\hyperref[eq:Analytical_references_integral_Gauss_times_free_field_GF_general]{\ref*{eq:Analytical_references_integral_Gauss_times_free_field_GF_general}}}). 
The remaining part w.r.t. $ {e^{ -{\frac{{( {\xi} - {x_s} )}^2}{{\sigma}^2}} }} $ is given by
\begin{align}
 & {\int_{{a}}^{{b} }}   {e^{ -{\frac{{( {\xi} - {x_s} )}^2}{{\sigma}^2}} }}   \cdot \left( -{\frac{1}{2L}} \right) \cdot \left( {{({\xi} - x)}^2}  + {\frac{{L^2}}{6}} \right) ~ {d{\xi}} \nonumber \\
 & ~ = -{\frac{1}{2L}} \left( {x^2} + {\frac{L^2}{6}} \right) {\int_{{a}}^{{b} }} {e^{ -{\frac{{( {\xi} - {x_s} )}^2}{{\sigma}^2}} }} ~ {d{\xi}} \nonumber \\
 & ~ \quad + {\frac{x}{L}} {\int_{{a}}^{{b} }} {e^{ -{\frac{{( {\xi} - {x_s} )}^2}{{\sigma}^2}} }} \cdot {\xi} ~ {d{\xi}} \nonumber \\
 & ~ \quad - {\frac{1}{2L}} {\int_{{a}}^{{b} }} {e^{ -{\frac{{( {\xi} - {x_s} )}^2}{{\sigma}^2}} }} \cdot {{\xi}^2} ~ {d{\xi}} .
 \label{eq:Analytical_references_integral_Gauss_times_rest_of_GF_for_per_BCs_general}
\end{align}
The only integral that has not yet been evaluated is
\begin{align}
 {\int_{{a}}^{{b} }} {e^{ -{\frac{{( {\xi} - {x_s} )}^2}{{\sigma}^2}} }} \cdot {{\xi}^2} ~ {d{\xi}} &= {\int_{{a} - {x_s} }^{{b} - {x_s} }} {e^{ -{\frac{{ {\tilde{\xi}}}^2}{{\sigma}^2}} }} \cdot {{\left( {\tilde{\xi}} + {x_s} \right)}^2} ~ {d{{\tilde{\xi}}}} ,
\end{align}
for which further
\begin{align}
 & {\int_{{a} - {x_s} }^{{b} - {x_s} }} {e^{ -{\frac{{ {\tilde{\xi}}}^2}{{\sigma}^2}} }} \cdot {{ {\tilde{\xi}}  }^2} ~ {d{{\tilde{\xi}}}} \nonumber \\
 & ~ = -{{\left[  {\frac{ {{\sigma}^2} {\tilde{\xi}} ~ {e^{ -{\frac{{ {\tilde{\xi}}}^2}{{\sigma}^2}} }}  }{2}}  \right]}_{{a} - {x_s}}^{{b} - {x_s}}} + {\frac{{\sigma}^2}{2}} {\int_{{a} - {x_s} }^{{b} - {x_s} }} {e^{ -{\frac{{ {\tilde{\xi}}}^2}{{\sigma}^2}} }} ~ {d{{\tilde{\xi}}}} \nonumber \\
 & ~ = -{\frac{ {{\sigma}^2} (b - {x_s}) ~ {e^{ -{\frac{{ (b - {x_s})}^2}{{\sigma}^2}} }}  }{2}}  + {\frac{ {{\sigma}^2} (a - {x_s}) ~ {e^{ -{\frac{{ (a - {x_s}) }^2}{{\sigma}^2}} }}  }{2}}  \nonumber \\
 & ~ \quad + {\frac{{{\sigma}^3} {\sqrt{\pi}} }{4}} { \left( {\mbox{erf}{\left( {\frac{ b - {x_s} }{{\sigma}}} \right)}} - {\mbox{erf}{\left( {\frac{ a - {x_s} }{{\sigma}}} \right)}} \right) } ,
\end{align}
\begin{align}
 & 2 {x_s} {\int_{{a} - {x_s} }^{{b} - {x_s} }} {e^{ -{\frac{{ {\tilde{\xi}}}^2}{{\sigma}^2}} }} \cdot {{ {\tilde{\xi}}  }} ~ {d{{\tilde{\xi}}}}  = -{x_s}{{\sigma}^2}  { \left( {e^{ -{\frac{{ (b - {x_s})}^2}{{\sigma}^2}} }} - {e^{ -{\frac{{ (a - {x_s})}^2}{{\sigma}^2}} }} \right) }
\end{align}
and
\begin{align}
 & {x_s^2} {\int_{{a} - {x_s} }^{{b} - {x_s} }} {e^{ -{\frac{{ {\tilde{\xi}}}^2}{{\sigma}^2}} }} ~ {d{{\tilde{\xi}}}} \nonumber \\
 & ~ = {x_s^2} {\frac{{\sigma}{\sqrt{\pi}}}{2}} { \left( {\mbox{erf}{\left( {\frac{ b - {x_s} }{{\sigma}}} \right)}} - {\mbox{erf}{\left( {\frac{ a - {x_s} }{{\sigma}}} \right)}} \right) }
\end{align}
follows. 
For the considered case of a Gauss distribution for which the position of the peak $ x_s $ is in the center of the interval according to $ a = {x_0} = {x_s} - {\frac{L}{2}} $, $ b = {x_0} + L = {x_s} + {\frac{L}{2}} $, it is 
\begin{align}
 \mbox{erf}\left( {\frac{a - {x_s}}{\sigma}} \right) &= \mbox{erf}\left( {{ -{\frac{L}{2{\sigma}}}}} \right) \nonumber \\
 &= -\mbox{erf}\left( {{ {\frac{L}{2{\sigma}}}}} \right) = \mbox{erf}\left( {\frac{b - {x_s}}{\sigma}} \right) , \\
 {e^{ -{\frac{{ (a - {x_s}) }^2}{{\sigma}^2}} }} &= {e^{ -{\frac{{ (b - {x_s}) }^2}{{\sigma}^2}} }} .
\end{align}
For the part
\begin{align}
 {\int_{{x_s} - {\frac{L}{2}}}^{{x_s} + {\frac{L}{2}}}} {e^{ -{\frac{{( {\xi} - {x_s} )}^2}{{\sigma}^2}} }} \cdot {\frac{|{\xi} - x|}{2}} ~ {d{\xi}} 
\end{align}
of ({\hyperref[eq:Analytical_references_integral_per_BCs_general]{\ref*{eq:Analytical_references_integral_per_BCs_general}}}), it follows then the same result as for ({\hyperref[eq:Analytical_references_integral_free_field]{\ref*{eq:Analytical_references_integral_free_field}}}) and the part ({\hyperref[eq:Analytical_references_integral_Gauss_times_rest_of_GF_for_per_BCs_general]{\ref*{eq:Analytical_references_integral_Gauss_times_rest_of_GF_for_per_BCs_general}}}) results in
\begin{align}
 & {\int_{{a}}^{{b} }}   {e^{ -{\frac{{( {\xi} - {x_s} )}^2}{{\sigma}^2}} }}   \cdot \left( -{\frac{1}{2L}} \right) \cdot \left( {{({\xi} - x)}^2}  + {\frac{{L^2}}{6}} \right) ~ {d{\xi}} \nonumber \\
 & ~ = -{\frac{1}{2L}} \left( {x^2} + {\frac{L^2}{6}} \right) {\sqrt{\pi}} {\sigma} ~ {\mbox{erf}{\left( {\frac{L}{2{\sigma}}} \right)}} \nonumber \\
 & ~ \quad  + {\frac{x}{L}} {\sqrt{\pi}} {\sigma} {x_s} ~ {\mbox{erf}{\left( {\frac{L}{2{\sigma}}} \right)}}
   + {\frac{{{\sigma}^2}}{4}} {e^{ -{\frac{L^2}{{(2{\sigma})}^2}} }} \nonumber \\
 & ~ \quad  - {\frac{{{\sigma}^3}{\sqrt{\pi}}}{4L}} ~ {\mbox{erf}{\left( {\frac{L}{2{\sigma}}} \right)}}
   - {\frac{{{x_s^2}{\sigma}}{\sqrt{\pi}}}{2L}} ~ {\mbox{erf}{\left( {\frac{L}{2{\sigma}}} \right)}}
\end{align}

For the part of ({\hyperref[eq:Analytical_references_integral_per_BCs_general]{\ref*{eq:Analytical_references_integral_per_BCs_general}}}) that involves the constant $ \overline{S} $, it is
\begin{align}
 & {\int_{{a}}^{ {b} }} {\overline{S}}  \cdot {\frac{1}{2L}} \cdot \left( {{({\xi} - x)}^2} - {L{|{\xi} - x|}} + {\frac{{L^2}}{6}} \right) ~ {d{\xi}} \nonumber \\
 & ~ =   {\frac{ {\overline{S}} }{2L}} \cdot \biggl( {\frac{1}{3}} \left( {{b}^3} -{{a}^3}  \right) - x \left( {{b}^2} -{{{a}}^2}  \right) \nonumber \\
 & ~ \quad + (b - a) \cdot \left( {x^2} + {\frac{L^2}{6}} \right)  \biggr) - {\frac{ {\overline{S}} }{2}} \cdot {\int_{{a}}^{b}} {|{\xi} - x|} ~ {d{\xi}} 
\end{align}
with
\begin{align}
  {\int_{{a}}^{{b}}} {|{\xi} - x|} ~ {d{\xi}} &= {\int_{{a} - x}^{{b} - x}} {|{\tilde{\xi}}|} ~ d{\tilde{\xi}} \nonumber \\
  &=  {{\left[ {\frac{ {\tilde{\xi}} \cdot |{\tilde{\xi}}| }{2}} \right]}^{{{b} - x}}_{{{a} - x}}} \nonumber \\
  &= {\frac{1}{2}} \left( ({b - x}) \cdot |{b - x}| - ({a - x}) \cdot |{a - x}| \right) .
\end{align}
Moreover, the specific choice of the integral bounds yields for $ {\overline{S}} $ itself
\begin{align}
 {\overline{S}} &= {\int_{{x_0}}^{{x_0} + L }} {e^{ -{\frac{{( {x} - {x_s} )}^2}{{\sigma}^2}} }} ~ dx \nonumber \\
 &= {\frac{{\sigma}{\sqrt{\pi}}}{2}} \left( {\mbox{erf}\left( {\frac{ {x_0} + L - {x_s} }{ {\sigma} }} \right)} - {\mbox{erf}\left( {\frac{ {x_0} - {x_s} }{ {\sigma} }} \right)} \right) \nonumber \\
 &= {{{\sigma}{\sqrt{\pi}}}} ~ {\mbox{erf}\left( {\frac{ L }{ 2{\sigma} }} \right)} .
 \label{eq:Test_example_1D_Gauss_per_BCs_expression_integral_over_S}
\end{align}

\subsection{Further Simulation Results}
\label{App_subsec:Further_simulation_results}

This appendix contains complementary results for the discussion in section {\hyperref[sec:Functionality_tests]{\ref*{sec:Functionality_tests}}} that were obtained for the alternative implementation of the multiplication step. 
Since the plots for these cases are very similar to corresponding cases in section {\hyperref[sec:Functionality_tests]{\ref*{sec:Functionality_tests}}} they were just included here for completeness.

\begin{figure*}[t]
\begin{minipage}[c]{0.325\textwidth}
\centering
\begin{tikzpicture}
\draw (0,0) node[inner sep=0]{\includegraphics[height=1.0\linewidth]{Poisson1D_source_2ndDerivGauss_px7.pdf}};
\end{tikzpicture}
\end{minipage}
\hfill
\begin{minipage}[c]{0.325\textwidth}
\centering
\begin{tikzpicture}
\draw (0,0) node[inner sep=0]{\includegraphics[height=1.0\linewidth]{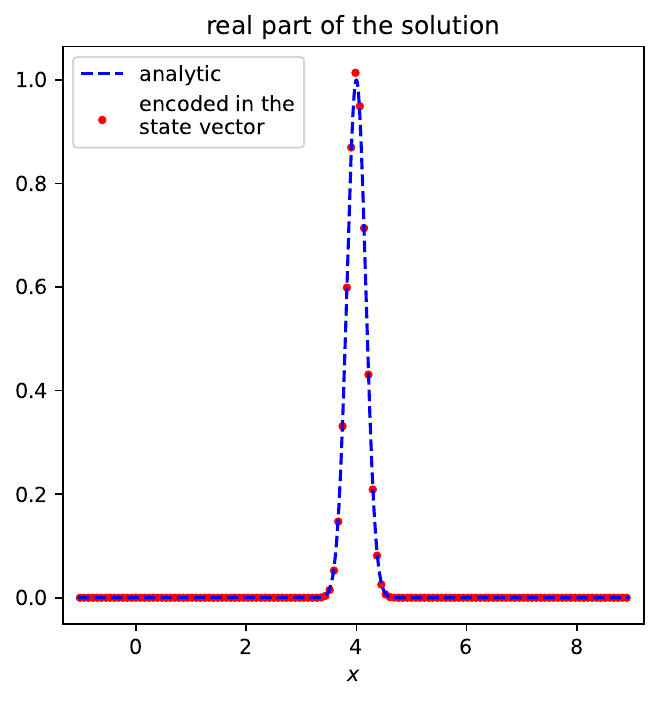}};
\end{tikzpicture}
\end{minipage}
\hfill
\begin{minipage}[c]{0.325\textwidth}
\centering
\begin{tikzpicture}
\draw (0,0) node[inner sep=0]{\includegraphics[height=1.0\linewidth]{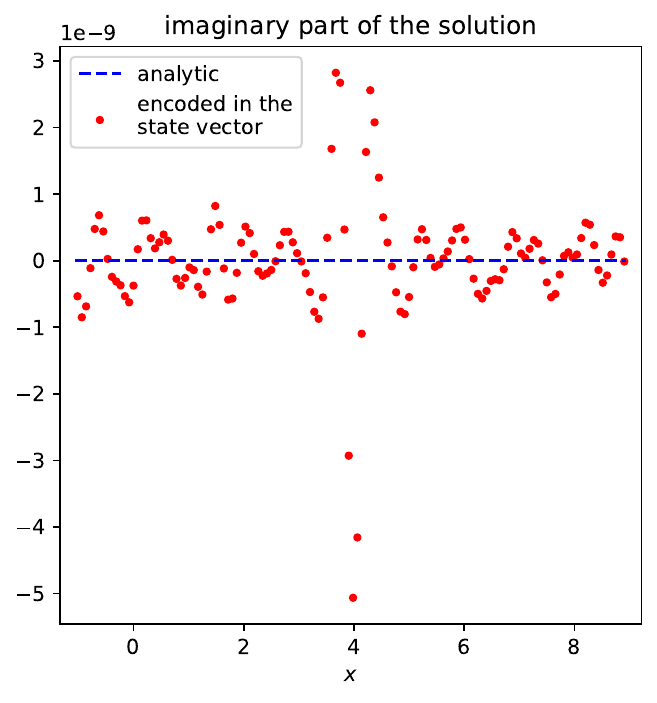}};
\end{tikzpicture}
\end{minipage}
\hfill
\begin{minipage}[c]{0.325\textwidth}
\centering
\begin{tikzpicture}
\draw (0,0) node[inner sep=0]{\includegraphics[height=1.0\linewidth]{Poisson1D_source_Gauss_px7.pdf}};
\end{tikzpicture}
\end{minipage}
\hfill
\begin{minipage}[c]{0.325\textwidth}
\centering
\begin{tikzpicture}
\draw (0,0) node[inner sep=0]{\includegraphics[height=1.0\linewidth]{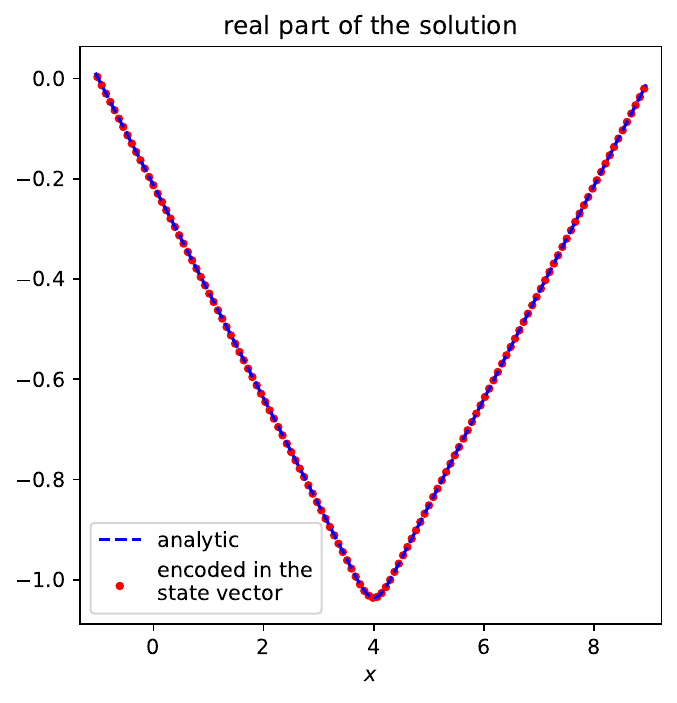}};
\end{tikzpicture}
\end{minipage}
\hfill
\begin{minipage}[c]{0.325\textwidth}
\centering
\begin{tikzpicture}
\draw (0,0) node[inner sep=0]{\includegraphics[height=1.0\linewidth]{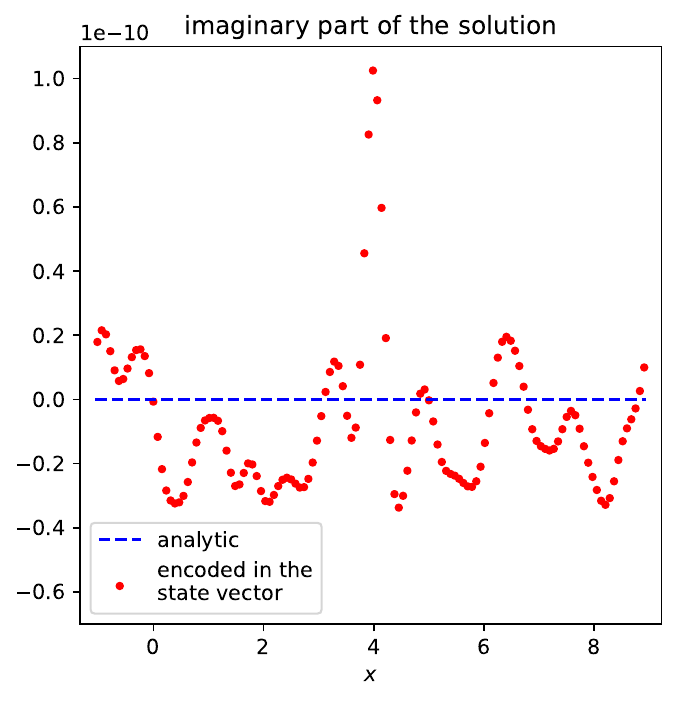}};
\end{tikzpicture}
\end{minipage}
\captionsetup{justification=raggedright, singlelinecheck=false}
\caption[]{Simulation results obtained via the second implementation variant for the test examples considered for free field conditions in one dimension, using $ 7 $ qubits to realize a resolution of $ 128 $ discretization points for the source domain of extension $ 10 $. 
In total $ 9 $ qubits were needed in these cases. 
The panels in the upper row are referred to the source ({\hyperref[eq:Test_example_1D_source_Gauss_2nd_deriv]{\ref*{eq:Test_example_1D_source_Gauss_2nd_deriv}}}) and those in the lower row to the source ({\hyperref[eq:Test_example_1D_source_Gauss]{\ref*{eq:Test_example_1D_source_Gauss}}}). 
The panels on the left show the sampled source distribution. 
The panels in the middle and on the right show the real and imaginary parts of the computed solution values $ \varphi $ in the computational subdomain of the source, respectively. 
The respective analytical solutions for these two cases are given by ({\hyperref[eq:Test_example_1D_source_Gauss_2nd_deriv_analytic_solution]{\ref*{eq:Test_example_1D_source_Gauss_2nd_deriv_analytic_solution}}}) and ({\hyperref[eq:Test_example_1D_source_Gauss_analytic_solution_free_field]{\ref*{eq:Test_example_1D_source_Gauss_analytic_solution_free_field}}}), where the additative constant was adjusted to match the real part of the computed solution value at the most left grid point of the source domain.}
\label{Fig:Test_examples_1D_free_field_variant_2_all}
\end{figure*}

\begin{figure*}[t]
\begin{minipage}[c]{0.325\textwidth}
\centering
\begin{tikzpicture}
\draw (0,0) node[inner sep=0]{\includegraphics[height=1.0\linewidth]{Poisson1D_source_Gauss_px8.pdf}};
\end{tikzpicture}
\end{minipage}
\hfill
\begin{minipage}[c]{0.325\textwidth}
\centering
\begin{tikzpicture}
\draw (0,0) node[inner sep=0]{\includegraphics[height=1.0\linewidth]{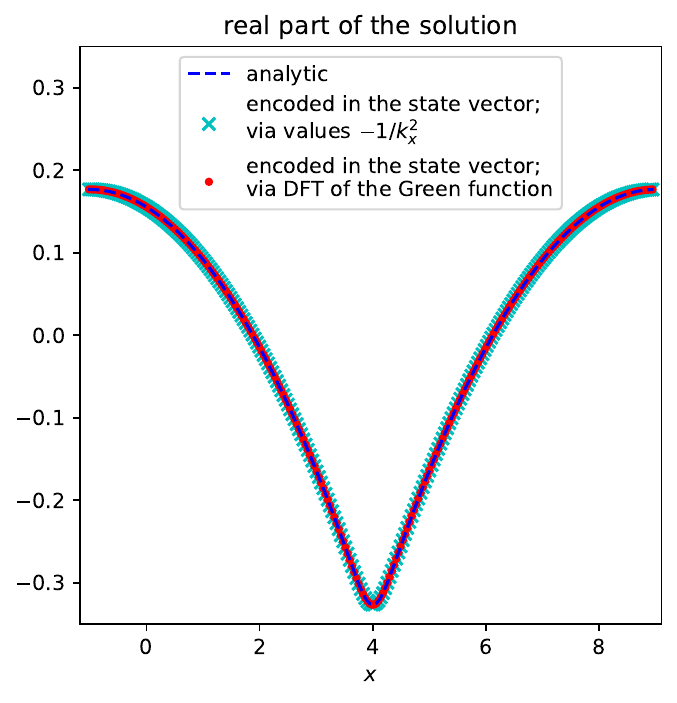}};
\end{tikzpicture}
\end{minipage}
\hfill
\begin{minipage}[c]{0.325\textwidth}
\centering
\begin{tikzpicture}
\draw (0,0) node[inner sep=0]{\includegraphics[height=1.0\linewidth]{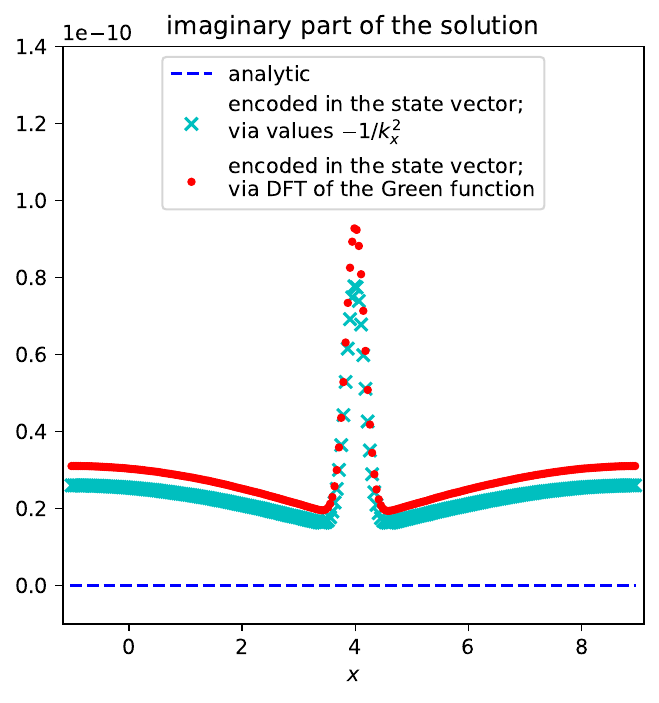}};
\end{tikzpicture}
\end{minipage}
\captionsetup{justification=raggedright, singlelinecheck=false}
\caption[]{Simulation results obtained via the second implementation variant for the test example considered for periodic BCs in one dimension, using $ 8 $ qubits to realize a resolution of $ 256 $ discretization points for the source domain of extension $ 10 $. 
In total $ 9 $ qubits were needed in this case. 
The panel on the left shows the sampled source distribution ({\hyperref[eq:Test_example_1D_source_Gauss_minus_offset]{\ref*{eq:Test_example_1D_source_Gauss_minus_offset}}}). 
The panels in the middle and on the right show the real and imaginary parts of the computed solution values $ \varphi $, respectively, where the computation of the multiplication step was done using the sampled analytical values ({\hyperref[eq:sequence_sampled_analytical_values]{\ref*{eq:sequence_sampled_analytical_values}}}) as well as the DFT of the corresponding Green function in position representation ({\hyperref[eq:GF_1D_per_BCs]{\ref*{eq:GF_1D_per_BCs}}}). 
The analytical solution for this case is given by  ({\hyperref[eq:Test_example_1D_source_Gauss_analytic_solution_per_BCs]{\ref*{eq:Test_example_1D_source_Gauss_analytic_solution_per_BCs}}}), where the additative constant was adjusted to match the real part of the computed solution value at the most left grid point of the source domain for the use of the DFT of the Green function for the multiplied matrix.}
\label{Fig:Test_example_Gauss_1D_per_BCs_variant_2_all}
\end{figure*}

\begin{figure*}[t]
\begin{minipage}[c]{0.240\textwidth}
\centering
\begin{tikzpicture}
\draw (0,0) node[inner sep=0]{\includegraphics[height=1.0\linewidth]{2DHockney_Poisson2D_source_Gauss2ndDeriv_px4_py4.pdf}};
\end{tikzpicture}
\end{minipage}
\hfill
\begin{minipage}[c]{0.240\textwidth}
\centering
\begin{tikzpicture}
\draw (0,0) node[inner sep=0]{\includegraphics[height=1.0\linewidth]{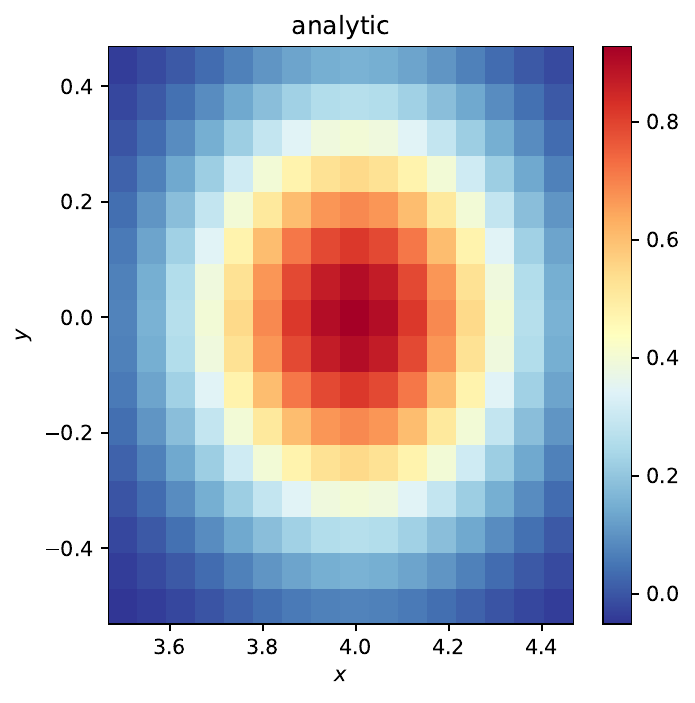}};
\end{tikzpicture}
\end{minipage}
\hfill
\begin{minipage}[c]{0.240\textwidth}
\centering
\begin{tikzpicture}
\draw (0,0) node[inner sep=0]{\includegraphics[height=1.0\linewidth]{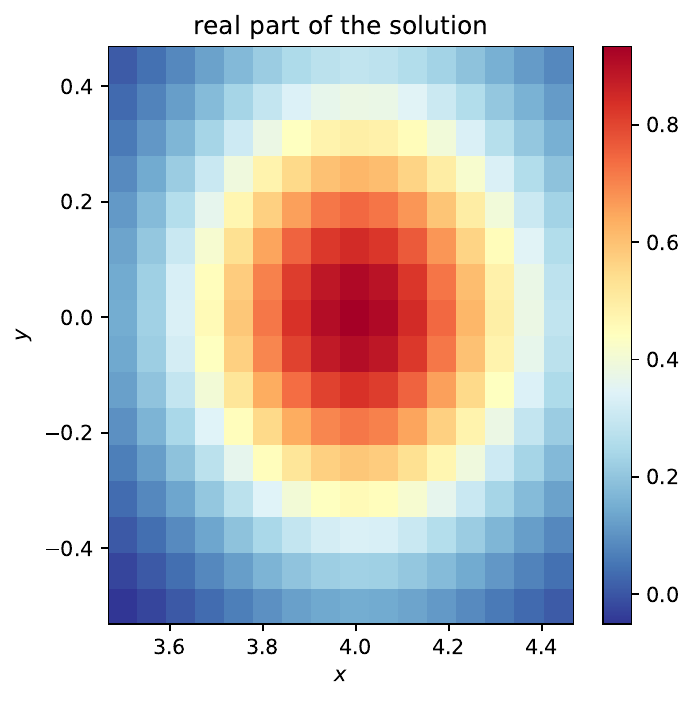}};
\end{tikzpicture}
\end{minipage}
\hfill
\begin{minipage}[c]{0.240\textwidth}
\centering
\begin{tikzpicture}
\draw (0,0) node[inner sep=0]{\includegraphics[height=1.0\linewidth]{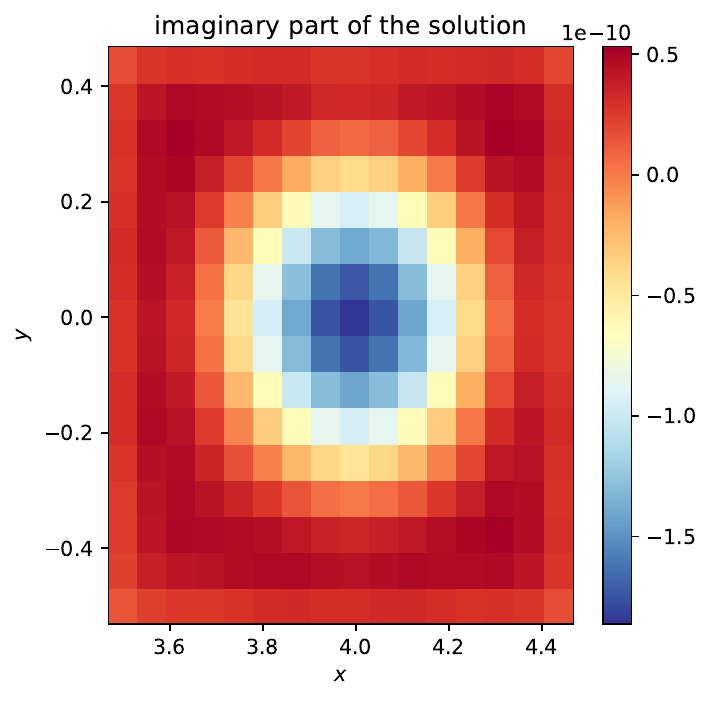}};
\end{tikzpicture}
\end{minipage}
\hfill
\begin{minipage}[c]{0.240\textwidth}
\centering
\begin{tikzpicture}
\draw (0,0) node[inner sep=0]{\includegraphics[height=1.0\linewidth]{2DHockney_Poisson2D_source_delta_px4_py4.pdf}};
\end{tikzpicture}
\end{minipage}
\hfill
\begin{minipage}[c]{0.240\textwidth}
\centering
\begin{tikzpicture}
\draw (0,0) node[inner sep=0]{\includegraphics[height=1.0\linewidth]{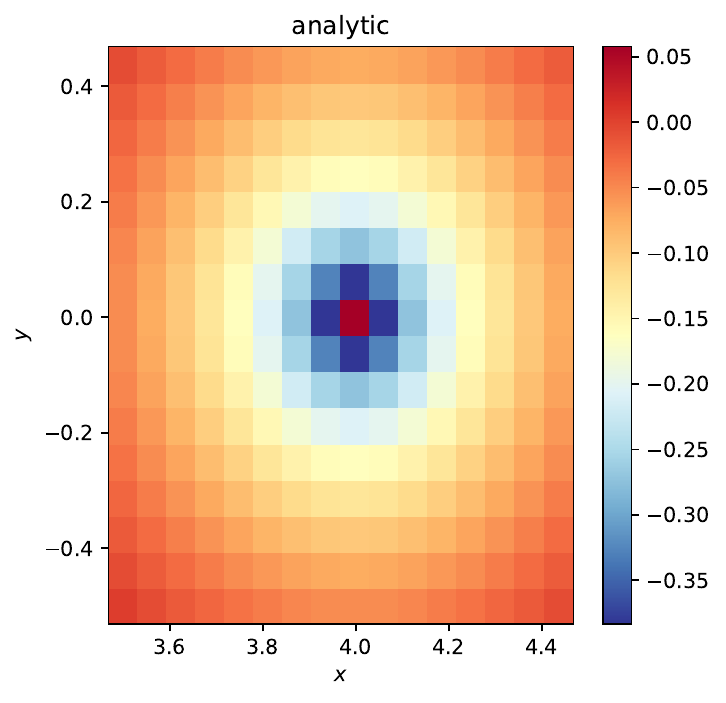}};
\end{tikzpicture}
\end{minipage}
\hfill
\begin{minipage}[c]{0.240\textwidth}
\centering
\begin{tikzpicture}
\draw (0,0) node[inner sep=0]{\includegraphics[height=1.0\linewidth]{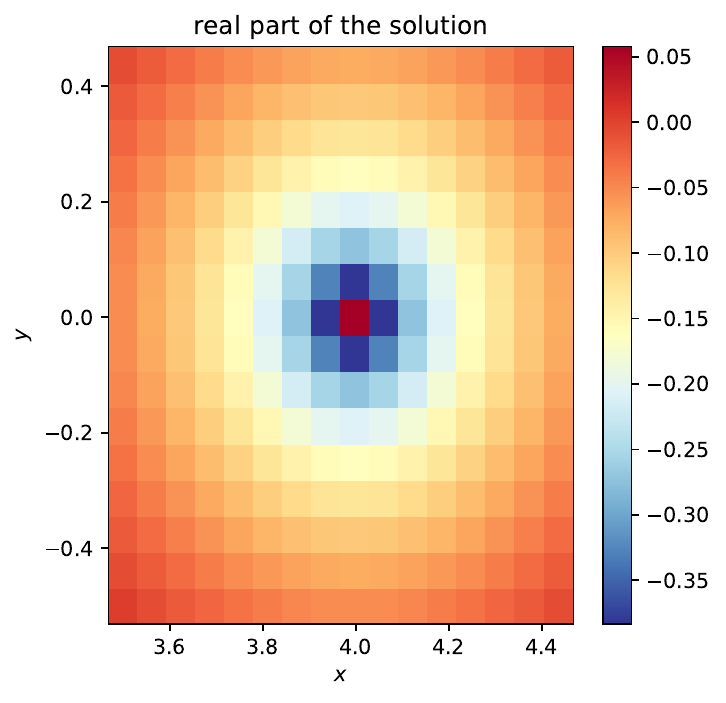}};
\end{tikzpicture}
\end{minipage}
\hfill
\begin{minipage}[c]{0.240\textwidth}
\centering
\begin{tikzpicture}
\draw (0,0) node[inner sep=0]{\includegraphics[height=1.0\linewidth]{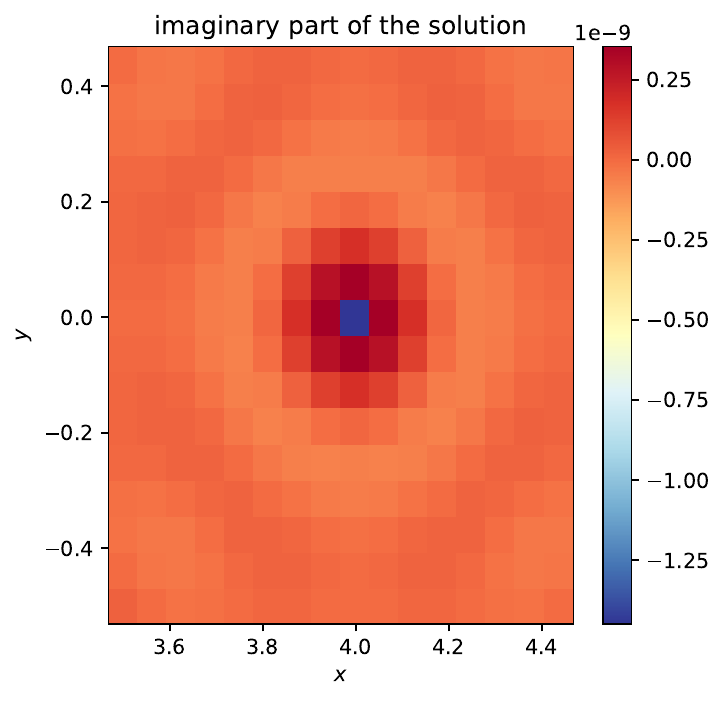}};
\end{tikzpicture}
\end{minipage}
\captionsetup{justification=raggedright, singlelinecheck=false}
\caption[]{Simulation results obtained via the second implementation variant for the test examples considered for free field conditions in two dimensions, using $ 8 $ qubits to realize a resolution of $ 16 $ discretization points per dimension for the source domain of extension $ 1 $ in each dimension. 
In total $ 11 $ qubits were needed in these cases. 
The panels in the upper row are referred to the source ({\hyperref[eq:Test_example_2D_source_Gauss_2nd_deriv]{\ref*{eq:Test_example_2D_source_Gauss_2nd_deriv}}}) and those in the lower row to the source ({\hyperref[eq:Test_example_2D_source_approximated_delta-distribution]{\ref*{eq:Test_example_2D_source_approximated_delta-distribution}}}). 
The panels most left and on the left in the middle show the sampled source distribution and sampled analytical solution, respectively. 
The panels on the right in the middle and most right show the real and imaginary parts of the computed solution values $ \varphi $ in the computational subdomain of the source, respectively. 
The respective analytical solutions for these two cases are given by ({\hyperref[eq:Test_example_2D_source_Gauss_2nd_deriv_analytic_solution]{\ref*{eq:Test_example_2D_source_Gauss_2nd_deriv_analytic_solution}}}) and ({\hyperref[eq:Test_example_2D_source_approximated_delta-distribution_analytic_solution]{\ref*{eq:Test_example_2D_source_approximated_delta-distribution_analytic_solution}}}), where the additative constant was adjusted to match the real part of the computed solution value at the grid point in the lower left corner of the source domain.}
\label{Fig:Test_examples_2D_free_field_variant_2_all}
\end{figure*}

\newpage
\renewcommand{\refname}{Bibliography}


\addcontentsline{toc}{section}{Bibliography}
\begin{thebibliography}{99}

\bibitem{Steijl_and_Barakos_Q_algs_for_CFD}
R. Steijl and G. N. Barakos, 
{{\emph{Parallel evaluation of quantum algorithms for computational fluid dynamics}}}, 
{\href{https://doi.org/10.1016/j.compfluid.2018.03.080}{{Computers and Fluids {\bf{173}}, 22-28 (2018)}}}.

\bibitem{Steijl_Q_algorithms_for_fluid_simulations}
R. Steijl, 
{{\emph{Quantum Algorithms for Fluid Simulations}}}, 
{\href{http://dx.doi.org/10.5772/intechopen.86685}{{http://dx.doi.org/10.5772/intechopen.86685 (2019)}}}.

\bibitem{Hockney_and_Eastwood_Computer_simulation_Book}
R. W. Hockney and J. W. Eastwood, 
{\href{https://doi.org/10.1201/9780367806934}{\emph{Computer Simulation Using Particles}}}  
(Taylor \& Francis, 1988).

\bibitem{Hu_et_al_Simulation_of_turbulent_boundary_layer_wall_pressure_fluctuations}
N. Hu, N. Reiche, and R. Ewert,  
{{\emph{Simulation of turbulent boundary layer wall pressure fluctuations via Poisson equation and synthetic turbulence}}}, 
{\href{https://doi.org/10.1017/jfm.2017.448}{{J. Fluid Mech. {\bf{826}}, 421-454 (2017)}}}.

\bibitem{Delfs_Lecture_notes}
J. Delfs, 
{\href{https://www.dlr.de/en/as/about-us/departments/technical-acoustics/lectures-in-aeroacoustics-tu-braunschweig}{\emph{Basics of Aeroacoustics}}} 
(lecture notes, 2023).

\bibitem{Wang_et_al_Q_alg_for_Poisson_eq}
S. Wang, Z. Wang, W. Li, L. Fan, Z. Wei, and Y. Gu, 
{{\emph{Quantum fast Poisson solver: The algorithm and complete and modular circuit design}}}, 
{\href{https://doi.org/10.1007/s11128-020-02669-7}{{Quantum Inf. Process. {\bf{19}}, 170 (2020)}}}.

\bibitem{Succi_et_al_Review}
S. Succi, W. Itani, K. Sreenivasan, and R. Steijl, 
{{\emph{Quantum computing for fluids: Where do we stand?}}}, 
{\href{https://iopscience.iop.org/article/10.1209/0295-5075/acfdc7}{{EPL {\bf{144}}, 10001 (2023)}}}. 

\bibitem{Nielsen_and_Chuang_QC_Book}
M. A. Nielsen and I. L. Chuang, 
{\href{https://doi.org/10.1017/CBO9780511976667}{\emph{Quantum Computation and Quantum Information}}}  
(10th anniversary edition, Cambridge University Press, 2010).

\bibitem{Pfeffer_Multi-dim_QFT_arxiv_v1}
P. Pfeffer, 
{{\emph{Multidimensional Quantum Fourier Transformation}}}, 
{\href{https://doi.org/10.48550/arXiv.2301.13835}{{arXiv:2301.13835v1 [quant-ph] (2023)}}}.

\bibitem{Cao_et_al_Q_alg_for_Poisson_eq}
Y. Cao, A. Papageorgiou, I. Petras, J. Traub, and S. Kais, 
{{\emph{Quantum algorithm and circuit design solving the Poisson equation}}}, 
{\href{https://iopscience.iop.org/article/10.1088/1367-2630/15/1/013021}{{New J. Phys. {\bf{15}}, 013021 (2013)}}}.

\bibitem{Harrow_et_al_HHL-alg}
A. W. Harrow, A. Hassidim, and S. Lloyd, 
{{\emph{Quantum Algorithm for Linear Systems of Equations}}}, 
{\href{https://doi.org/10.1103/PhysRevLett.103.150502}{{Phys. Rev. Lett. {\bf{103}}, 150502 (2009)}}}.

\bibitem{Mandelt_Buxade_et_al_Hybrid_Newton_method_arxiv_v1}
M. Mandelt Buxad\'{e}, S. Langer, and P. Bekemeyer, 
{{\emph{Solving Nonlinear Partial Differential Equations via a Hybrid Newton Method Using Quantum Linear System Solver}}}, 
{\href{https://doi.org/10.48550/arXiv.2603.23258}{{arXiv:2603.23258v1 [quant-ph] (2026)}}}.

\bibitem{Budinski_Q_alg_by_streamfunction_vorticity_LBM}
L. Budinski, 
{{\emph{Quantum algorithm for the Navier Stokes equations by using the streamfunction vorticity formulation and the lattice Boltzmann method}}}, 
{\href{https://doi.org/10.48550/arXiv.2103.03804}{{arXiv:2103.03804v2 [quant-ph] (2022)}}}.

\bibitem{Over_et_al_Q_alg_for_AD_eq}
P. Over, S. Bengoechea, P. Brearley, S. Laizet, and T. Rung, 
{{\emph{Quantum algorithm for the advection-diffusion equation by direct block encoding of the time-marching operator}}}, 
{\href{https://doi.org/10.1103/d8hb-fv93}{{Phys. Rev. A {\bf{112}}, L010401 (2025)}}}.

\bibitem{Bengoechea_et_al_Q_algs_BCs}
S. Bengoechea, P. Over, and T. Rung, 
{{\emph{Quantum Time-Marching Algorithms for Solving Linear Transport Problems Including Boundary Conditions}}}, 
{\href{https://doi.org/10.1002/nme.70326}{{International Journal for Numerical Methods in Engineering {\bf{127}}, No. 8, e70326 (2026)}}}.

\bibitem{Steijl_VKI_lecture_notes_2026}
R. Steijl, 
{\emph{Quantum CFD Approaches}} 
(lecture notes of the lecture course {\emph{Introduction to Quantum Computing in Fluid Dynamics (STO-AVT-377)}}, held at the von Karman Institute of Fluid Dynamics, 6 - 10 July 2026).

\bibitem{Dewitte_et_al_QFT_harmonic_balance_solver}
L. Dewitte, J. Roland, and F. Eulitz, 
{{\emph{Application of the quantum Fourier transform in a harmonic balance solver for Burgers’ equation}}}, 
{\href{https://doi.org/10.1016/j.compfluid.2025.106619}{{Computers and Fluids {\bf{295}}, 106619 (2025)}}}.

\bibitem{qiskit}
A. Javadi-Abhari, M. Treinish, K. Krsulich, C. J. Wood, J. Lishman, J. Gacon, S. Martiel, P. D. Nation, L. S. Bishop, A. W. Cross, B. R. Johnson, and J. M. Gambetta,
{{\emph{Quantum computing with Qiskit}}}, 
{\href{https://doi.org/10.48550/arXiv.2405.08810}{{arXiv:2405.08810v3 [quant-ph] (2024)}}}.

\bibitem{Marshall_Periodic_Green_functions}
S. L. Marshall,  
{{\emph{A periodic Green function for calculation of
coloumbic lattice potentials}}}, 
{\href{https://iopscience.iop.org/article/10.1088/0953-8984/12/21/304}{{J. Phys.: Condens. Matter {\bf{12}}, 4575–4601 (2000)}}}.

\bibitem{Shende_et_al_Synthesis_of_q_circuits}
V. V. Shende, S. S. Bullock, and I. L. Markov, 
{{\emph{Synthesis of quantum-logic circuits}}}, 
{\href{https://doi.org/10.1109/TCAD.2005.855930}{{IEEE Transactions on Computer-Aided Design of Integrated Circuits and Systems {\bf{25}}, 6, pp. 1000-1010 (2006)}}}.

\bibitem{Childs_and_Wiebe_LCU}
A. M. Childs and N. Wiebe, 
{{\emph{Hamiltonian simulation using linear combinations of unitary operations}}}, 
{\href{https://doi.org/10.26421/QIC12.11-12-1}{{Quantum Information and Computation {\bf{12}}, No. 11 \& 12 (2012)}}}.

\bibitem{Koesel_et_al_Resource_implications_QCFD_arxiv_v1}
H. A. Kösel, R. Ewert, and J. W. Delfs, 
{{\emph{Resource Implications of Different Encodings for Quantum Computational Fluid Dynamics}}}, 
{\href{https://doi.org/10.48550/arXiv.2604.05577}{{arXiv:2604.05577v1 [quant-ph] (2026)}}}.

\bibitem{Griffiths_and_Niu_MQFT}
R. B. Griffiths and C.-S. Niu, 
{{\emph{Semiclassical Fourier Transform for Quantum Computation}}}, 
{\href{https://doi.org/10.1103/PhysRevLett.76.3228}{{Phys. Rev. Lett. {\bf{76}}, 3228 (1996)}}}.

\bibitem{Baeumer_et_al_QFT_using_dynamic_circuits}
E. Bäumer, V. Tripathi, A. Seif, D. Lidar, and D. S. Wang, 
{{\emph{Quantum Fourier Transform Using Dynamic Circuits}}}, 
{\href{https://doi.org/10.1103/PhysRevLett.133.150602}{{Phys. Rev. Lett. {\bf{133}}, 150602 (2024)}}}.

\bibitem{Baeumer_et_al_QFT_using_dynamic_circuits_Sup_Mat}
Supplemental Material of E. Bäumer, V. Tripathi, A. Seif, D. Lidar, and D. S. Wang,  
{{\emph{Quantum Fourier Transform Using Dynamic Circuits}}}, 
{\href{https://journals.aps.org/prl/abstract/10.1103/PhysRevLett.133.150602#supplemental}{{Phys. Rev. Lett. {\bf{133}}, 150602 (2024)}}}.

\bibitem{Koopman_and_Bisseling_Minimizing_communication_in_FFT_arxiv_v2}
T. Koopman and R. H. Bisseling, 
{{\emph{Minimizing communication in the multi-dimensional FFT}}}, 
{\href{https://doi.org/10.48550/arXiv.2203.11795}{{arXiv:2203.11795v2 [cs.DC] (2023)}}}.

\bibitem{Kay_quantikz_arxiv_v7}
A. Kay,
{{\emph{Tutorial on the Quantikz Package}}}, 
{\href{https://arxiv.org/abs/1809.03842v7}{{arXiv:1809.03842v7 [quant-ph] (2023)}}}.


\end{thebibliography}
\end{document}